\documentclass[aps,prx,twocolumn,longbibliography,superscriptaddress,showpacs]{revtex4-1}
\usepackage{graphicx}
\usepackage{latexsym}
\usepackage{amssymb}
\usepackage{amsmath}
\usepackage{amsfonts}
\usepackage{mathtools}
\usepackage{dsfont}
\usepackage{bm}
\usepackage{multirow}
\usepackage{color}
\usepackage{xcolor}
\newcommand{\ket}[1]{\vert#1\rangle}
\newcommand{\bra}[1]{\langle#1\vert}
\usepackage[colorlinks=true, citecolor={blue!80!black}, urlcolor={blue!50!black}, linkcolor = {blue!80!black}]{hyperref}

\usepackage[percent]{overpic}
 \usepackage[utf8]{inputenc}

\renewcommand{\Im}{\mathop{\mathrm{Im}}}
\newcommand{\SO}{\mathrm{SO}}
\newcommand{\rO}{\mathrm{O}}

\newcommand{\U}{\mathrm{U}}

\newcommand{\CLR}[1]{\mathrm{Cl}_{#1}(\mathbb{R})}

\renewenvironment{psmallmatrix}
  {\left(\begin{smallmatrix}}
  {\end{smallmatrix}\right)}

\DeclareMathOperator{\Pf}{Pf}

\newcommand{\tr}{\intercal}

\DeclareMathAlphabet{\mathbbold}{U}{bbold}{m}{n}

\DeclareMathOperator{\Tr}{Tr} 

\def\GL{{\rm GL}}

\def\i{\textbf{i}}
\def\T{\textsf{T}}

\def\ft{\mathfrak{t}}
\def\Jp{\mathcal{J}_{\rm pb}}
\def\chit{\chi_{\rm t}}

\definecolor{forestgreen}{rgb}{0.13, 0.55, 0.13}
\definecolor{gr}{rgb}{0,0.82,0.18}

\def\hgamma{\hat{\gamma}}
\def\hrho{\hat{\rho}}
\def\tH{t_\mathcal{H}}
\def\hupsilon{\hat{\eta}}
\newcommand{\tG}[1]{\Gamma^{(#1)}}

\begin{document}

\title{Criticality and entanglement in non-unitary quantum circuits \\ and tensor networks of non-interacting fermions}

\author{Chao-Ming Jian}
\affiliation{Department of Physics, Cornell University, Ithaca, New York 14853, USA}

\author{Bela Bauer}
\affiliation{Microsoft Station Q, University of California, Santa Barbara, California 93106-6105, USA}

\author{Anna Keselman}
\affiliation{Microsoft Station Q, University of California, Santa Barbara, California 93106-6105, USA}
\affiliation{Kavli Institute for Theoretical Physics, University of California, Santa Barbara, CA 93106-4030}

\author{Andreas W. W.  Ludwig}
\affiliation{Department of Physics, University of California, Santa Barbara, CA 93106, USA}

\date{\today}

\begin{abstract}

Models for non-unitary quantum dynamics, such as quantum circuits that include projective measurements, have recently been shown to exhibit rich quantum critical behavior. There are many complementary perspectives on this behavior. For example, there is a known correspondence between $d$-dimensional local non-unitary quantum circuits and tensor networks on a $D=(d+1)$-dimensional lattice.
Here, we show that in the case of systems of non-interacting fermions, there is furthermore a full
correspondence between non-unitary circuits in $d$ spatial dimensions and unitary
non-interacting fermion problems with static Hermitian
Hamiltonians in $D=(d+1)$ spatial dimensions. This provides a powerful new perspective for understanding entanglement phases and critical behavior exhibited by non-interacting circuits. Classifying the symmetries of the corresponding non-interacting Hamiltonian, we show that
a large class of random circuits, including the most generic circuits with randomness in space and time, are in correspondence with Hamiltonians with static spatial disorder in the ten Altland-Zirnbauer symmetry classes.
We find the criticality that is known to occur in all of these classes to be the origin of the critical entanglement properties of the corresponding random non-unitary circuit. To exemplify this, we numerically study the quantum states at the boundary of Haar-random Gaussian fermionic tensor networks of dimension $D=2$ and $D=3$.
We show that the most general such tensor network ensemble corresponds to a unitary problem of non-interacting fermions with static disorder in
Altland-Zirnbauer symmetry class DIII, which for both $D=2$ and $D=3$ is known to exhibit a stable critical metallic phase.
Tensor networks and corresponding random non-unitary circuits
in the other nine Altland-Zirnbauer symmetry classes can be obtained
from the DIII case by implementing Clifford 
algebra extensions for classifying spaces.

\end{abstract}

\maketitle

\tableofcontents

\section{Introduction}

Inspired by the fundamental question of how equilibrium statistical mechanics emerges in closed quantum systems~\cite{srednicki1994chaos} and in which case such an equilibrium may not occur~\cite{Basko06a,Basko06b,Oganesyan07}, the last decade has seen an explosion of research on the dynamics of quantum many-body systems far from equilibrium. In many cases, it has turned out that many-body entanglement is a useful way of characterizing the behavior of the system~\cite{Eisert2010}.

Fruitful settings to study the dynamics of quantum entanglement include the evolution 
following a quantum quench~\cite{Calabrese_2005,fagotti2008evolution,Hubeny_2007,Eisler_2007}, the dynamics under random unitary evolution~\cite{nahum2017quantum,nahum2018operator,zhou2019emergent}, and, starting with Refs.~\onlinecite{li2018quantum,skinner2019measurement}, the evolution under non-unitary circuits~\cite{li2019measurement,pai2019localization,chan2019unitary,jian2020measurement,nahum2020entanglement,bao2020theory,gullans2020dynamical,gullans2020scalable,lang2020entanglement,Choi_2020}.
The latter have been found to exhibit rich phenomenology, including entanglement transitions -- quantum phase transitions that are primarily characterized by a change in entanglement structure.
In particular, Refs.~\onlinecite{li2018quantum,skinner2019measurement} have shown that the states that emerge at late times from unitary circuits interspersed with projective measurement can exhibit a phase transition between volume and area law entanglement.


A closely related family of models are tensor networks~\footnote{For a recent review of tensor networks and their use as variational algorithms, see, e.g., Ref.~\onlinecite{orus2019tensor}.}. Indeed, any quantum circuit can, for a given set of measurement outcomes, be interpreted as the contraction of a tensor network; conversely, using the polar decomposition, the many-body transfer matrix describing the contraction of a tensor network can be re-interpreted as a quantum circuit consisting of unitary and non-unitary evolution. The universal behaviors exhibited in non-unitary circuits and tensor networks are therefore closely related to each other, and entanglement transitions similar to those in random measurement circuits have been observed in tensor networks~\cite{vasseur2019entanglement}.

\begin{figure}
    \centering
    \includegraphics[width=\columnwidth]{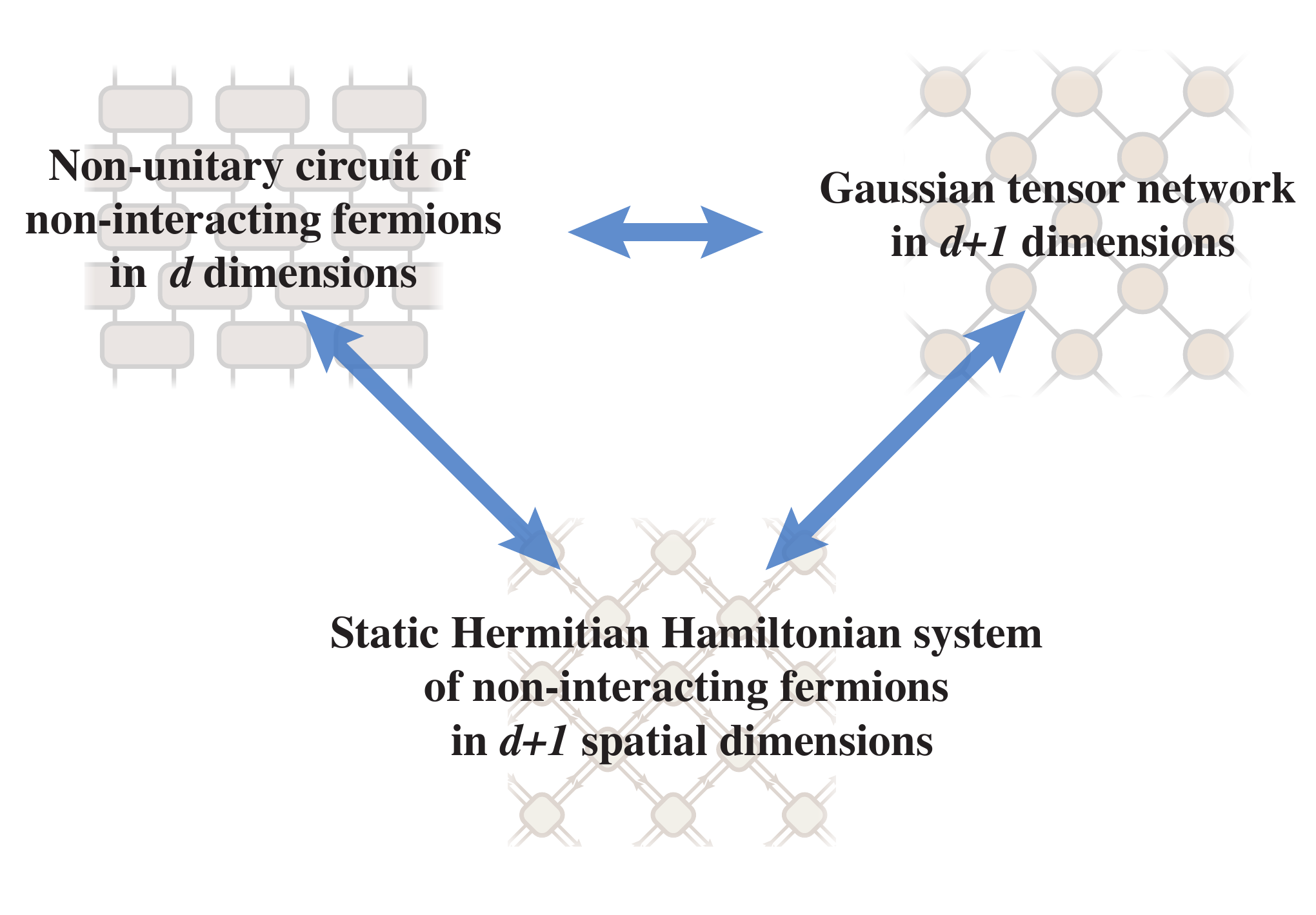}
    \caption{A diagram showing the correspondences among non-unitary quantum circuits of non-interacting fermions in $d$ dimensions, Gaussian tensor networks in $D=d+1$ dimensions and static Hermitian Hamiltonian systems of non-interacting fermions in $D=d+1$ spatial dimensions.}
    \label{fig:correspondences}
\end{figure}

As we will show in this paper, this correspondence between tensor networks in $D=(d+1)$ dimensions and non-unitary quantum circuits acting on quantum systems in $d$ dimensions can be extended further when considering the case of non-interacting fermions. In this setting, not only is there a correspondence between tensor networks in $D=(d+1)$ dimensions and non-unitary quantum circuits acting on quantum systems in $d$ dimensions, but there is a further correspondence with static Hermitian Hamiltonian problems 
(and thus unitary time evolutions)
in $D=(d+1)$ spatial dimensions (see Fig.~\ref{fig:correspondences}). This will allow us to relate critical phenomena that we numerically observe in random tensor networks for non-interacting fermions to critical behavior in random non-unitary circuits and random Hamiltonian systems. In the following, we will refer to quantum circuits acting on $d$-dimensional quantum systems as $d$-dimensional quantum circuits.

The class of tensor networks that forms part of this correspondence are Gaussian fermionic tensor networks (GTNs)~\cite{kraus2010,evenbly2010entanglement,schuch2012,dubail2015,haegeman2013,fishman2015,evenbly2016,haegeman2018,jahn2019holography,schuch2019}. Such Gaussian fermionic tensor networks are constructed from Gaussian fermionic states, which are the most general class of states that obey Wick’s theorem, i.e. all their equal-time correlation functions are completely characterized by the equal-time two-point correlation function~\cite{bravyi2004}. Slater-determinant states form a subset of Gaussian states.
Gaussian fermionic tensor networks share many properties with conventional tensor networks, but the Gaussian structure leads to an exponential improvement in the scaling with entanglement entropy of both memory and computation time. In addition to their numerical usefulness~\cite{schuch2019}, they thus serve as a natural playground to explore the physics of random tensor networks in a more tractable setting. The quantum circuits that correspond to such Gaussian tensor networks are non-unitary quantum circuits of non-interacting fermions. For brevity, we will refer to these as non-unitary Gaussian circuits (NGCs). Given the correspondence between such NGCs and GTNs, we will often use these terms together. When we further consider Gaussian fermionic tensor networks consisting of random tensors, the corresponding quantum circuits will be subject to space-time disorder.
We emphasize that the class of GTN/NGCs is a very broad class and encompasses all non-unitary dynamics of non-interacting fermions~\cite{bravyi2004}. As such, it includes models previously discussed in the literature, such as those of Refs.~\onlinecite{nahum2020entanglement,chen2020emergent,lang2020entanglement}, as well as discrete-time variants of those in Refs.~\onlinecite{alberton2020trajectory,cao2018entanglement}. We emphasize, and explain in detail below (cf end of  Sec.~\ref{LabelSubSubSectionCriticalEntanglementPhase}), that the class of models considered in this work are more general than the loop-model-based circuits considered by other authors and can have manifestly different behavior of physical observables 
(such as, e.g., of disorder-averaged moments of correlation functions).

The correspondence between GTN/NGCs and non-interacting fermions undergoing a unitary time-evolution in the presence of static (quenched) disorder is an important result of this work. This correspondence relies on two important ingredients. The first ingredient is to construct a transfer matrix that captures the single-particle action of the NGC in an enlarged Hilbert space where the density matrix that the NGC acts on is 
treated as a vector (incorporating both, bra and ket).
Such a single-particle
transfer matrix, evolving the {\it density matrix} by one time step, exists
for each disorder realization of the NGC and preserves the locality of the 
circuit.
The second, complementary, ingredient is to express the unitary disordered fermion problem, at fixed energy, in discrete space as a general Chalker-Coddington network model~\cite{Chalker1988}.
This Chalker-Coddington network model also admits a transfer matrix description for each disorder realization. The desired correspondence between random GTN/NGCs and disordered unitary fermion problems is then established by identifying the transfer matrix description of the NGC in the enlarged Hilbert space
representing the density matrix
with that of the Chalker-Coddington network model. The spacetime randomness in the $d$-dimensional NGC corresponds to the spatial quenched disorder in the $D=(d+1)$-dimensional unitary fermion problem. As will be shown 
later (see in particular Sec. \ref{sec:transfer-matrix}),
enlarging the Hilbert space for the NGC so that its transfer matrix acts on the density matrix
is crucial to ensure that the mapping between GTNs/NGCs and non-interacting unitary fermion problems exists in both directions. The enlarged Hilbert space leads to the $D$-dimensional unitary fermion problems in the correspondence having more symmetries than those apparent in the corresponding GTN, 
and this is
crucial for the correct understanding of the underlying physics.

With this correspondence and the appropriate symmetry identification in hand, important results for disordered non-interacting fermions such as the ten-fold Altland-Zirnbauer symmetry classification~\cite{ZirnbauerSusySymmSpacesMJathPhys1996,altland1997nonstandard} 
as well as the well-studied phase diagrams and the well-developed understanding of their critical behavior can be directly reinterpreted in the language of quantum circuits. In particular, our results imply that the critical phases and critical points that emerge at late times in the evolution under non-unitary Gaussian circuits (and correspondingly GTNs) exhibit conformal symmetry and share properties such as multifractality, possible
logarithmic corrections 
etc., with well-known
models of disordered fermions. Furthermore, the existence of topologically distinct gapped phases in random non-interacting fermion systems 
implies the existence
of distinct area law phases in GTN/NGCs, with critical points separating these phases. It is also worth noting that we find no robust volume-law phases
(in line with Ref.~\onlinecite{fidkowskiHaahHastingsHowDynamicalMemoriesForget-arXiv2008.10611}),
except for rather fine-tuned choices of GTNs/NGCs such as those corresponding to
unitary circuits.

\subsection*{Overview of main results}

The present work begins by elucidating the relation between quantum circuits with measurements, circuits with non-unitarity arising from other mechanisms, and tensor networks. We then proceed to introduce Gaussian fermionic tensor networks. Our random tensor network construction proceeds analogously to Ref.~\onlinecite{vasseur2019entanglement}. One typical example of the systems we consider is given
by Gaussian tensor networks on the square lattice, where each tensor is chosen i.i.d. from an orthogonal Haar-random ensemble. This turns out to be the most generic ensemble possible, and the related quantum circuits preserve no quantum numbers except fermion parity. We numerically observe that the contraction of these tensors gives rise to a state exhibiting many features of quantum criticality, including power-law decay of correlations 
(with a particular form of logarithmic corrections which we specify, arising from marginally irrelevant corrections to scaling),
a logarithmic growth of all Renyi entanglement entropies with (sub)system size, and 
scaling of the mutual information between two intervals with the cross-ratio, a hallmark feature of
underlying conformal symmetry.

This GTN model in $D=d+1$ dimensions admits a single-particle transfer matrix formalism where the transfer matrix can also be understood as the 1st-quantized description of the corresponding NGC. In particular, this transfer matrix captures the NGC-generated evolution of density matrices in $d$ spatial dimensions  (i.e. it acts on a space that has twice the dimension of the space of ``ket''-vectors on which the circuit itself acts). Interestingly, we show that, despite the absence of symmetries other than fermion parity in the GTN/NGC itself, the transfer matrices of the GTN/NGC can be identified with those of the Chalker-Coddington network model in $D$ spatial dimensions in Altland-Zirnbauer
symmetry class DIII, which is characterized \cite{SchnyderRyuFurusakiLudwigPRB2008,RyuSchnyderFurusakiLudwig-NewJPhysics12-2010-065010}
by a time-reversal symmetry 
that squares to $-1$, a particle-hole symmetry that squares to $+1$, and  chiral symmetry. The origin of these symmetries is related to the fact that the NGC-generated evolution of density matrices in $d$ spatial dimensions maintains the purity and the Hermiticity of the density matrices. This symmetry class DIII is 
known~\cite{WegnerBetaFunctions-NPB316-1989-663}\footnote{
See also App.~\ref{LabelSubsectionRenormalizationGroupCalculation}.}
to exhibit a disordered metallic phase in spatial dimension $D=d+1=2$ which the aforementioned numerically-observed entanglement criticality naturally corresponds to. 
The stability of the disordered metallic phase implies that the numerically-observed critical entanglement properties should be those of an entire {\it critical entanglement  phase}.
We also show 
that there is a transition from the critical entanglement phase into an area law phase when sufficiently strong dimerization/staggering is turned on. In the language of Chalker-Coddington network models, this transition 
is known as a metal-insulator transition
into one of two gapped phases (one of them topological). This transition is known to be continuous, and driven by proliferation of topological defects in the theory describing the metal~\footnote{See e.g. Ref.~\onlinecite{FulgaAkhmerovBeenakker-DIII-2012}. 
In the long-wavelength description the topological defects arise since the first homotopy group of the target space of the corresponding Non-Linear Model field theory (see, e.g., Appendix~\ref{app:marginal}) is equal to $Z_2$~\cite{RyuSchnyderFurusakiLudwig-NewJPhysics12-2010-065010}}.

We can repeat our construction for all the other nine Altland-Zirnbauer symmetry classes using the 
tools of Clifford algebra extensions~\cite{Kitaev_2009} in any dimension $D$. 
Any criticality known in all these cases for unitary evolution with static disorder is the origin of critical entanglement properties of the corresponding GTN/NGC.
For example, symmetry class AIII emerges from DIII by imposing a global U(1) conservation law~\cite{FosterLudwigHubbardPRB2008,SchnyderRyuFurusakiLudwigPRB2008,ChongWang-Senthil-PRB2014,LudwigNobelSymposium2015} for the circuit evolution, which 
in $D=2$ leads
to continuously varying critical behavior associated with a line of fixed points of the random quantum circuit~\cite{GadeWegnerNPB360-1991-213,GadeWegnerNPB398-1993-499,GLL-NPB583-2000-475,BocquetChalker-NetworkChiralClasses-PRB67-054204,RyuMudryLudwigFurusaki-GlobalPhaseDiagram-PRB85-2012-235115}~\footnote{Each fixed point on this line can be taken into a gapped phase through
a continuous transition (while remaining in symmetry class AIII) driven by proliferation of topological
defects. For a description in terms of the long-wavelength Non-Linear-Sigma Model~(NLSM)
field theory of this transition see e.g. Ref.~\onlinecite{KoenigOstrovskyProtopopovMirlinMetalInsulatorChiralClasses-PRB85-2012-195130}; in the
long-wavelength description the topological defects arise since the first homotopy group of the target space of the NLSM field theory is non-trivial (see e.g. Ref.~\onlinecite{RyuSchnyderFurusakiLudwig-NewJPhysics12-2010-065010}). Numerically, this transition
into the gapped phase is observed in Ref.~\onlinecite{BocquetChalker-NetworkChiralClasses-PRB67-054204}.}. Interestingly, we also show that symmetry class BDI can emerge from symmetry class AIII by imposing a further constraint on the GTN/NGC.
For details, see Sec.~\ref{sec:class_AIII}.
Symmetry class BDI is also known to exhibit a line of critical fixed points~\cite{GLL-NPB583-2000-475,RyuMudryLudwigFurusaki-GlobalPhaseDiagram-PRB85-2012-235115}.
It turns out that the circuit in Ref.~\onlinecite{chen2020emergent} is in symmetry class BDI 
with corresponding universal properties.

Finally, we numerically consider the case of Gaussian fermionic tensor networks in $D=3$ dimensions, and observe a logarithmic violation of the area law in a critical entanglement phase. This behavior is a reflection of the known~\cite{WegnerBetaFunctions-NPB316-1989-663} stable metallic phase of the corresponding Hamiltonian problem in symmetry class DIII with static disorder in $D=3$ spatial dimensions.
A similar area law violation was found for non-unitary Clifford circuits in the same spatial dimension in Ref.~\onlinecite{turkeshi2020measurement}.
We note that metallic phases are known~\cite{WegnerBetaFunctions-NPB316-1989-663} to occur in all symmetry classes of disordered non-interacting fermions for $D \geq 3$, and thus such entanglement phases with logarithmic area law violations will occur generically in those cases.

The remainder of this paper is structured as follows: In Sec.~\ref{sec:TN_QC_relation}, we introduce tensor networks in more detail and discuss the relation between them and non-unitary quantum circuits. We also discuss the relation between non-unitary quantum circuits and quantum systems whose non-unitarity arises from measurements. In Sec.~\ref{sec:gtn}, we introduce Gaussian fermionic tensor networks. In Sec.~\ref{sec:numerics}, we introduce the Haar-random Gaussian fermion tensor network, the 
numerical setup, and the various signatures of criticality that we observe. In Sec.~\ref{sec:transfer-matrix}, we introduce the transfer matrix formalism, establish the mapping between GTNs/NGCs and Chalker-Coddington network models, and provide an analytical understanding of the numerically observed entanglement criticality via the theory describing the metallic phase of disordered fermions in Altland-Zirnbauer symmetry class DIII. In Sec.~\ref{sec:nonunitary-to-unitary}, we discuss the construction of models in all ten Altland-Zirnbauer symmetry classes. In Sec.~\ref{sec:3d}, we discuss the case of GTNs in $D=3$ dimensions. {Finally, in Sec.~\ref{sec:outlook}  we provide an outlook for future directions.}

\section{Tensor networks and quantum circuits} \label{sec:TN_QC_relation}

\subsection{General tensor network on the square lattice}
\begin{figure}
\centering
\begin{minipage}{0.49\columnwidth}
\begin{overpic}[width=1\columnwidth]{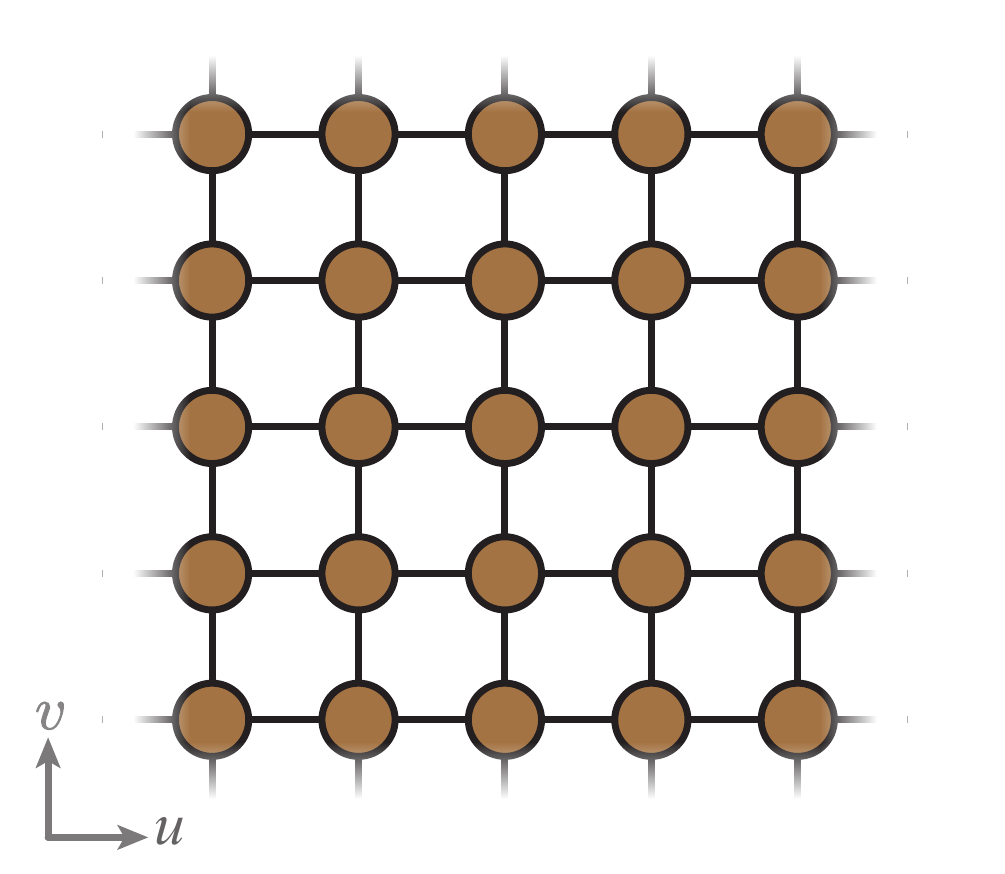} \put (0,77) {\footnotesize{(a)}} \end{overpic}
\end{minipage}
\begin{minipage}{0.49\columnwidth}
\begin{overpic}[width=1\columnwidth]{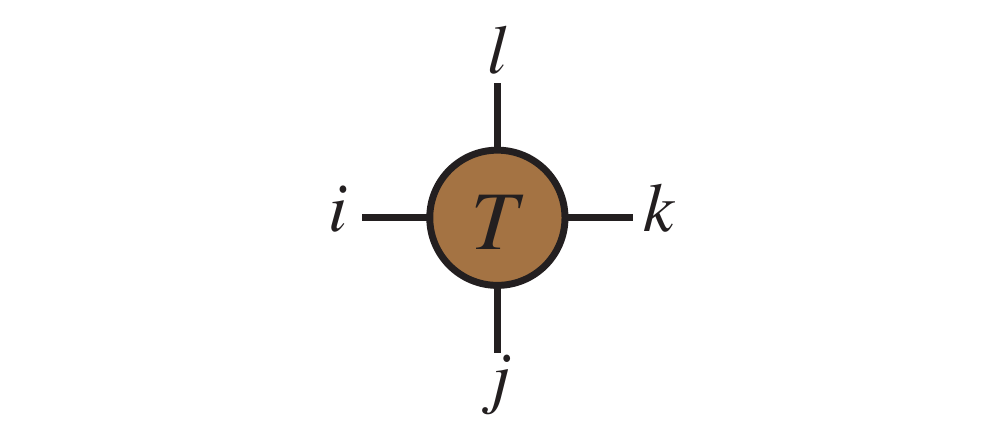} \put (0,30) {\footnotesize{(b)}} \end{overpic}\\
\begin{overpic}[width=1\columnwidth]{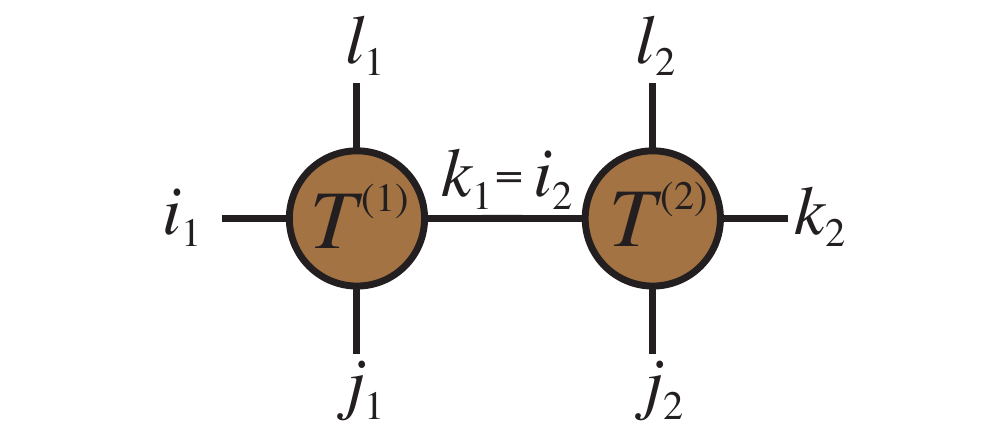} \put (0,30) {\footnotesize{(c)}} \end{overpic}
\end{minipage}
\caption{(a) A generic tensor network on the square lattice is depicted. The horizontal and vertical directions of the tensor network are labeled as $u$ and $v$. (b) The graphical representation of the four-leg tensor $T_{ijkl}$ is shown. (c) The contraction between two tensors $T^{(1)}$ and $T^{(2)}$ is graphically represented by connecting the contracted legs.
}
\label{fig:TN}
\end{figure}

A general tensor network on a square lattice is depicted in Fig.~\ref{fig:TN}~(a). Each individual four-leg tensor $T\in \mathbb{C}^{M^4}$, shown in Fig.~\ref{fig:TN}~(b), defines a state in a $M^4$-dimensional Hilbert space:
\begin{equation}
\ket{T} =\sum_{ijkl} T_{ijkl} \, |ijkl\rangle,
\end{equation}
where each of the indices $i,j,k,l=1,2,3,...,M$ labels the states within the $M$-dimensional Hilbert space associated with a given leg of the tensor $T$. Here, $M$ is the bond dimension of each leg of the tensor $T$. The contraction of two tensors $T_{1,2}$, as shown in Fig.~\ref{fig:TN}~(c), yields a new tensor
\begin{equation}
\sum_{i_2, k_1 = 1}^{M} \delta_{k_1, i_2}  T^{(1)}_{i_1j_1k_1l_1} T^{(2)}_{i_2j_2k_2l_2}.
\end{equation}
Equivalently, we can think of this tensor as being defined by $P_{12} \big( |T^{(1)}\rangle \otimes |T^{(2)}\rangle \big)$, where $P_{12}$ is a projection operator that acts on the tensor product of the Hilbert spaces associated with two contracted legs, and projects onto the maximally entangled state $\frac{1}{\sqrt{M}} \sum_{k_1 i_2} \delta_{k_1, i_2} |k_1\rangle \otimes |i_2\rangle$~\cite{verstraete2004renormalization}. Similarly, all the contractions in the tensor network shown in Fig.~\ref{fig:TN}~(a) can be viewed as the projections onto maximally entangled states on the contracted legs.

\begin{figure}
\centering
\begin{overpic}[width=0.53\columnwidth]{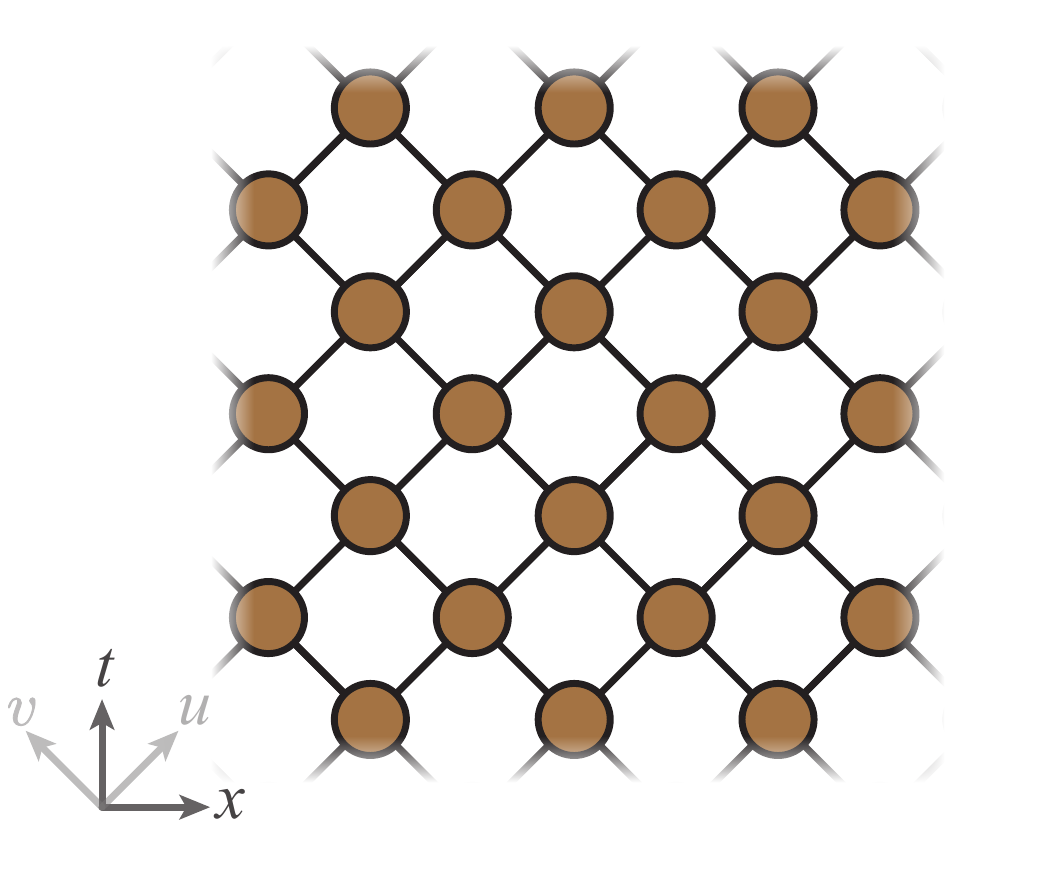} \put (0,77) {(a)} \end{overpic}
\begin{overpic}[width=0.45\columnwidth]{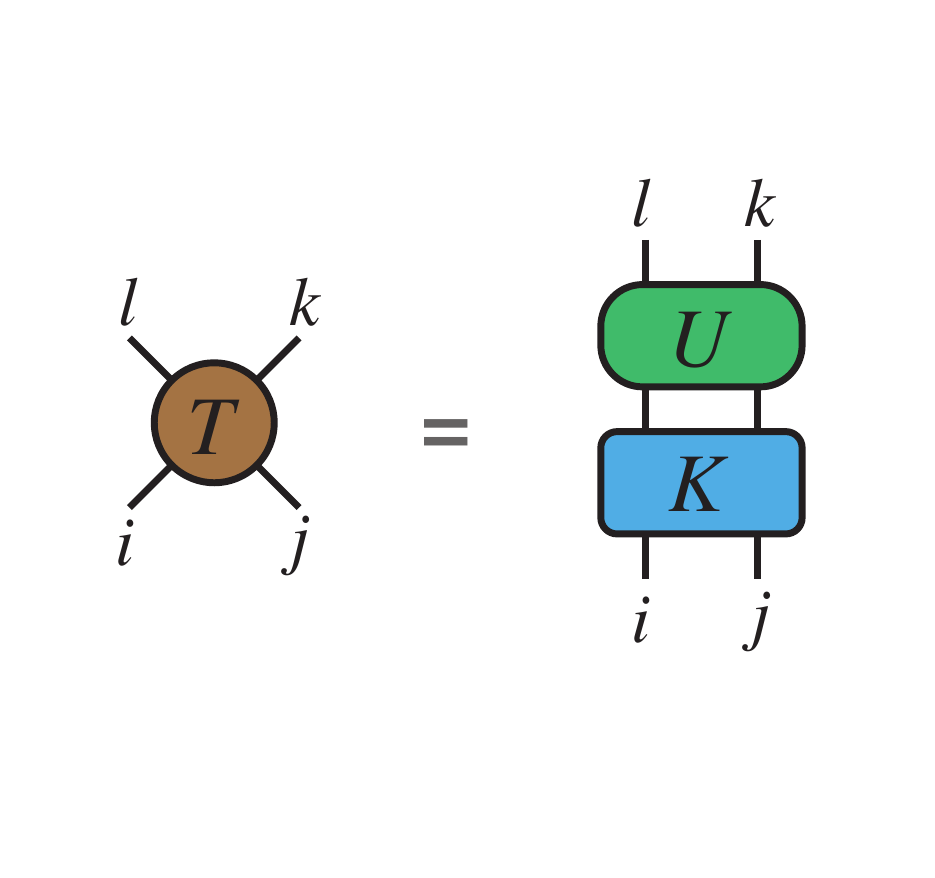} \put (0,90) {(b)} \end{overpic}
\caption{(a) We can rotate the square-lattice tensor network shown in Fig.~\ref{fig:TN}~(a) by $45^\circ$ and view it as a quantum circuit acting on a one-dimensional qudit chain along the $x$-direction. The $t$-direction is viewed as the physical time direction of the quantum circuit. (b) Each four-leg tensor can be viewed as a (non-unitary) quantum gate acting on two neighboring sites on the one-dimensional qudit chain. By the polar decomposition, this quantum gate can be factored into the product of a unitary operator $U$ and the positive-semidefinite Hermitian operator $K$.}
\label{fig:circuit}
\end{figure}

When we rotate the square lattice tensor network (counter-clockwise) by $45^\circ$, we can view it as a quantum circuit that acts on a qudit chain with each qudit carrying an $M$-dimensional local Hilbert space (see Fig.~\ref{fig:circuit}~(a)).  Each tensor can be viewed as an operator acting on two neighboring qudits. The matrix elements of the operator associated with the tensor $T$ (as shown in Fig.~\ref{fig:circuit}~(b)) are given by $T_{ijlk}$, where the pair of tensor indices $i$ and $j$ are viewed as the column
indices of the matrix, and the pair $l$ and $k$ as the row
indices. When $T$ is viewed as an operator, it has a polar decomposition $T=U K$ (graphically represented in Fig.~\ref{fig:circuit}~(b)) where $U$ is a unitary operator and $K$ is a positive-semidefinite Hermitian operator. Physically, we can view $U$ as the real-time evolution operator under some Hermitian Hamiltonian and $K$ as the imaginary-time evolution operator under some other Hermitian Hamiltonian~\footnote{To be more precise, since $K$ is positive-semidefinite, its corresponding Hamiltonian may have positively infinite-energy eigenstates that correspond to the zero eigenvalues of the operator $K$}. With this operator interpretation of each four-leg tensor, the whole tensor network shown in Fig.~\ref{fig:circuit}~(a) can be interpreted as a non-unitary quantum circuit that evolves the qudit-chain quantum states with Hermitian Hamiltonians but in a mixture of real and imaginary time (or simply as a non-unitary quantum circuit that evolves the qudit-chain quantum states only in real time but using non-Hermitian Hamiltonians). In the remainder of this paper, unless specified otherwise, Hermiticity is always implicitly assumed when we talk about the Hamiltonian of a system.

It is conceptually straightforward to generalize the construction above to fermionic tensor networks, where each tensor represents a state in a fermionic Hilbert space. For general overviews of how to take into account the fermionic exchange sign in contractions, see Refs.~\onlinecite{corboz2010mera,corboz2010peps,kraus2010,barthel2009contraction}. In Sec.~\ref{sec:gtn}, we will specialize to the case of {\it Gaussian} fermionic tensor networks and discuss the technical issues arising from their fermionic nature.

\subsection{Relation to quantum systems undergoing unitary evolution and generalized measurements}
\label{sec:UnitaryTimeEvolutionAndMeasurements}

A particularly interesting physical scenario where non-unitary circuits and non-unitary evolutions arise is 
given by a quantum system that undergoes both unitary evolution and projective measurements~\cite{li2018quantum,li2019measurement}. In the presence of measurements, the evolution of the quantum system is characterized by an ensemble of {\it quantum trajectories} with each quantum trajectory labeled by a different set of measurement outcomes. In the following, we first briefly introduce our notion of quantum trajectory before we discuss the connection to tensor networks. 
A more detailed description of quantum trajectories can be found, for example, in Ref.~\onlinecite{Wiseman1996}.

For our purposes, a single measurement is described by a set of Kraus operators $C_m$, where $m$ labels the different measurement outcomes. Given some input density matrix $\rho$, Born's rule gives the probability of the $m$'th measurement outcome as $p_m = \Tr(C_m \rho C_m^\dagger)$; normalization of probability requires $\sum_{m} C^\dagger_m C_m = \mathds{1}$, i.e.
the operators $C^\dagger_m C_m$ form a positive operator-valued measure~\cite{Wiseman1996}. For the $m$'th measurement outcome, the density matrix after the measurement is given by $\rho' = C_m \rho C_m^\dagger / p_m$; 
if the input density matrix represents a pure state, $\rho = \ket{\psi}\bra{\psi}$, 
so does the density matrix $C_m \ket{\psi}\bra{\psi} C_m^\dagger / p_m$ after measurement. In the case of a conventional projective measurement, the $C_m$ are projectors, i.e. $C_m^2 = C_m$.
We emphasize that we are free to interpret \emph{any} set of operators $\lbrace C_m \rbrace $ that satisfies the condition $\sum_m C_m^\dag C_m = \mathds{1}$ as a generalized measurement within this formalism, and any such measurement can be physically implemented (See, e.g., Ref.~\onlinecite{watrous2018theory}.).

It is clear that we can interpret a circuit composed of unitary evolution and projective measurements within this formalism. Given a set of measurement outcomes $\vec{m}$ (each entry corresponding to the outcome of one projective measurement),
we can define an operator $C_{\vec{m}}$ 
as the product  (over time steps) of the unitary evolution 
operators followed by the projection operators corresponding to the measurement outcomes in each step (noting that the succession of a measurement with $N$ and one with $M$ outcomes can be thought of as a measurement with $N \cdot M$ outcomes). While the set of measurement outcomes will grow exponentially with the volume of the circuit and the Born-rule probability $p_{\vec{m}}$ of a given set of outcomes $\vec{m}$ becomes exponentially small, the entire set of operators $\lbrace C_{\vec{m}} \rbrace$ will still satisfy the conditions above, in particular the $C_{\vec{m}}^\dagger C_{\vec{m}}$ also form a positive operator-valued measure.

As explained in the previous section, a generic tensor network can be viewed as a non-unitary quantum circuit comprised of real-time and imaginary-time evolution. In this work, we will not focus on individual tensor networks but rather the average behavior in certain random ensembles of tensor networks with each realization of the tensor network taking an equal weight in the average. Each such ensemble of tensor networks provides an ensemble of non-unitary circuits $\{C_m\}$. If the condition $\sum_m C_m^\dag C_m = \mathds{1}$ is satisfied, the corresponding ensemble of tensor networks can describe a physical quantum system undergoing both unitary evolution and generalized measurement.
However, an important yet subtle distinction arises due to the Born-rule probability $p_m = \Tr (C_m \rho C_m^\dagger)$ of the system choosing a particular trajectory in the case of measurements, which depends on the initial density matrix $\rho$. To establish a precise correspondence between non-unitary circuits where the non-unitarity arises due to measurements and those where non-unitarity arises from some other mechanism, the ensemble in the latter case may have to be reweighed according to the Born-rule probability~\footnote{Indeed, due to this, it is very unlikely that circuits involving imaginary-time evolution or with pre-determined measurement outcomes can experimentally be efficiently implemented, unless one were willing to engage in anthropic computing; see Ref.~\onlinecite{aaronson2005quantum}}.

It appears possible that certain universal behavior exhibited in these two situations is closely related. 
Ref.~\onlinecite{nahum2020measurement} speculated about potential differences
in circuits with Haar random evolution.
As we will see below, certain classes of tensor networks are amenable to a rather complete numerical and analytical treatment, and as such provide valuable insights into the non-unitary dynamics of quantum many-body systems. Therefore, for the purpose of this paper, we employ the language of tensor networks; however, as discussed in App.~\ref{app:random_GTN_vs_MeasurementProblem}, our tensor network construction could be adapted as a measurement circuit. Comparison between the universal behavior in the tensor-network ensemble and the Born-rule ensemble will be reserved for future investigations.

\section{Fermionic Gaussian tensor network}
\label{sec:gtn}

\subsection{Definition}
\label{sec:gtnA}

In this work, we focus on random ensembles of {\it Gaussian} fermionic tensor networks (GTNs)~\cite{kraus2010,evenbly2010entanglement,schuch2012,dubail2015,haegeman2013,fishman2015,evenbly2016,haegeman2018,jahn2019holography,schuch2019}, which are a special type of fermionic tensor networks that describe Gaussian states. In a GTN on the square lattice, each four-leg tensor defines a fermionic Gaussian state in a Hilbert space associated to $4\chi$ Majorana fermion modes. The $4\chi$ Majorana fermion modes are divided into 4 groups of $\chi$ Majorana fermion modes each, with each group residing on one of the legs of the four-leg tensor. We refer to the number of Majorana modes $\chi$ on each leg of the tensors as the Majorana bond number of the GTN; it is related to the bond dimension $M$ introduced previously via $M = \sqrt{2}^\chi$, i.e. GTNs are an exponentially more compact representation. In particular, for conventional tensor networks, the maximal amount of entanglement that can be captured is $\mathcal{O}(\log M)$ (and the computational effort thus exponential in the amount of entanglement), whereas here it is $\mathcal{O}(\chi)$ (and the computational effort thus polynomial in the amount of entanglement).

In a GTN, each tensor is itself a fermionic Gaussian state. Such a state is completely determined by its two-point fermion correlation functions. To be more precise, 
let us denote the Majorana modes associated to a four-leg tensor with Majorana bond number $\chi$ for each leg as $\hgamma_{i=1,2,...,4\chi}$. The fermionic Gaussian state associated to the four-leg tensor is completely determined by the $4\chi\times 4\chi$ covariance matrix~\cite{bravyi2004}
\begin{equation} \label{eqn:cov-matrix}
\Gamma_{ij} = \Big\langle \frac{\i }{2} [\hgamma_i, \hgamma_j] \Big\rangle = \langle \i \hgamma_i \hgamma_j\rangle -\i \delta_{ij},
\end{equation}
where the expectation value is taken in the fermionic Gaussian state. Multi-point fermion correlation functions in the fermionic Gaussian state can be constructed from $\Gamma_{ij} $ via Wick's theorem~\footnote{See Ref.~\onlinecite{Bravyi2017} for a recent review of this formalism.}. Hence, we can use the covariance matrices $\Gamma_{ij}$ to represent the four-leg tensors in the GTN. However, we emphasize that the meaning of indices of the covariance matrix $\Gamma_{ij}$ which labels the Majorana fermion operators associated with a given tensor is different from the meaning of the indices of the four-leg tensor $T_{ijkl}$ which labels the states in a (sub-)Hilbert space. The covariance matrix $\Gamma_{ij}$ for 
a {\it pure} Gaussian state satisfies
\begin{align}
\Gamma^\tr = - \Gamma,~~~~\Gamma^*=  \Gamma,~~~~\Gamma^2= -\mathds{1}. 
\label{eqn:GammaCondition}
\end{align}
The first two conditions are simply derived from the properties of Majorana fermion modes while the third condition is the consequence of a pure Gaussian state. The fermion parity of the pure Gaussian state is given by the Pfaffian $\Pf(\Gamma) = \pm 1$ of the covariance matrix $\Gamma$. The space of all $4\chi\times 4\chi$ covariance matrices $\Gamma$ satisfying the conditions in Eq.~\eqref{eqn:GammaCondition} is given by the symmetric space $\frac{{\rm O}(4\chi)}{\U(2\chi)}$. The space of $\Gamma$ further restricted to the sector with a fixed fermion parity $\Pf(\Gamma)$ is given by the symmetric space $\frac{{\rm SO}(4\chi)}{\U(2\chi)}$. In particular, the four-leg tensor that can be interpreted as a quantum gate that conserves the fermion parity are the ones with a fixed fermion parity $\Pf(\Gamma) =1$.  In principle, one can also consider mixed Gaussian states, which can also be fully specified by their covariance matrix. The covariance matrices of a mixed Gaussian states still satisfy the first two conditions in Eq.~\eqref{eqn:GammaCondition}, but the third condition is relaxed to $\Gamma^2 \succeq -\mathds{1}$, meaning that no eigenvalues of the Hermitian matrix $\Gamma^2$ are smaller than $-1$. This work will mostly focus on the square-lattice GTNs where every four-leg tensor is associated with a pure fermionic Gaussian state. We refer to this type of tensor network as the pure-state GTN.
We also impose an extra requirement that each tensor in the pure-state square-lattice GTN has a fixed fermion parity $+1$ so that when the pure-state square-lattice GTN is interpreted as a quantum circuit, each quantum gate in the circuit respects the fermion parity and, hence, can be viewed as a bosonic operator (in the sense that it does not change the fermion parity of the state it acts on).

The pure Gaussian state $|\Gamma \rangle$ that is 
associated~
\footnote{specifically, so that the 2-point fermion correlation function is then \text{
$\langle {\hat\gamma}_i {\hat \gamma}_j\rangle=
\langle \Gamma |{\hat\gamma}_i {\hat \gamma}_j
|\Gamma\rangle={\rm Tr} ({\hat \rho}_{\Gamma}
{\hat\gamma}_i {\hat \gamma}_j
)$}, with \text{${\hat \rho}_{\Gamma}=|\Gamma\rangle \langle \Gamma|$} the pure state density matrix}

with a four-leg tensor, with each leg having Majorana bond number $\chi$, and which has a $4\chi\times 4\chi$ covariance matrix $\Gamma_{ij}$, can
be determined via the equation
\begin{align}
\left(\hgamma_i - \i \sum_j \Gamma_{ij} \hgamma_j \right) |\Gamma\rangle = 0,~~~~i=1,2,...,4\chi.
\label{eqn:GS_Def_Prop}
\end{align}
When we view $\hgamma$ as a 4$\chi$-component column vector of Majorana operators, the equation above can be conveniently written as $\left(\hgamma - \i \, \Gamma \hgamma \right) |\Gamma\rangle = 0$. 

\subsection{Contraction of Gaussian tensors}
\label{sec:gtn_contraction}
Since Gaussian tensor networks are just a special case of general fermionic tensor networks, the contraction of two tensors can similarly be viewed as a projection onto a maximally-entangled-pair state on the legs that are being contracted. Crucially, the result of such a contraction of two Gaussian tensors is again a Gaussian tensor; if the two input states are pure, so is the contracted state.

It is worth noting that the contraction of two Gaussian tensors can also be viewed as applying a quantum operation defined by one Gaussian state to the other Gaussian state. Here, by quantum operation we mean any completely positive trace-non-increasing linear operation on density matrices, i.e. the most general operation that transforms a valid (pure or mixed) quantum state into another valid quantum state. Indeed, as proven in Refs.~\onlinecite{terhal2002classical,bravyi2004}, \emph{any} Gaussian map, i.e. completely positive linear map that transforms Gaussian states into Gaussian states, can be described as the contraction of (possibly non-pure) tensors in this formalism. As such, any kind of circuit for non-interacting fermions can be translated into the contraction of a GTN. Therefore, the circuits discussed in Refs.~\onlinecite{chen2020emergent,lang2020entanglement} and the quantum measurement circuits without annihilation in Ref.~\onlinecite{nahum2020entanglement} can be viewed as particular examples of Gaussian tensor networks.

\begin{figure}
    \centering
    \begin{overpic}[width=0.9\columnwidth]{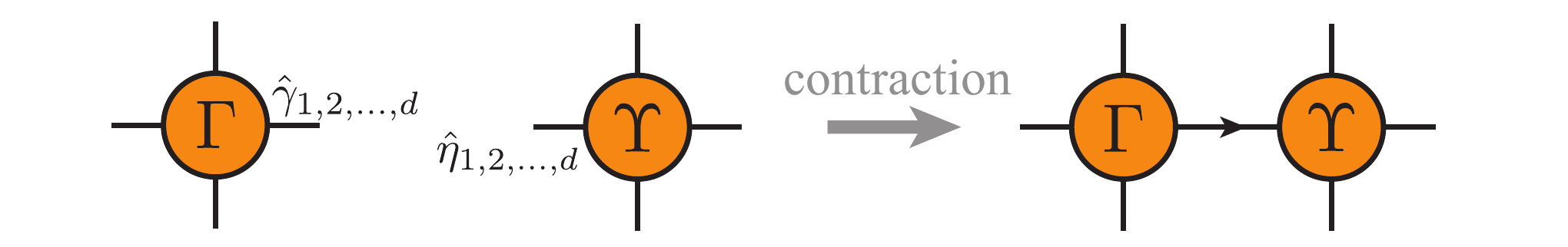} \put (-7,10) {\footnotesize{(a)}} \end{overpic} 
    \\
    ~
    \\
    \begin{overpic}[width=0.9\columnwidth]{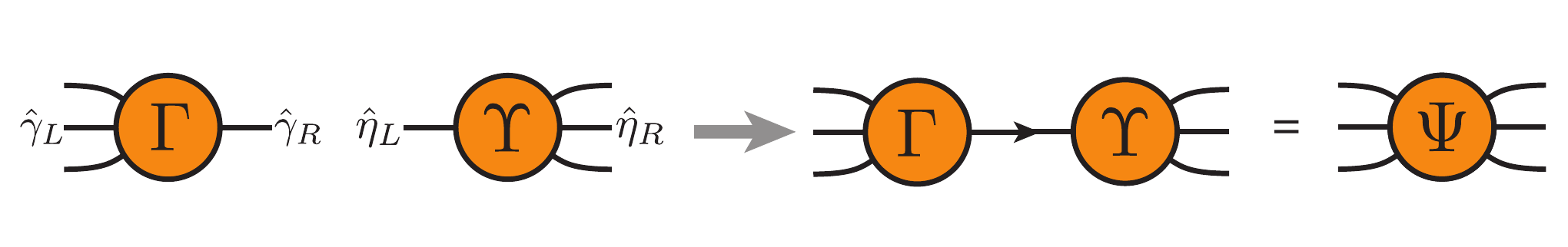} \put (-7,10) {\footnotesize{(b)}} \end{overpic}
    \caption{(a) The Gaussian tensors $\Gamma$ and $\Upsilon$ are contracted. The Majorana modes on the contracted legs of the Gaussian tensors $\Gamma$ and $\Upsilon$ are denoted as $\hgamma_{1,2,..,d}$ and $\hupsilon_{1,2,...,d}$ respectively. (b) We can view both of the Gaussian tensors $\Gamma$ and $\Upsilon$ as two-leg tensors when implementing the contraction between them.}
    \label{fig:FermionGTN_Contraction}
\end{figure}

To illustrate in greater detail the procedure of contraction, 
let us discuss the contraction of two four-leg tensors as shown in Fig.~\ref{fig:FermionGTN_Contraction}~(a). Consider two Gaussian tensors represented by the covariance matrices $\Gamma$ and $\Upsilon$. When a Majorana-bond-number-$\chi$ leg of the tensor $\Gamma$ that carries Majorana fermion modes $\hgamma_{i=1,2,..\chi}$ is contracted with a Majorana-bond-number-$\chi$ leg of the tensor $\Upsilon$ that carries Majorana fermion modes $\hupsilon_{i=1,2,..\chi}$, the contraction yields a new Gaussian state
\begin{align}
\ket{\Psi} = P_{12} \left( \ket{\Gamma} \otimes \ket{\Upsilon} \right),
\end{align}
where the projection operator $P_{12}$ is given by
\begin{align}
P_{12}= \prod_{i=1}^\chi \frac{1+ \i \hgamma_i \hupsilon_i}{2}.
\end{align}
The contraction is graphically represented in Fig.~\ref{fig:FermionGTN_Contraction}~(a). Notice that the contracted leg in Fig.~\ref{fig:FermionGTN_Contraction}~(a) has a direction which indicates the ordering of Majorana modes $\hgamma_i$ and $\hupsilon_i$ in the projection operator, i.e. the contraction with the reversed direction is implemented by the projection $ \prod_{i=1}^\chi (1 + \i \hupsilon_i \hgamma_i )/2$. The choice of direction is necessary for each contracted bond of a fermionic tensor network.

The Gaussian state $\ket{\Psi}$ is again fully characterized by a covariance matrix $\Psi$. Here, we view $\ket{\Psi}$ as Gaussian state residing in the Hilbert space given by only the Majorana modes on the un-contracted legs. For an explicit expression for $\Psi$, it is convenient to relabel the Majorana modes on $\Gamma$ and $\Upsilon$ as shown in Fig.~\ref{fig:FermionGTN_Contraction}~(b): in particular, the modes on $\Gamma$ are grouped into $\hgamma_L$, which remain open, and $\hgamma_R$, which are to be contracted; for $\Upsilon$, the modes $\hupsilon_L$ are to be contracted, while $\hupsilon_R$ remain open. We can reorganize the two covariance matrices in a block form,
\begin{align}
\Gamma &= \left(\begin{array}{cc}
    \Gamma_{LL} &\Gamma_{LR} \\
    -\Gamma_{LR}^\tr &\Gamma_{RR}
\end{array} \right)
&\Upsilon &= \left(\begin{array}{cc}
    \Upsilon_{LL} &\Upsilon_{LR} \\
    -\Upsilon_{LR}^\tr &\Upsilon_{RR}
\end{array} \right).
\label{eqn:Gamma_Upsilon}
\end{align}
Here, $\Gamma_{LL}$ and $\Upsilon_{RR}$ are $3\chi \times 3\chi$ matrices describing the correlations between the $3\chi$ modes $\hgamma_L$ and $\hupsilon_R$, respectively, and similarly $\Gamma_{RR}$ and $\Upsilon_{LL}$ are $\chi \times \chi$ matrices, and the off-diagonal matrices are $\chi \times 3\chi$ or $3\chi \times \chi$ rectangular matrices.
The covariance matrix $\Psi$, which is a $6\chi \times 6\chi$ matrix that describes the correlations of the $6\chi$ Majorana operators $\hgamma_L$ and $\hupsilon_R$, is then given by~\cite{bravyi2004}
\begin{multline}
     \Psi = 
     \left(
     \begin{array}{cc}
          \Gamma_{LL}  & 0  \\
         0 & \Upsilon_{RR}
    \end{array}
    \right)
   + \\
     \left(
     \begin{array}{cc}
          \Gamma_{LR}  & 0  \\
         0 & \Upsilon_{RL}
    \end{array}
    \right)
    \left(
     \begin{array}{cc}
          \Gamma_{RR}  & \mathds{1}  \\
         -\mathds{1} & \Upsilon_{LL}
    \end{array}
    \right)^{-1}
     \left(
     \begin{array}{cc}
          \Gamma_{LR}  & 0  \\
         0 & \Upsilon_{RL}
    \end{array}
    \right)^{\T}.
    \label{eqn:SchurContraction}
\end{multline}
Notice that $\Psi$ depends on $\Gamma$ and $\Upsilon$ in a non-linear way. Furthermore, one can check that, if $\left(\Gamma \right)^2 = \left(\Upsilon \right)^2 = -\mathds{1}$, then $\left( \Psi \right)^2 = -\mathds{1}$, \i.e. the contraction between pure-state tensors results in a pure-state tensors. This expression can be evaluated in $\mathcal{O}(\chi^3)$ time.

Having introduced the contraction between two tensors, a GTN can be built by contracting the involved tensors one by one using Eq.~\eqref{eqn:SchurContraction}. The order of tensor contractions does not affect the final result as long as the same tensor network geometry is maintained. This independence of ordering can understood as follows. A GTN that consists of tensors $\{\Gamma^{(n)}\}_n$ produces a Gaussian state $P \left( \otimes_n |\Gamma^{(n)} \rangle\right) $ where $P$ is the product of projections onto the maximally-entangled-pair state on each pair of contracted legs in the GTN. The projections on different pairs of contracted legs commute with each other, which implies that the final GTN is independent of the order in which the projectors are applied. In practice, the order may make a difference to the computational cost, see App.~\ref{app:numerical-details}.

\section{Numerical results}
\label{sec:numerics}

\subsection{Setup}
\label{sec:numerics-setup}

We now turn to a numerical investigation of the properties of Gaussian tensor networks with random tensors. We will find that for a generic choice of random ensemble (which we introduce below), the state obtained by contracting the tensor network exhibits signatures of quantum criticality and scale invariance, namely a logarithmic divergence of the bipartite entanglement entropy with the subsystem size and power-law decay of two-point correlation functions.

We consider an ensemble of tensor networks where each tensor is independently drawn from an identical probability distribution of what we call Haar-random Gaussian pure states. To construct such a state on $2n$ Majorana fermions, we start from the reference pure state
\begin{equation}
\label{LabelEqState}
\Omega_{2n} = \begin{pmatrix}
0 &\openone_{n} \\
-\openone_{n} &0
\end{pmatrix}
\end{equation}
where $\openone_{n}$ is the $n \times n$ identity matrix. Then, we obtain a random special orthogonal matrix  $O \in \SO(2n)$ following the approach of Ref.~\onlinecite{ozols2009generate}.
The desired Haar-random Gaussian pure state is then given by as
\begin{equation}
\Gamma = O \, \Omega_{2n}\, O^\tr.
\label{eqn:random_Gamma_generator}
\end{equation}
This will generate pure states with a fixed parity $\Pf(\Gamma)=+ 1$. The ensemble of $\Gamma$ generated by the Haar-random matrix $O\in {\rm SO}(2n)$ is equivalent to the random ensemble of $\Gamma$ in the symmetric space $\frac{{\rm SO}(2n)}{\U(n)}$ with a uniform probability measure. In principle, one can extend the ensemble to that of states with random fermion parity $\Pf(\Gamma)=\pm 1$ by considering Haar-random matrix $O\in {\rm O}(2n)$ in Eq.~\eqref{eqn:random_Gamma_generator} instead; however, we find that the numerical results presented in this section, which are obtained in an ensemble of fixed parity, are, within sampling error, identical to the ones obtained for random fermion parities.
For the square-lattice GTN with Majorana bond number $\chi$, we perform this procedure with $2n=4\chi$ for each four-leg tensor independently.

\begin{figure}
    \centering
    \includegraphics[width=1\columnwidth]{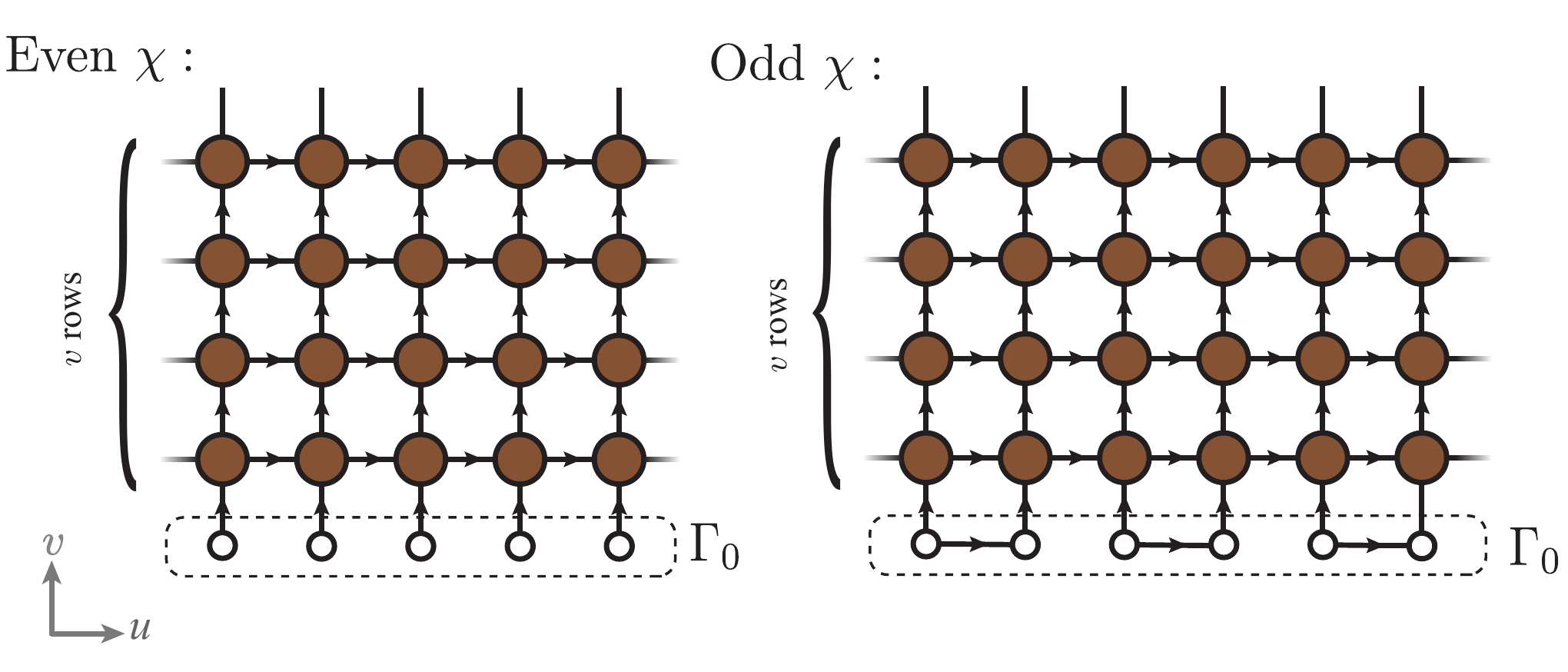}
    \caption{In the numerical study, we consider the square-lattice random GTN with periodic boundary condition in the $u$-direction. At $v=0$, we start with the initial Gaussian state represented by the Gaussian tensor $\Gamma_0$ which is a product state. The contraction of the tensor $\Gamma_0$ with $v$ (``depth") rows of the random GTN yields the Gaussian state represented by the tensor $\Gamma_v$.
    }
    \label{fig:Numerics_Illutstration}
\end{figure}

In the contraction of the tensor network (see Fig. \ref{fig:Numerics_Illutstration}), we start from an initial state at $v=0$ given by the covariance matrix
\begin{equation} \label{eqn:initialstate}
\Gamma_0 = \bigoplus_{k=1}^{L\chi/2} \Omega_{2},
\end{equation}
which is unentangled for even $\chi$ and entangled only between adjacent sites for odd $\chi$ (there is no fully unentangled state for odd $\chi$; note also that we require $L\chi$ even to construct pure-state GTNs). The initial state $\Gamma_0$ of the GTN for even and odd $\chi$ are pictorial represented by the hollow circles shown in Fig.~\ref{fig:Numerics_Illutstration}. We have chosen to fix the parity of the initial state to $\Pf(\Gamma_0)=+1$; similar as for the parity of the tensors on each site of the network, we have confirmed that results are indistinguishable for an initial state of the other parity. We contract the initial state $\Gamma_0$ with rows of tensors as shown in Fig.~\ref{fig:Numerics_Illutstration}. 
We denote by $\Gamma_v$ the state that is the result after the contraction of $v$ (``depth'') rows of tensors with the initial state $\Gamma_0$; i.e.,  that state is defined by the open legs at the top of the network with depth $v$. We are interested in the behavior for $v \rightarrow \infty$. We apply periodic boundary conditions in the spatial direction, i.e. the $u$-direction, of the network.

Here, we've defined a Haar-random ensemble of pure-state square-lattice GTNs. This ensemble of GTNs yields an ensemble of random quantum circuits via the correspondence between GTNs and non-unitary Gaussian circuits in Sec.~\ref{sec:TN_QC_relation}. As shown in App.~\ref{app:random_GTN_vs_MeasurementProblem}, this quantum circuit ensemble can, in principle, characterize the dynamics of a physical system undergoing both unitary evolution and generalized measurements. However, one needs to be careful that, for the physical system undergoing both unitary evolution and generalized measurements, the probability for a specific quantum circuit (corresponding to fixed measurement outcomes) in the circuit ensemble to appear needs to follow the Born's rule, which implies that different quantum circuits in general appear with different probabilities. In contrast, in all the following discussions of the Haar-random ensemble of GTNs, each GTN/NGC appear with equal probabilities.

\subsection{Correlation functions}
\label{Label-SubSection-CorrelationFunctions}

We first analyze the two-point correlation function $\langle \i \gamma_k \gamma_l \rangle$ (i.e., the equal-time Green's function) in the state $\Gamma_v$, which resides on the one-dimensional lattice of open legs on the top the GTNs shown in Fig.~\ref{fig:Numerics_Illutstration}. Each site of this lattice corresponds to an open leg and therefore contains $\chi$ Majorana modes. The quantity of interest, corresponding to the average
of the square of the two-point  correlation function for Majorana modes that are $r$ sites away from each other, is given by
\begin{equation} \label{eqn:Cd}
    C(r) = \frac{1}{\chi^2 L} \sum_{m,n=1}^\chi \sum_{p=1}^L \overline{ \langle \i \hat{\gamma}_{p,m} \hat{\gamma}_{p+r,n} \rangle^2 },
\end{equation}
where by $\hat{\gamma}_{p,m}$ we denote the $m$'th Majorana mode ($m=1,\ldots,\chi$) on the $p$'th site of the lattice, we apply periodic boundary conditions, i.e. $\hat{\gamma}_{p+L,n} = \hat{\gamma}_{p,n}$, and the bar indicates ensemble-averaging. The individual expectation values in a fixed realization of disorder (i.e. before disorder averaging) correspond exactly to elements of the covariance matrix $\Gamma_v$: $\langle \i \hat{\gamma}_{p,m} \hat{\gamma}_{p+r,n} \rangle = (\Gamma_v)_{(p-1)\chi+m,(p+r-1)\chi+n}$ (except for $r=n=0$).

\begin{figure}
  \centering
  \begin{overpic}[width=0.97\columnwidth]{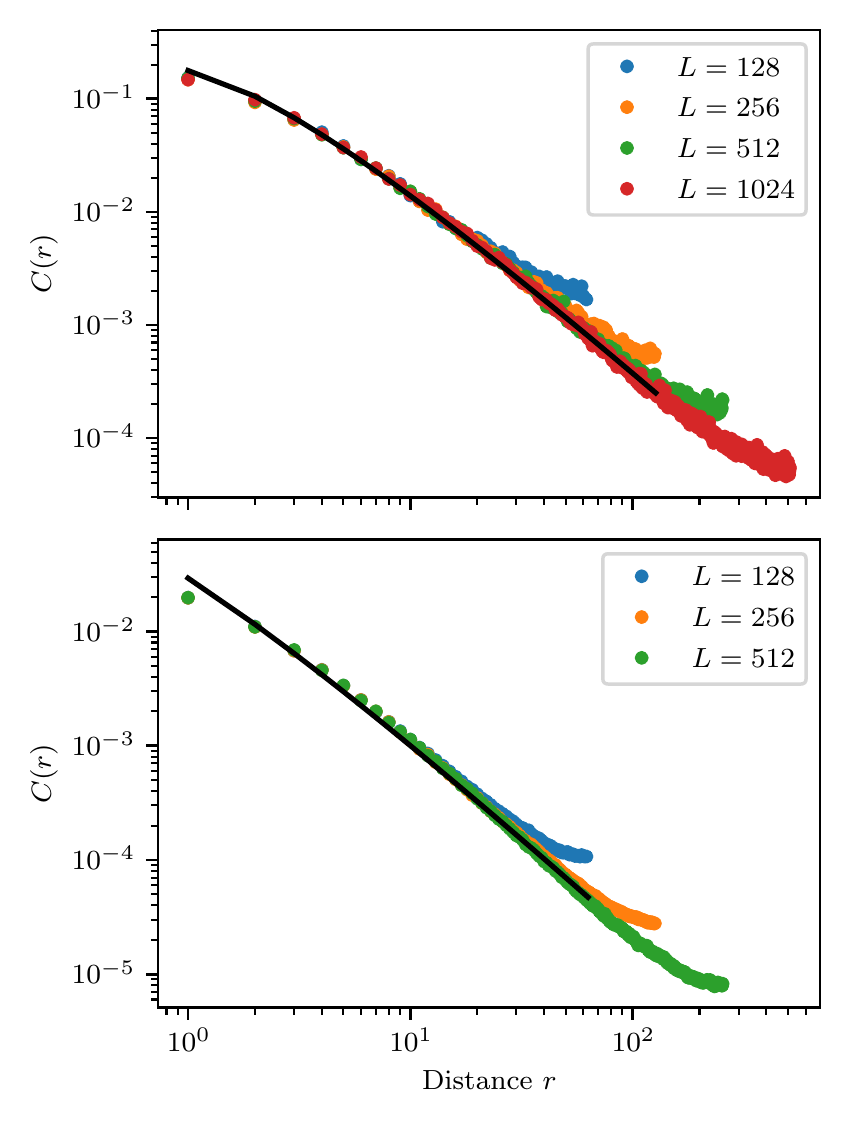} \put (0,95) {(a)} \put (0,50) {(b)} \end{overpic}
  \caption{Correlation function $C(r)$ (cf. Eq.~\eqref{eqn:Cd}) obtained from averaging over 20 disorder realizations for Majorana bond number $\chi=1$ (a) and $\chi=6$ (b) and $r$ up to $L/2$. Points indicate raw data, while the black line indicates a fit of the data for the largest system size to Eq.~\eqref{eqn:corr-decay}.
    \label{fig:correlations}
  }
\end{figure}

Our results are shown in Fig.~\ref{fig:correlations}, where we take the depth $v=500$ and average over 20 realizations. In random systems, the mean and typical correlations can differ widely near critical points
because a correlation function (as opposed to, e.g., a free energy) is in general not self-averaging~\cite{DERRIDA-PhysRepts1984}. In particular, different disorder moments of a correlation function can scale with independent critical exponents~\cite{LudwigHierarchiesNPB1990}, a phenomenon called multifractality which is ubiquitous in disordered non-interacting fermion systems~\cite{CastellaniPelitiMultifractality-JPhysA1986,JanssenMetzlerZirnbauer-PhysRevB59-15836} (for a relatively recent discussion see e.g. Refs.~\onlinecite{GruzbergLudwigMirlinZirnbauerSymmetriesMultifractalSpectraPRL2011,ObuseBeraLudwigGruzbergEversPointContact}). An extreme version of this phenomenon is known to occur in one-dimensional quantum systems with (static) spatial randomness where disorder moments of
correlation functions can be dominated entirely by rare event (Griffiths) physics leading to completely different functional forms of mean and typical correlations, such as e.g. in the random singlet phase~\cite{fisher1994random,motrunich2001griffiths}.
We have checked for this
numerically
and find that in our case, the mean and typical correlations differ only by a prefactor. The reason for the self-averaging of these correlations will be given in App.~\ref{sec:app-field-theory}.

The correlations shown in Fig.~\ref{fig:correlations} clearly decay with a power law, consistent with a critical system. To quantify this more precisely, we perform a fit to
\begin{equation} \label{eqn:corr-decay}
C(r) = A \,\, \frac{\left(1 + \lambda_0 \log r\right)^2}{r^2},
\end{equation}
where $A$ and $\lambda_0$ are fit coefficients. The correction 
to a pure power-law decay in the numerator arises from the presence of a marginally irrelevant operator, whose coupling constant is denoted by $\lambda_0$. For details, see Appendix~\ref{app:marginal}. We find excellent agreement with this form with a constant $\lambda_0$ that depends on $\chi$, as shown in Fig~\ref{fig:correlations}.

\subsection{Entanglement entropy}
\label{sec:numerics-entanglement}
\begin{figure}[htp]
    \centering
    \begin{overpic}[width=0.97\columnwidth]{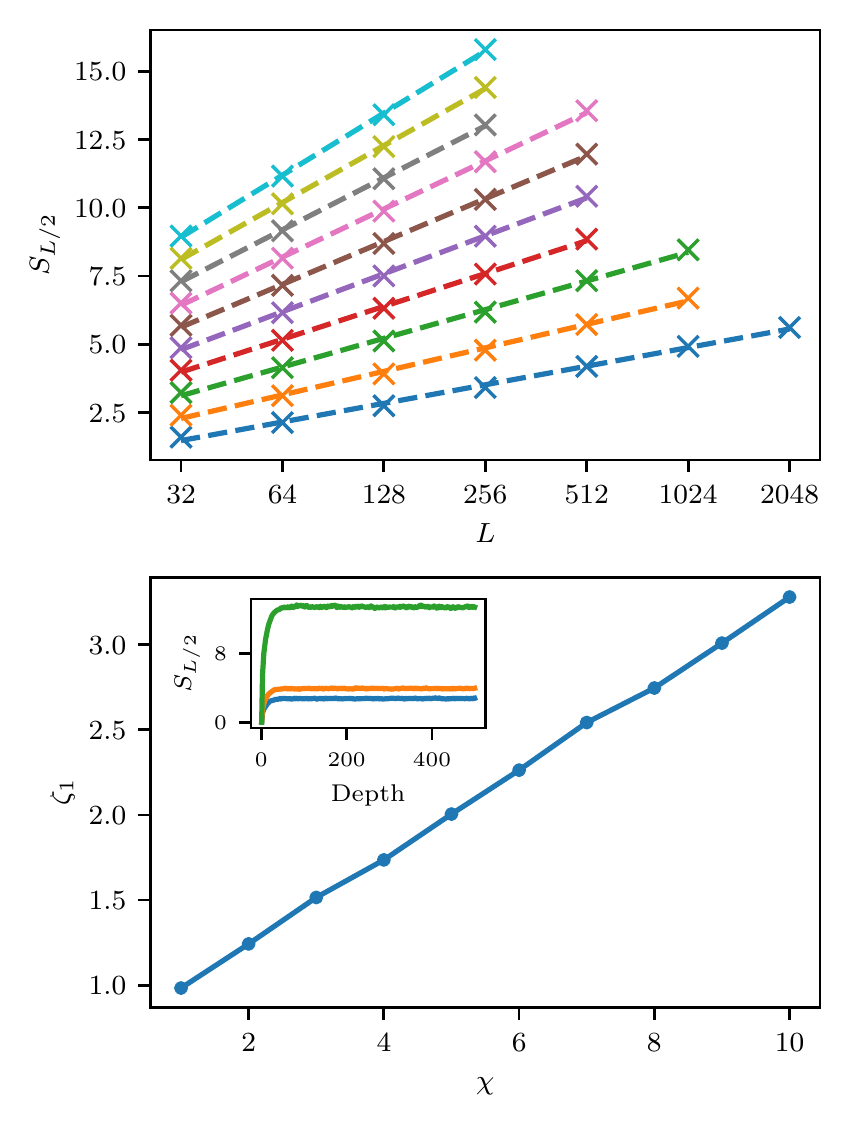} \put (0,95) {(a)} \put (0,47) {(b)} \end{overpic}
    \caption{
    (a) Scaling of bipartite entropy of half of the system with total system size in the Haar-random ensemble. Majorana bond numbers are $\chi=1$ through $\chi=10$ from bottom to top. Crosses indicate the numerical data, while the dashed lines indicate fits to the form $S_{L/2} = \zeta_1 \log (L/L_0)$.
    (b) Dependence of the entropy scaling prefactor $\zeta_1$ on the Majorana bond number $\chi$.
    \emph{Inset of} (b): Convergence of the entropy at the center of the system with depth $v$ for $L=128$ and $\chi=1,2,10$.
    \label{fig:entropyscaling}}
\end{figure}

As a second quantity of interest, we compute the von Neumann entanglement entropy for a contiguous block of $L/2$ sites in the state $\Gamma_v$ for periodic boundary condition along the $u$-direction. We denote this quantity as $S_{L/2}$. In the inset in Fig.~\ref{fig:entropyscaling}~(b), we show the dependence of $S_{L/2}$ on the depth $v$ for a few characteristic values of $L$ and $\chi$. We find that it very quickly reaches a plateau. To obtain averaged quantities, we average over $v$ for $v$ greater than some cutoff (usually $v=250$) as well as several completely independent simulations (here 100).

The averaged $S_{L/2}$ is shown for Majorana bond numbers between $\chi=1$ and $\chi=10$ and system sizes ranging from $L=32$ to $L=2048$ (for the smallest bond number) in Fig.~\ref{fig:entropyscaling}~(a). As indicated by the dashed lines, we find excellent agreement with the scaling form
\begin{equation} 
    S_{L/2} = \zeta_1 \log \left( L/L_0 \right),
    \label{eqn:vN-scaling}
\end{equation}
where we take both $\zeta_1$ and $L_0$ as fit parameters. In Fig.~\ref{fig:entropyscaling}~(b), we show the dependence of $\zeta_1$ extracted from our fits on Majorana bond number $\chi$, which we find to be nearly linear.

This scaling form of the entanglement entropy is familiar from a variety of other systems, where it may occur for completely unrelated physical reasons. Therefore, we intentionally introduce the new letter $\zeta_1$ for the prefactor of the logarithm to avoid any possible confusion with these better-known cases. These include ground states of non-random critical Hamiltonian systems in one dimension~\cite{vidal2003,calabrese2004}, where this form follows from conformal symmetry and the coefficient of the logarithm is related to the central charge $c$ of the corresponding conformal field theory (CFT) via $\zeta_1 = c/3$. It also appears in scenarios not related to conformal symmetry, such as random-singlet phases~\cite{refael2004entanglement} or, under particular circumstances, in ferromagnets and other symmetry-broken systems~\cite{popkov2005logarithmic} (in which case the prefactor is non-universal). Finally, this scaling of the entanglement entropy was found at the critical points that occur
in the interacting random circuit and tensor network 
models~\cite{li2018quantum,li2019measurement,vasseur2019entanglement,jian2020measurement,skinner2019measurement,LiChenLudwigFisher-arXiv2020HybridCircuits-conformal}, which also exhibit conformal symmetry and where the coefficient $\zeta_1$ of the logarithm is twice the scaling dimension of a boundary operator.

\begin{figure}
    \centering
    \includegraphics{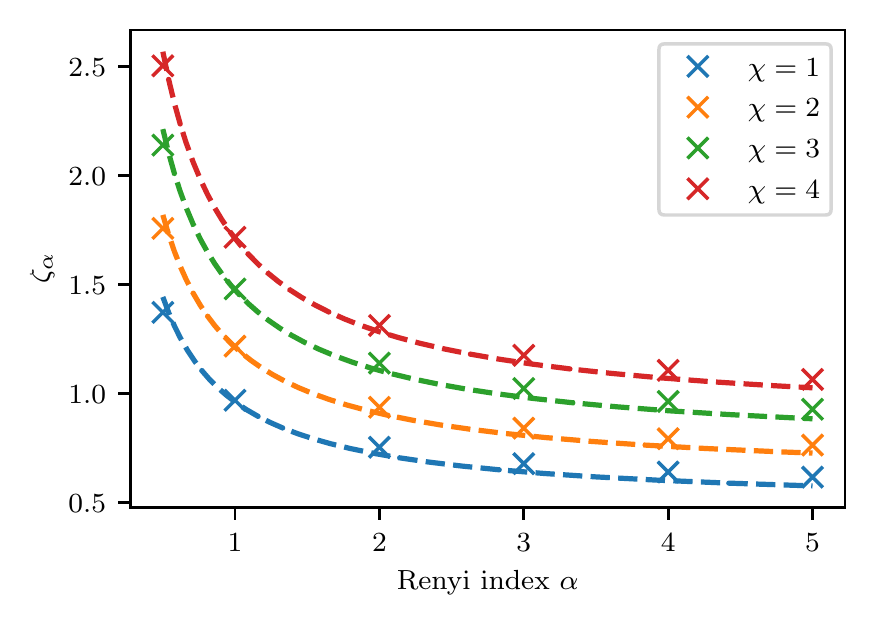}
    \caption{Coefficient of the logarithmic term in the entanglement entropy for Renyi entropies of different index $\alpha$; see Eqs.~\eqref{eqn:renyi} and \eqref{eqn:renyiB}. Crosses represent raw data, and dashed lines fits to the form $\zeta_\alpha= B(1+1/\alpha)$, where $B$ is the only fit coefficient.}
    \label{fig:renyi}
\end{figure}

We can repeat the analysis of the entanglement entropy for Renyi entropies, defined by
\begin{equation} \label{eqn:renyi}
S_\alpha(\rho) = \frac{1}{1-\alpha} \Tr \left[ \log \left(\rho^\alpha \right) \right],
\end{equation}
for our random GTN/NGCs.
For $\alpha=1$, the von Neumann entropy is obtained. For a one-dimensional non-random gapless Hamiltonian system, conformal field theory predicts~\cite{calabrese2009entanglement} that the ground state entanglement entropy of an interval of length $L$ (embedded in a much larger system) scales as $\zeta_\alpha \log (L/L_0)$ for all $\alpha$, where $\zeta_\alpha$ is given by
\begin{equation} \label{eqn:renyiB}
\zeta_\alpha = {\zeta_1\over 2}\left(1+\frac{1}{\alpha} \right),
\end{equation}
and 
where $\zeta_1 = c/3$ of Eq.~\eqref{eqn:vN-scaling} with $c$ the central charge.
For the measurement-induced transition in {\it interacting} random circuits, Ref.~\onlinecite{zabalo2020critical} observed that an additional constant term appears in the
$\alpha$-dependence of $\zeta_\alpha$ as compared to Eq.~\eqref{eqn:renyiB}.
We can numerically test the $\alpha$-dependence of $\zeta_\alpha$ in the random GTN. To this end, we compute the Renyi entropy with index $\alpha$ ranging from 0.5 to 5 for half of the system analogous to the calculation shown in Fig.~\ref{fig:entropyscaling} and perform a fit to extract the prefactor of the logarithmic scaling. Our results are shown in Fig.~\ref{fig:renyi}. We observe excellent agreement with the form of the $\alpha$-dependence displayed in Eq.~\eqref{eqn:renyiB} (without additional constant). We reiterate that $\zeta_1$ is not related to the central charge of a non-random CFT.

In addition to the entanglement entropy, we also examine the full entanglement spectrum. We find results consistent with the Gaussian unitary ensemble, see App.~\ref{sec:entanglement-spectrum}.

\subsection{Mutual information}

\begin{figure}
    \centering
    \includegraphics{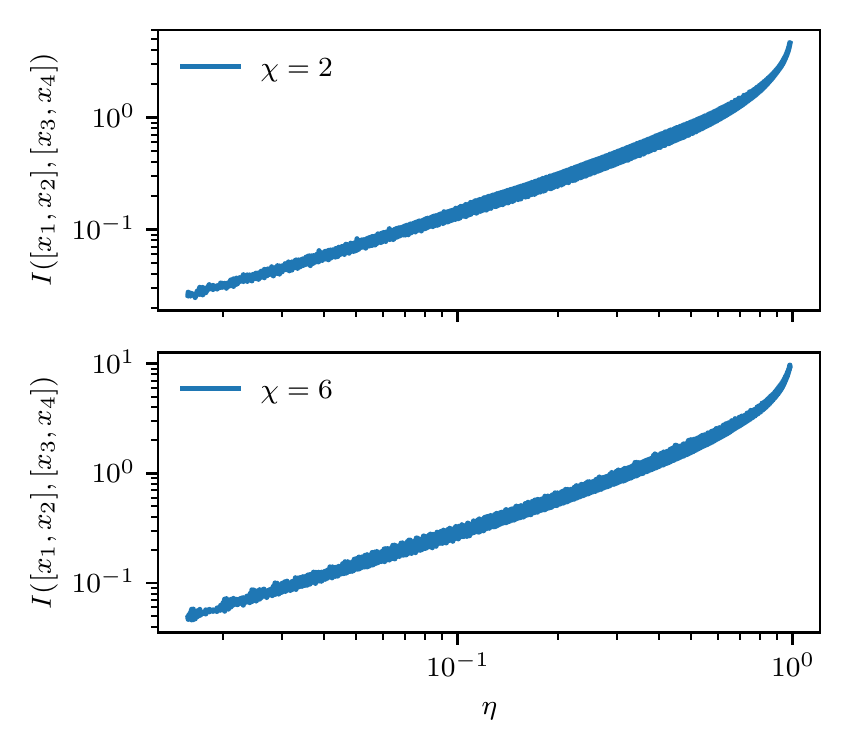}
    \caption{Mutual information between the disjoint segments $[x_1,x_2]$ and $[x_3,x_4]$ versus the cross-ratio $\eta$ of Eq.~\eqref{eqn:crossratio}. In both panels $L=100$. We exclude any data points where the intervals $[x_1,x_2]$ or $[x_3,x_4]$ are shorter than 4 sites, or the ends of the intervals are closer than 4 sites.}
    \label{fig:crossratio}
\end{figure}

Following Ref.~\onlinecite{li2019measurement}, we also compute the mutual information of two disjoint segments $[x_1,x_2]$ and $[x_3,x_4]$, which is given by
\begin{multline}
I([x_1,x_2],[x_3,x_4]) = S([x_1,x_2]) + S([x_3,x_4]) \\ - S([x_1,x_2] \cup [x_3,x_4]),
\end{multline}
where $S([x,y])$ denotes the von Neuman entanglement entropy between sites $x$ through $y$ and the rest of the system, and by $\cup$ we denote the union of two segments.
We plot this as a function of the cross-ratio of the segment endpoints. Defining the chord distance between points $x$ and $y$ as
\begin{equation}
r_{xy} = \frac{L}{\pi} \sin \left( \frac{\pi}{L} \left| x - y \right| \right),
\end{equation}
the cross-ratio is given by
\begin{equation}
\eta = \frac{r_{x_2 x_1} r_{x_4x_3}}{r_{x_3x_1} r_{x_4x_2}}. \label{eqn:crossratio}
\end{equation}
Our results are shown in Fig.~\ref{fig:crossratio}. We observe that all the mutual information data nicely collapse onto a function that depends only on a single variable $\eta$. This behavior of the mutual information is strongly suggestive of a two-dimensional conformal field theory description of the random GTN/NGC system. Moreover, we observe a power-law behavior of $I([x_1,x_2],[x_3,x_4])$ at small cross-ratio $\eta$.

We can examine the behavior for small $\eta$ in more detail by taking the limit where the intervals $[x_1,x_2]$ and $[x_3,x_4]$ are chosen to be short compared to their separation and to the system size. More precisely, we take $|x_1 - x_2| = |x_3 - x_4| = d$ and $|x_2 - x_3| = r$, and focus on the parameter regime with $d\ll r \ll L $ such that $\eta = d^2/r^2$. In this case, the mutual information $I([x_1,x_1+d],[x_1+r,x_1+r+d])$ with a fixed $d$ will decay with the distance $r$ between the intervals with the same functional form as the correlation function, Eq.~\eqref{eqn:corr-decay}. In particular, the corrections to the asymptotic power-law decay have the same form.

\section{Transfer matrix formalism and analytical understanding of the entanglement criticality }
\label{sec:transfer-matrix}

To facilitate an analytical approach to understand these numerical results, we now introduce a transfer matrix formalism for the contraction of pure-state GTNs.
Using this transfer matrix formalism, we will then map any lattice pure-state GTNs (with no uncontracted legs in the bulk) to a corresponding network model of unitary scattering problems on the same lattice. This type of network model that we obtain from the GTN turns out to be exactly what is commonly known as a Chalker-Coddington network model which was originally introduced to study non-interacting fermion systems with static/quenched disorder. This connection will allow us to understand the criticality observed in the previous section in terms of stable critical phases 
or critical points in disordered systems of non-interacting fermions. In the following, we will refer to the critical behavior obtained in Sec. \ref{sec:numerics} as the \emph{entanglement criticality} 
of the Haar-random GTN (and its corresponding NGC).
Interestingly and surprisingly, even though 
the entanglement criticality
is obtained in a most generic Haar-random pure-state GTN without any symmetry constraint, it shares the same description as the disordered metallic phase in Altland-Zirnbauer symmetry class DIII in two spatial dimensions.

\subsection{ Transfer Matrix: Definition and Properties}
\label{sec:TransferM_Intro}

\begin{figure}
    \centering
    \begin{overpic}[width=\columnwidth]{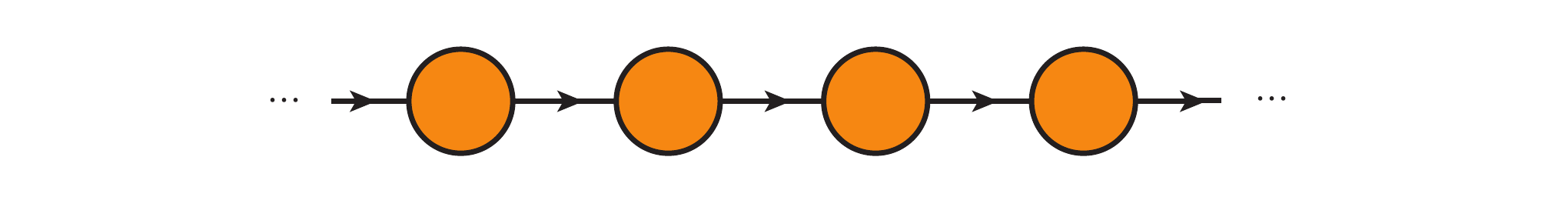} \put (10,10) {\footnotesize{(a)}} \end{overpic}
    \\
    \begin{overpic}[width=0.45\columnwidth]{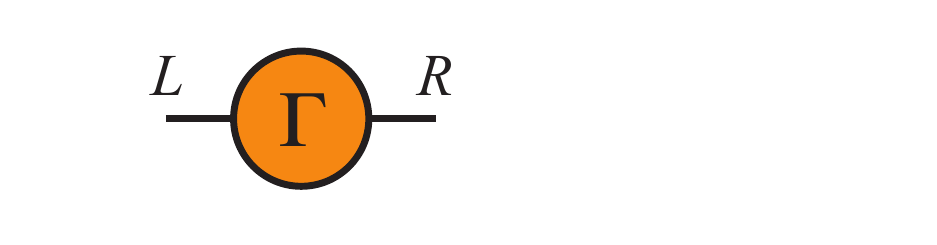} \put (0,20) {\footnotesize{(b)}} \end{overpic}
    \begin{overpic}[width=0.45\columnwidth]{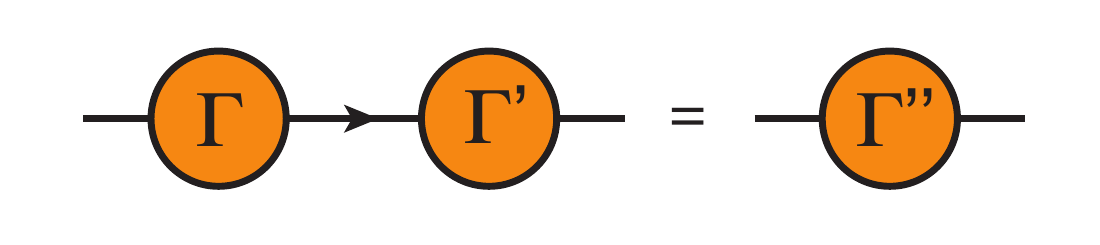} \put (0,20) {\footnotesize{(c)}} \end{overpic}
    \caption{(a) A one-dimensional GTN is shown. (b) A two-leg tensor $\Gamma$ in the one-dimensional GTN is shown. Its two legs are labeled by $L$ and $R$ respectively. (c) The contraction of the two two-leg tensors $\Gamma$ and $\Gamma'$ yields a third two-leg tensor $\Gamma''$. This contraction can be equivalently captured using the transfer matrix formalism.}
    \label{fig:1dGTN}
\end{figure}

To introduce the transfer matrix approach, we first consider the contraction of a one-dimensional GTN as shown in Fig.~\ref{fig:1dGTN}, where each tensor has two legs with the same Majorana bond number $\chit$. It turns out that the contraction of a GTN on a higher-dimensional lattice can always be reduced to this case (with a possibly system-size-dependent $\chit$) while still respecting locality; we will expand on this reduction to a one-dimensional GTN in Sec.~\ref{sec:TransferM_Sq}.

\subsubsection{Transfer matrix in one-dimensional geometry}
In the one-dimensional pure-state GTN shown in Fig.~\ref{fig:1dGTN}~(a), every tensor has two legs, each with a Majorana bond number $\chit$.
As discussed in Sec.~\ref{sec:TN_QC_relation}, we can view the one-dimensional pure-state GTN shown in Fig.~\ref{fig:1dGTN}~(a) as a quantum circuit with its time direction going from the left to the right of the GTN. This quantum circuit acts on a Hilbert space associated with $\chit$ Majorana fermion operators, which we denote as $\hat{\alpha}_{i=1,2,...,\chit}$. The fact that this tensor network is Gaussian implies that its corresponding quantum circuit always evolves a single Majorana fermion operator to another single Majorana fermion operator. Each two-leg tensor $\Gamma$ (for example the one in Fig.~\ref{fig:1dGTN}~(b)) in the one-dimensional pure-state GTN corresponds to a (non-unitary) quantum gate $g_\Gamma$ that induces a linear transformation of the Majorana fermion operators via 
\begin{equation}
\hat{\alpha}_i \rightarrow g_\Gamma \hat{\alpha}_i g_\Gamma^{-1} = \sum_j \ft_p[\Gamma]_{ij}\hat{\alpha}_j.
\end{equation}
Here, we have introduced the $\chit\times \chit$ matrix $\ft_p[\Gamma]$, which will be referred to as the P-sector transfer matrix of $\Gamma$. Since $g_\Gamma$ is a quantum gate of non-interacting fermions, the P-sector transfer matrix $\ft_p[\Gamma]$, which can be viewed as the first-quantized (or single-particle) version of $g_\Gamma$, contains the full information about $g_\Gamma$ and hence of the tensor $\Gamma$. In the following, we will only consider pure-state Gaussian tensors $\Gamma$ with fixed fermion parity $\Pf(\Gamma) = 1$ such that the corresponding quantum gate $g_{\Gamma}$ is a bosonic operator, i.e. $g_{\Gamma}$ is a quantum gate that preserves fermion parity.

As is shown in Fig.~\ref{fig:1dGTN}~(b), for each two-leg tensor $\Gamma$, we use the labels $L$ and $R$ to distinguish the two legs and their associated Majorana modes. We can write the covariance matrix $\Gamma$ shown in Fig.~\ref{fig:1dGTN}~(b) in a block form,
\begin{equation}
\Gamma =  \begin{pmatrix} \Gamma_{LL} & \Gamma_{LR} \\ \Gamma_{RL} & \Gamma_{RR}  \end{pmatrix},
\end{equation}
where the block $\Gamma_{LR}$ captures the correlations between the $\chit$ Majorana modes residing on the left leg and those on the right leg, and likewise for the other blocks. Since the Majorana bond number, i.e. the number of Majorana modes, associated with each leg is $\chit$, each block in $\Gamma$ is a $\chit\times \chit$ square matrix. The P-sector transfer matrix $\ft_p[\Gamma]$ of the tensor $\Gamma$ then turns out to be
given by
\begin{align}
\ft_p[\Gamma] = \Gamma_{LR}^{-1} \left(\mathds{1}  - \i \, \Gamma_{LL} \right) = \left(\mathds{1} + \i \, \Gamma_{RR} \right) \Gamma_{LR}^{-1},
\label{eqn:transferM_Def}
\end{align}
where the second equality is guaranteed by the pure-state condition $\Gamma^2 = -\mathds{1}$. The detailed derivation of the Eq.~\eqref{eqn:transferM_Def} is summarized in App.~\ref{app:TranferM_pGTN_Derivation}.

A key property of the transfer matrix, defined in this way, is that matrix multiplication of two such transfer matrices yields a result consistent with the contraction of the corresponding tensors. Specifically, consider, as shown in  Fig.~\ref{fig:1dGTN}~(c), the contraction of the two tensors $\Gamma$ and $\Gamma'$, which yields the tensor $\Gamma''$. This tensor contraction is equivalently described by the product of the two quantum gates $g_\Gamma$ and $g_{\Gamma'}$ that are associated with the two tensors $\Gamma$ and $\Gamma'$ respectively, i.e. $g_{\Gamma'} \cdot g_{\Gamma} = g_{\Gamma''} $ with $g_{\Gamma''}$ the quantum gate associated with the tensor $\Gamma''$. We therefore conclude that the transfer matrix corresponding to $\Gamma''$ is given by the product of those corresponding to $\Gamma$ and $\Gamma'$:
\begin{align}
    \ft_p[\Gamma'] \cdot \ft_p[\Gamma] = \ft_p [\Gamma''].
    \label{eqn:transferM_Multiply}
\end{align}
This result can also be checked explicitly using Eqs.~\eqref{eqn:SchurContraction} and \eqref{eqn:transferM_Def}.

In Eq.~\eqref{eqn:transferM_Def}, we've assumed that $\Gamma_{LR}$ is invertible, which is true for a generic pure-state tensor $\Gamma$ in the symmetric space $\frac{\SO(2\chit)}{\U(\chit)}$ of all possible $2\chit\times 2\chit$ pure-state covariance matrices. The exceptions to the assumption merely form a measure-zero subspace of $\frac{\SO(2\chit)}{\U(\chit)}$~\footnote{This subspace includes the gates that project the Majorana modes onto a state that is unentangled from the rest of the system.}. In the following, unless otherwise specified, we will assume the generic situation where $\Gamma_{LR}$ is invertible.

Given the conditions that $\Gamma^2 = -\mathds{1}$ and $\Gamma= -\Gamma^\T$, we notice that the P-sector transfer matrix satisfies the property that 
\begin{align}
    \ft_p[\Gamma]^\T \cdot \ft_p[\Gamma] = \mathds{1},
\end{align}
which means that the P-sector transfer matrix $\ft_p[\Gamma]$ belongs to the complexified special orthogonal group:
\begin{equation}
\ft_p[\Gamma] \in \SO(\chit)_\mathbb{C}.
\end{equation}
In the special case where $\Gamma_{LL}=\Gamma_{RR} = 0$, the quantum gate $g_\Gamma$ associated with the two-leg tensor $\Gamma$ becomes unitary and the P-sector transfer matrix $\ft_p[\Gamma]$ becomes real, i.e. $\Gamma \in \SO(\chit)$. 

As we have discussed before, 
the P-sector transfer matrix contains the full information of the tensor $\Gamma$.
Furthermore, for any element $\ft_p \in \SO(\chit)_\mathbb{C}$, there exists a pure-state two-leg Gaussian tensor $\Gamma$ such that the P-sector transfer matrix of $\Gamma$ is given by $\ft_p$ via Eq.~\eqref{eqn:transferM_Def}
(except for those $\ft_p$ with ${\rm Re}(\ft_p)$ non-invertible, a situation encountered only in a measure-zero subset of $\SO(\chit)_\mathbb{C}$).
Therefore, the elements of the symmetric space $\frac{\SO(2\chit)}{\U(\chit)}$ of all possible $2\chit\times 2\chit$ covariance matrices $\Gamma$, viewed as two-leg tensors, are in one-to-one correspondence with the elements of the space (and also group) $\SO(\chit)_\mathbb{C}$ of the P-sector transfer matrices (except for subsets of zero measure). 

\subsubsection{P- and H-sector transfer matrices}
As discussed earlier, from the perspective of quantum circuits, the (non-unitary) quantum gate $g_\Gamma$ associated with the two-leg tensor $\Gamma$ evolves the Majorana fermion operators in Hilbert space by $\hat{\alpha}_i \rightarrow g_\Gamma \hat{\alpha}_i g_\Gamma^{-1} = \sum_j \ft_p[\Gamma]_{ij}\hat{\alpha}_j$ leading to the definition of the P-sector transfer matrix $\ft_p[\Gamma]$. Similarly, we can define the $\chit\times \chit$ H-sector transfer matrix $\ft_h[\Gamma]$ by the evolution
\begin{equation}
\hat{\alpha}_i \rightarrow g_\Gamma^{-1\dag} \hat{\alpha}_i g_\Gamma^{\dag} = \sum_j \ft_h[\Gamma]_{ij}\hat{\alpha}_j
\end{equation}
It is easy to show that $\ft_h[\Gamma] = \ft_p[\Gamma]^*$. We can further introduce the full transfer matrix $\ft[\Gamma]$ for both sectors:
\begin{align}
    \ft[\Gamma] = \left( 
     \begin{array}{cc}
         \ft_p[\Gamma] & 0 \\
         0 & \ft_h[\Gamma] 
     \end{array} \right).
     \label{eqn:TransferM_Def_full}
\end{align}
Obviously, under tensor contraction, the H-sector transfer matrix $\ft_h$ and the full transfer matrix $\ft$ obey the same multiplication rule as the P-sector transfer matrix, i.e. Eq.~\eqref{eqn:transferM_Multiply}.

For a mixed-state Gaussian tensor $\Gamma$, i.e. $\Gamma^2 \neq -\mathds{1}$, only the full transfer matrix $\ft[\Gamma]$ remains well-defined (see App.~\ref{app:TranferM_mGTN_Derivation} for  a detailed discussion of the full transfer matrix for mixed-state Gaussian tensors). The decoupling of the full transfer matrix into the block diagonal form Eq.~\eqref{eqn:TransferM_Def_full} with the P-sector and the H-sector transfer matrices can be viewed as a special property of the pure-state tensor $\Gamma$ with $\Gamma^2 = -\mathds{1}$. As such, it is natural to interpret $\ft[\Gamma]$ as the evolution of the density matrix,
and the P-sector transfer matrix as a single-particle (or first quantized) description of the evolution of ket vectors in the Hilbert space. Likewise, the H-sector transfer matrix should be viewed as a single-particle (or first quantized) description of the evolution of bra vectors in the Hilbert space. This interpretation of the P-sector, H-sector and full transfer matrix follows naturally from the detailed derivation of them in App. \ref{app:TranferM_pGTN_Derivation}.

For a pure-state tensor $\Gamma$, given that its P-sector transfer matrix $\ft_p[\Gamma]$ already contains all the information of the two-leg tensor $\Gamma$, the introduction of the H-sector transfer matrix $\ft_h[\Gamma]$ and the full transfer matrix $\ft[\Gamma]$ may naively seem redundant. However, as we will see in the later subsections, the full transfer matrix will serve as an important tool for us to establish an exact mapping between the pure-state GTN (together with its corresponding non-unitary quantum circuit) and a network model of unitary scattering centers which can be viewed as
arising from a unitary Hamiltonian system of non-interacting fermions. Also, we will see that the structure of the full transfer matrix of the GTN ensures the unitarity of the corresponding network model. More interestingly, the structure of the full transfer matrix guarantees the symmetries of the network model even when no symmetry constraints is imposed on its correspond pure-state GTN. 

\subsubsection{Transfer matrix of the square-lattice pure-state GTN}
\label{sec:TransferM_Sq}

\begin{figure}
    \centering
    \includegraphics[width=1\columnwidth]{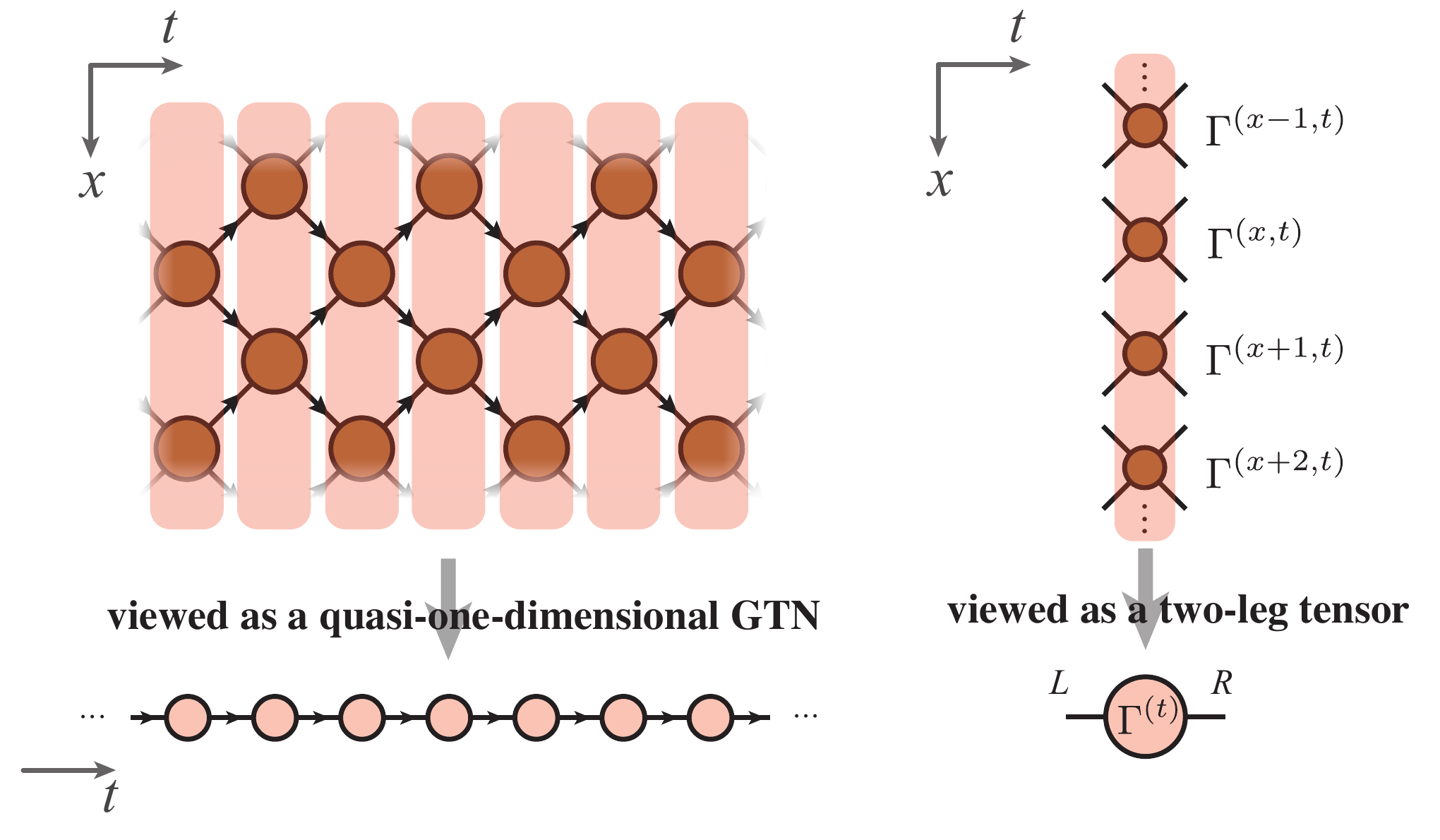}
    \caption{We can view the square-lattice GTN as a quasi-one-dimensional GTN along the $t$ direction. Each two-leg tensor $\tG{t}$ in the quasi-one-dimensional GTN consists of all four-leg tensors $\tG{x,t}$ in the square-lattice GTN that share the same $t$ coordinate. }
    \label{fig:GTN_dimensional_reduction}
\end{figure}

We now turn to the two-dimensional GTN introduced in Sec.~\ref{sec:gtn}, and show how to reduce it to a quasi-one-dimensional system that can be tackled with the transfer matrix tools introduced in the previous sections while preserving locality.
As shown in Fig.~\ref{fig:GTN_dimensional_reduction}, a square-lattice GTN can be viewed as a quasi-one-dimensional GTN in the $t$-direction. In this mapping, an entire column of tensors in the square-lattice GTN is represented by a single tensor, as illustrated in the right panel of Fig.~\ref{fig:GTN_dimensional_reduction}. This tensor is described by a covariance matrix $\tG{t}$ of size $4 L_x \chi \times 4 L_x \chi$, where $L_x$ is the extent of the tensor network in the $x$ direction, that is given by the direct sum of the covariance matrices of all $\tG{x,t}$ with the same coordinate $t$:
\begin{align}
    \tG{t} =\begin{psmallmatrix}
\ddots & & & \\
& \tG{x,t}& & \\
& & \tG{x+1,t}& \\
& & & \ddots
\end{psmallmatrix}.
\end{align}

This block structure is a direct consequence of the locality of the tensor network, and is preserved after rewriting the contraction in the transfer matrix language. This can be seen explicitly by applying Eq.~\eqref{eqn:transferM_Def}, which respects the block-form of the covariance matrix and thus yields a P-sector transfer matrix $\ft_p[\tG{t}]$ of size $2 L_x \chi \times 2 L_x \chi$ and of the form
\begin{align}
    \ft_p[\tG{t}] =\begin{psmallmatrix}
\ddots & & & \\
& \ft_p[\tG{x,t}]& & \\
& & \ft_p[\tG{x+1,t}]& \\
& & & \ddots
\end{psmallmatrix}.
\label{eqn:sq_lattice_tp_block}
\end{align}
Here, to apply Eq.~\eqref{eqn:transferM_Def} to each four-leg tensor $\tG{x,t}$ shown in Fig.~\ref{fig:GTN_dimensional_reduction}, we've grouped the two legs on the left and the two on the right together, respectively.
Similarly, the H-sector transfer matrix $\ft_h[\tG{t}]$ and the full transfer matrix $\ft[\tG{t}]$ of the quasi-one-dimensional GTN (obtained from the square-lattice GTN) are given by the direct sums of $\ft_h[\tG{x,t}]$ and $\ft[\tG{x,t}]$, respectively. 

At this point, we've concluded that, at each given time coordinate $t$, the P-sector transfer matrix $\ft_p[\tG{t}]$ has a block-diagonal form with each diagonal block having dimension $2\chi \times 2\chi$. However, it is important to note that the transfer matrix for the full tensor network does not have this block structure, since the positions of the diagonal blocks in the transfer matrices $\ft_p[\tG{t}]$ and $\ft_p[\tG{t+1}]$ at consecutive times $t$ and $t+1$ are shifted relative to each other by $\chi$ in their row and column indices. Therefore, the product $\prod_{t=1}^{L_t} \ft_p[\tG{t}]$ that describes the entire square-lattice GTN does not ``decouple" into small blocks, and can thus be a generic element of $\SO(\chit)_\mathbb{C}$.

\subsection{Mapping to Unitary Scattering Problems with TR, PH and Chiral Symmetries}

In this subsection, we use the transfer matrix formalism to establish an exact mapping the between a single pure-state Gaussian tensor (together with its corresponding quantum gate) and a unitary scattering problem with a static and non-interacting Hamiltonian. Applying this mapping to a lattice of such tensors (free of uncontracted legs in the bulk) yields a network model of unitary scatterers that resides on the same lattice. Such network models are commonly known as Chalker-Coddington network models and were introduced as lattice models for problems of non-interacting fermions subject to
static/quenched disorder.

We will show that even in the absence of any constraints on the pure-state Gaussian tensors $\Gamma$,
the corresponding scattering problems always have time-reversal (TR) symmetry, particle-hole (PH) symmetry and chiral symmetry, which corresponds to symmetry class DIII in the Altland-Zirnbauer ten-fold symmetry classification. Applying this to the random GTNs numerically studied in Sec.~\ref{sec:numerics} allows us to identify the criticality observed there with the known disordered metallic phase in symmetry class DIII in two spatial dimensions. We can thus compute properties of the entanglement criticality from the theory of this metallic phase. Importantly, this implies that the entanglement criticality observed in Sec.~\ref{sec:numerics} should be viewed as a critical entanglement phase that is stable against sufficiently weak deformation of the Haar-random ensemble. At the end of this section, we will show that, by deforming the Haar-randomness of the ensemble and by introducing ``staggering" in the square-lattice pure state GTN, one can access a transition from critical entanglement phase to a area-law entanglement phase. This transition is the same as the known metal-to-insulator transition in symmetry class DIII in two spatial dimensions.

\subsubsection{Mapping a single pure-state Gaussian tensor to a unitary scattering problem with a single scattering center}
\label{sec:scattering_single_tensor}

\begin{figure}
    \includegraphics[width=1\columnwidth]{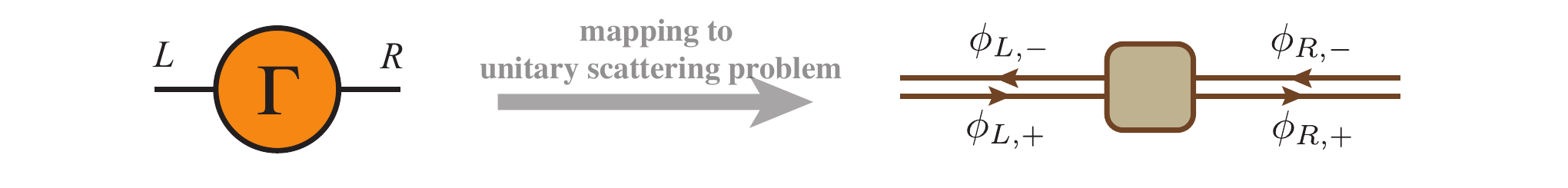}
    \caption{A generic pure-state Gaussian tensor $\Gamma$ can be mapped to a unitary scattering problem with its transfer matrix given by $\ft[\Gamma]$. Every Majorana mode associated
    with the Gaussian tensor $\Gamma$ corresponds to a pair of counter-propagating modes in the scattering problem. The scattering problem respects unitarity, TR symmetry, PH symmetry and chiral symmetry.}
    \label{fig:scattering_1d}
\end{figure}

We start by considering a single two-leg pure-state Gaussian tensor $\Gamma$ with the Majorana bond number on each leg given by $\chi_s$. It is straightforward to verify that its full transfer matrix $\ft[\Gamma]$ satisfies the condition
\begin{align}
    \ft[\Gamma]^\dag \cdot \Jp \cdot \ft[\Gamma] = \Jp
    \label{eqn:pGTN_TransM_Jconserve}
\end{align}
where
\begin{align}
    \Jp = \begin{pmatrix}
    0 &  \mathds{1} \\
    \mathds{1} & 0
    \end{pmatrix}.
    \label{eqn:DIII_conserved_current}
\end{align}
To understand the physical implication of this relation, it is instructive to interpret the transfer matrix $\ft[\Gamma]$ as describing a scattering problem with two $\chi_s$-component modes with amplitudes $\phi_{L, p}$ and $\phi_{L,h}$ on the left hand side of the scattering center and two other $\chi_s$-component modes with amplitudes $\phi_{R, p}$ and $\phi_{R,h}$ on the right hand side. The relation between the modes on the left and the right of the scattering center is given by the transfer matrix
\begin{align}
    \left(
    \begin{array}{c}
         \phi_{R, p}  \\
         \phi_{R, h}
    \end{array}
    \right)
    = \ft[\Gamma]
    \left(
    \begin{array}{c}
         \phi_{L, p}  \\
         \phi_{L, h}
    \end{array}
    \right),
    \label{eqn:1d_scacttering_transferM}
\end{align}
We can interpret the operator $\Jp$ as the (single-particle) probability current operator and view Eq.~\eqref{eqn:pGTN_TransM_Jconserve} as the conservation of the probability current in the scattering problem. With the probability current conserved, this scattering problem is unitary and, hence, should be viewed as arising from a static (Hermitian) Hamiltonian.

Note that the modes $\phi_{L/R, p/h}$ are not eigenstates of the probability current operator $\Jp$. The eigenstates of $\Jp$ are instead given by (see Fig.~\ref{fig:scattering_1d})
\begin{eqnarray}
\phi_{L,\pm} &=& \frac{1}{\sqrt{2}} (\phi_{L,p} \pm  \phi_{L,h}) \label{eqn:phi_L_pm}\\
\phi_{R,\pm} &=& \frac{1}{\sqrt{2}}(\phi_{R,p} \pm  \phi_{R,h})
\label{eqn:phi_R_pm}
\end{eqnarray}
Here, the $\phi_{L,\pm}$ and $\phi_{R,\pm}$ each form $\chi_s$ pairs of counter-propagating modes. Based on the physical meaning of the full transfer matrix $\ft[\Gamma]$ in the GTN context, we can associate each such pair of counter-propagating modes with a Majorana mode of the tensor $\Gamma$. In this example, the modes labeled $L$ and $R$ correspond to the two legs of the tensor $\Gamma$; a similar assignment of pairs of counter-propagating scattering modes to legs of the tensor can be made in the more general case of tensors with more than two legs.

Another perspective is gained by thinking of the modes $\phi_{L,+}$ and $\phi_{R, -}$ as the in-states, namely the modes traveling towards the scattering center and the modes $\phi_{L,-}$ and $\phi_{R, +}$ as the out-states, i.e. the modes traveling away from the scattering center. This allows us to define the scattering $S$-matrix of this scattering problem by
\begin{align}
    \left(
    \begin{array}{c}
         \i \phi_{L, -}  \\
          \phi_{R, +}
    \end{array}
    \right)
    \equiv S
    \left(
    \begin{array}{c}
         \i \phi_{L, +}  \\
         \phi_{R, -}
    \end{array}
    \right),
    \label{eqn:1d_Smatrix_Def}
\end{align}
where the $S$-matrix is a $2\chi_s \times 2\chi_s$ matrix that relates the in-states to the out-states. The factors of $\i$ in the definition of the $S$-matrix above are merely gauge choices. For a scattering problem arising from a static Hamiltonian, this $S$-matrix is expected to be unitary. The $S$-matrix can be obtained as follows. We plug the expression Eq.~\eqref{eqn:transferM_Def} of $\ft[\Gamma]$ into Eq.~\eqref{eqn:1d_scacttering_transferM} and apply the basis transformations shown in Eq.~\eqref{eqn:phi_L_pm} and Eq.~\eqref{eqn:phi_R_pm} to obtain a linear relation among $\phi_{L,\pm}$ and $\phi_{R,\pm}$. By re-arranging this linear relation into the form shown in Eq.~\eqref{eqn:1d_Smatrix_Def}, we obtain the $S$-matrix in this scattering problem:  
\begin{align}
    S = \i \, \Gamma,
    \label{eqn:1d_scacttering_Smatrix}
\end{align}
which, given that $\Gamma^\T = -\Gamma$ and $\Gamma^2 = -\mathds{1}$, is indeed unitary.

We can now discuss the symmetries in this unitary single-particle scattering problem. For any two-leg pure-state tensor $\Gamma$, the associated scattering problem has time-reversal (TR), particle-hole (PH) and chiral symmetries. At the level of the transfer matrix of a single scattering center, these symmetries correspond to the following three conditions:
\begin{align}
\begin{split} \text{TR symmetry:}~~&\Theta_{tr}^\dag \cdot \ft[\Gamma]^* \cdot \Theta_{tr} = \ft[\Gamma] \\
    &~~~~~~\text{with}~~
\Theta_{tr} = 
    \left( 
    \begin{array}{cc}
    0 &  -\mathds{1} \\
     \mathds{1} & 0 
    \end{array}
    \right)
\end{split} 
\label{eqn:full_transferM_TR_sym}
\\
\begin{split} \text{PH symmetry:}~~&\Xi_{ph}^\dag \cdot \ft[\Gamma]^* \cdot \Xi_{ph} = \ft[\Gamma] \\
    &~~~~~~\text{with}~~
\Xi_{ph} = 
    \left( 
    \begin{array}{cc}
    0 &  \mathds{1} \\
     \mathds{1} & 0 
    \end{array}
    \right),
\end{split} 
\label{eqn:full_transferM_PH_sym}
\\
\begin{split} \text{Chiral symmetry:}~~&\Sigma_{c} \cdot \ft[\Gamma] \cdot \Sigma_{c} = \ft[\Gamma] \\
    &~~~~~~\text{with}~~
\Sigma_c = 
    \left( 
    \begin{array}{cc}
     \mathds{1} & 0 \\0 &  -\mathds{1} 
    \end{array}
    \right).
\end{split}
\label{eqn:full_transferM_chiral_sym}
\end{align}
These conditions are automatically satisfied by any pure-state Gaussian tensor $\Gamma$. Here, we note that the TR symmetry squares to $-1$ ($\Theta_{tr}^2 =-1$) and the PH symmetry squares to $+1$ ($\Xi_{ph}^2 = +1$). The chiral symmetry can be viewed as a product of the TR and the PH symmetries. Therefore, this scattering problem belongs to 
symmetry class DIII in the ten-fold Altland-Zirnbauer symmetry classification. As we point out earlier, the transfer matrix corresponding to the scattering problem, $\ft[\Gamma]$, just like the P-sector transfer matrix $\ft_p[\Gamma]$, should be identified as an element of the  complexified special orthogonal group $\SO(\chi_s)_\mathbb{C}$. Based on Eq.~\eqref{eqn:1d_scacttering_Smatrix}, the scattering $S$-matrix should be identified as a point in the symmetric space $\frac{\SO(2\chi_s)}{\U(\chi_s)}$ of the pure-state Gaussian tensors $\Gamma$. The group formed by the full transfer matrices $\ft[\Gamma]$ (which are also transfer matrix in the scattering problem) and the symmetric space of the scattering $S$-matrix $S$ we obtained here for symmetry class DIII are consistent with the classification given in Refs.~\onlinecite{SchnyderRyuFurusakiLudwig, Ludwig2013}. 

As we now see, the identification of a conserved probability current and the identification of all the symmetries in the scattering problem fundamentally rely on the existence of both the P sector and the H sector in the full transfer matrix $\ft[\Gamma]$. In the scattering problem, this is evident from the fact that all three symmetries act only within the pairs of counter-propagating modes, where each such pair corresponds to one Majorana mode of the Gaussian tensor $\Gamma$.

\subsubsection{Mapping a lattice GTN to a lattice network model of scattering problems}
\label{sec:GTN_CC_mapping}

The mapping between a single two-leg pure-state Gaussian tensor $\Gamma$ and a unitary scattering problem introduced above can be straightforwardly extended to a correspondence between a lattice of Gaussian tensors (with no uncontracted legs in the bulk) and a lattice of unitary scattering problems defined at scattering centers located at the sites/vertices of the lattice. For the simplest case, consider the one-dimensional ``tensor network" shown in Fig.~\ref{fig:1dGTN}~(c) which consists of two pure-state Gaussian tensors $\Gamma$ and $\Gamma'$. Such a tensor network can be mapped to a one-dimensional network model of scattering problems that consists of two scattering centers. The transfer matrices at these two scattering centers are given by $\ft[\Gamma]$ and $\ft[\Gamma']$ respectively. Global unitarity follows immediately from the fact that $\ft[\Gamma''] = \ft[\Gamma'] \cdot \ft[\Gamma]$ is a valid transfer matrix that obeys probability conservation, i.e.
\begin{multline}
\left(\ft[\Gamma']\cdot \ft[\Gamma]\right)^\dag \cdot \Jp \cdot \left(\ft[\Gamma']\cdot \ft[\Gamma]\right) \\=
\ft[\Gamma'']^\dag \cdot \Jp \cdot \ft[\Gamma''] = \Jp.
\end{multline}
By analogous arguments, TR, PH and chiral symmetries of the transfer matrix of an individual tensor are inherited by the transfer matrix describing the entire tensor network.

\begin{figure}
    \centering
    \begin{overpic}[width=1\columnwidth]{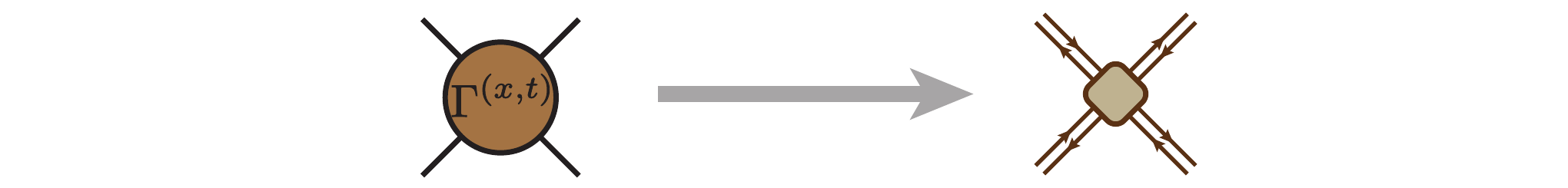} \put (5,10) {\footnotesize{(a)}} \end{overpic}
    \\
    ~~
    \\
    \begin{overpic}[width=1\columnwidth]{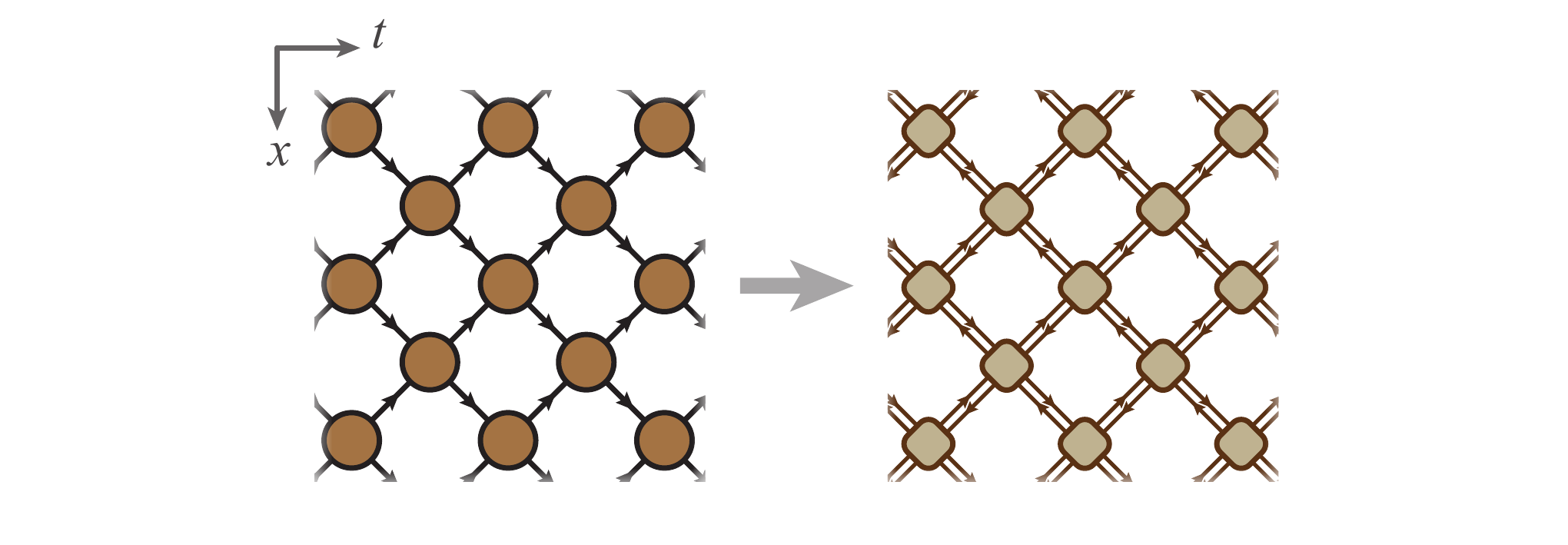} \put (5,30) {\footnotesize{(b)}} \end{overpic}
    \caption{(a) Each four-leg tensor in the square-lattice GTN is mapped to a unitary scattering problem with TR, PH and chiral symmetry on a four-leg geometry. (b) Applying this mapping to each four-leg tensor, we map the square-lattice pure-state GTN to a network model of unitary scatters on the square lattice. The network model also respects TR, PH and chiral symmetries.}
    \label{fig:scattering_Sq}
\end{figure}

Clearly, this argument can be iterated for a one-dimensional chain of two-leg tensors. To generalize to more complex geometries, we need to consider the case of tensors with more than two legs. This generalization is again straightforward: as we have discussed in the previous section, the scattering problem for an individual tensor is conveniently constructed in terms of pairs of counter-propagating modes, where each pair corresponds to one Majorana mode of the original tensor. Just like we have grouped the Majorana modes of the original tensor into legs, each carrying $\chi$ modes, we can group such pairs of counter-propagating modes into legs of the tensor. Thus, for a tensor with $r$ legs of Majorana bond number $\chi$, there will be $r$ sets of $\chi$ pairs of counter-propagating modes. Noting again that the symmetries act only within these pairs, it is clear that the same symmetry properties that hold for the two-leg tensors also hold for tensors with an arbitrary number of legs. This is illustrated for the case of a four-leg tensor in Fig.~\ref{fig:scattering_Sq}~(a).

Applying this mapping to every four-leg tensor in the pure-state square-lattice GTN, we obtain a network model of 
scattering problems
on the square-lattice with a local scattering center at each site of the square lattice as show in Fig.~\ref{fig:scattering_Sq}~(b). To ensure global unitarity of the square-lattice network model, we simply need to verify that the conserved probability currents at each scattering center are globally compatible with each other. One way to see the global compatibility of the probability current is to view the square-lattice network model (and its corresponding square-lattice GTN) as a network (and a GTN) on a quasi-one-dimensional geometry along the $t$-direction. We can identify the globally conserved probability current as the probability current along the $t$-direction in this quasi-one-dimensional geometry. Therefore, the square-lattice network model of scattering problems obtained 
from 
the square-lattice pure-state GTN is a unitary model that can be viewed as arising from a static non-interacting (Hermitian) Hamiltonian. This network model naturally inherits the TR, the PH and the chiral symmetries from the scattering problems at each site of the square lattice and thus belongs to symmetry class DIII in the Altland-Zirnbauer ten-fold symmetry classification. We emphasize that the mapping we describe is applicable to any realization of the square-lattice GTN with no constraints on each constituent pure-state Gaussian tensor.

\subsection{Critical Entanglement Phase as Symmetry-Class-DIII Disordered Metallic Phase}

\subsubsection{Identification of critical entanglement phase}
\label{LabelSubSubSectionCriticalEntanglementPhase}

In Sec.~\ref{sec:GTN_CC_mapping}, we have introduced a mapping
between any pure-state square-lattice GTN and a unitary network model of scatterers on the square lattice. The latter is an example of what is commonly known as a Chalker-Coddington network model. Such models were originally introduced to study the physics of non-interacting fermions subjected to static/quenched disorder. Such disordered non-interacting fermion problems obviously admit, in any fixed realization of disorder, descriptions by static (and Hermitian) Hamiltonians. In general, for a Hamiltonian problem of non-interacting fermion in $D$ spatial dimensions, the corresponding Chalker-Coddington network model also resides in $D$ spatial dimensions. This is because, in the absence of interactions, the fermion modes with different real (or Matsubara) frequency decouple, and the Chalker-Coddington network models are used to describe a slice of fixed real (or Matsubara) frequency. The quenched disorder in the non-interacting fermion problem is captured by randomness at each scattering center~\footnote{Some of that randomness may be transferred, depending on the case, to randomness on links} in the Chalker-Coddington network. If the disordered non-interacting fermion problem respects certain symmetries, the scatterers in the Chalker-Coddington network should respect the same set of symmetries. A more detailed review of the Chalker-Coddington model can be found in App.~\ref{app:CC_intro}.

We are now in a position to relate the criticality observed in Sec.~\ref{sec:numerics} to known results.
We have mapped the most generic Haar-random ensemble of square-lattice pure-state GTNs into a random/disordered ensemble of Chalker-Coddington network models
on the square lattice, where each realization of disorder of the network model preserves TR symmetry (with $\Theta_{\rm tr}^2 =-1$), PH symmetry (with $\Xi_{\rm ph}^2 =1$) and chiral symmetry. Therefore, by the mapping between Chalker-Coddington network models and static Hamiltonians in the same dimension, the problem of Haar-random pure-state square-lattice GTNs can be viewed as the problem of unitary systems of disordered non-interacting fermions in symmetry class DIII and in two spatial dimensions. It is known that this latter model exhibits a metallic
phase~\cite{ZirnbauerSusySymmSpacesMJathPhys1996,ZinnJustin2002QuantumFieldTheoryCriticalPhen}.
The critical entanglement phase observed in Sec.~\ref{sec:numerics} should thus naturally be identified with this critical phase.

Therefore, properties of the critical entanglement phase that we observe numerically should be described by the theory of the corresponding metallic phase.
The renormalization-group fixed point governing the universal behavior of this phase turns out to be a two-dimensional conformal field theory of free scalar fields. A more detailed description of this fixed point theory is provided in App.~\ref{app:CC_intro}. The numerically obtained logarithmic scaling of the half-system entanglement entropy shown in Eq.~\eqref{eqn:vN-scaling} (and in Fig.~\ref{fig:entropyscaling}) and the scaling collapse of mutual information as a function of the cross-ratio shown in Fig.~\ref{fig:crossratio} are both non-trivial numerical verifications of the the criticality and of the conformal symmetry of the critical entanglement phase. 

Stronger evidence consistent with the specific metallic phase
is observed from the numerically obtained second disorder moment of the two-point Majorana fermion correlation function
shown in Fig.~\ref{fig:correlations}, which fits nicely with the particular scaling form of Eq.~\eqref{eqn:corr-decay}, which can be derived from the metallic fixed point that describes the symmetry-class-DIII disordered metallic phase in two spatial dimensions. The $1/r^2$ decay of Eq.~\eqref{eqn:corr-decay} is given by the equal-time correlations at the absorbing boundary of the disordered metallic phase (see App.~\ref{app:CC_intro} for details on this boundary condition) and the factor of $(1+\lambda_0 \log L)^2$ in Eq.~\eqref{eqn:corr-decay} results from the marginally irrelevant operator known to exist at this fixed point
(for details, see App.~\ref{app:marginal}).

It is important to note that this two-dimensional fixed point describing the metallic phase in symmetry class DIII has no relevant or marginally relevant perturbations allowed by symmetry. Therefore, since the symmetries are always present within the GTN construction, the entanglement criticality observed in Sec.~\ref{sec:numerics} is really a \emph{critical entanglement phase} that extends beyond the Haar-random pure-state GTN and is stable to \emph{any} weak perturbation of this GTN ensemble.
For further details, see App.~\ref{app:CC_intro}.

The theory of the metallic fixed point also predicts that the $N$'th disorder moment of the square of the two-point Majorana fermion correlation function $\overline{ \langle \i \hat{\gamma}_{p,m} \hat{\gamma}_{p+r,n} \rangle^{2N} }$ exhibits a $1/r^{2N}$ power-law decay. This is discussed at the end of Appendix \ref{LabelSubsection-AbsorbingBoundary}. The presence of the marginally
irrelevant operator will lead, on top of this power law, to logarithmic corrections to scaling (analogous to  those displayed in Eq.~\eqref{eqn:corr-decay} for $N=1$).
In contrast, all the corresponding disorder moments of the two-point Majorana fermion correlation functions in 
loop-model-based
circuit models (for example, models studied in Ref. \onlinecite{nahum2020entanglement}, including the 3-dimensional variant, 
in Ref. \onlinecite{Sang2021LoopModel}, and  also
in Ref. \onlinecite{lang2020entanglement}) would be independent of the order of the moment~\footnote{This follows because in any realization of disorder of the loop model, the elements of the covariance matrix~\eqref{eqn:GammaCondition}, which are exactly the Majorana correlators in question, are equal either to 0 or $\pm 1$. The non-zero entries correspond to pairs of Majorana modes connected by a loop. Therefore, any even power of the correlation function will be equal to 0 or $+1$, and all even moments must be exactly the same.}
(and are, depending on the particular loop-model, subject to corresponding logarithmic corrections to scaling).

\subsubsection{Transition from critical entanglement phase to area-law entanglement phase}

The study of the two-dimensional Chalker-Coddington network model in symmetry class DIII in Ref. \onlinecite{FulgaAkhmerovBeenakker-DIII-2012} shows that the transitions out of the critical metallic phase discussed in the previous section to a gapped phase can be induced by turning on a ``staggering pattern" of sufficient strength on the square lattice. In the language of GTNs or quantum circuits, this transition is a transition from the critical entanglement phase to an area-law entanglement phase. In the following, we will demonstrate that the same phenomenon can be observed by introducing a staggering deformation to the Haar-random pure-state square-lattice GTN  with 
Majorana bond number $\chi = 1$.

\begin{figure}
\centering
\begin{overpic}[width=0.4\columnwidth]{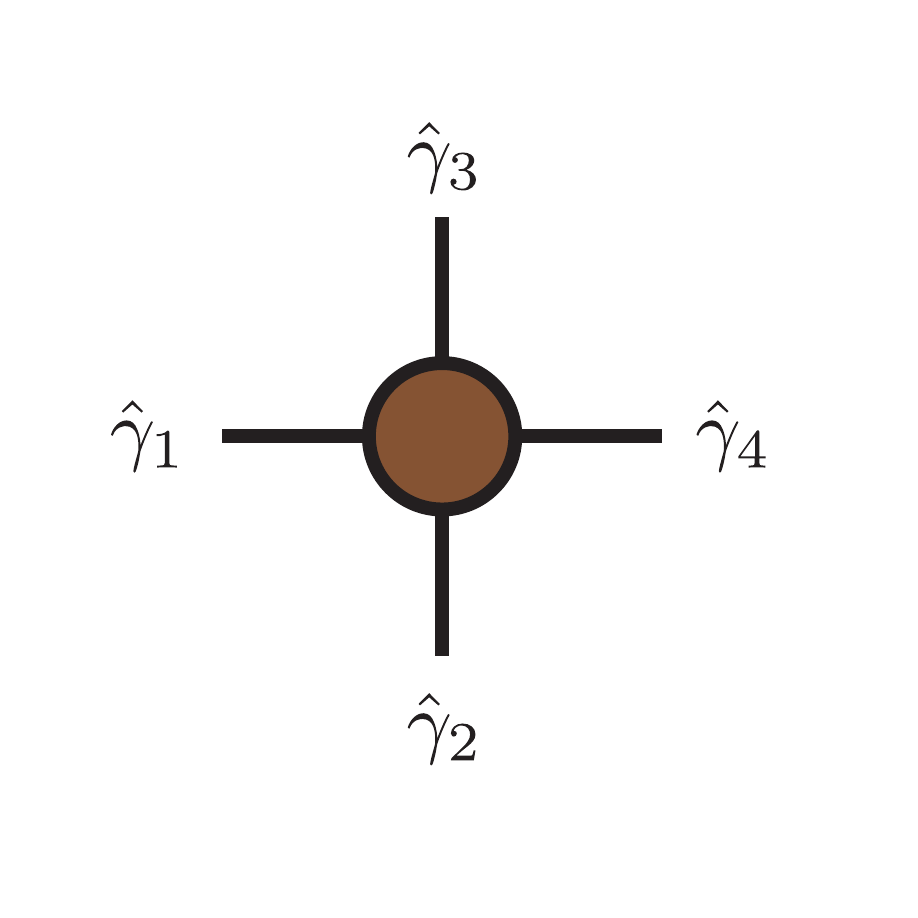} \put (0,100) {\footnotesize{(a)}} \end{overpic}
\begin{overpic}[width=0.5\columnwidth]{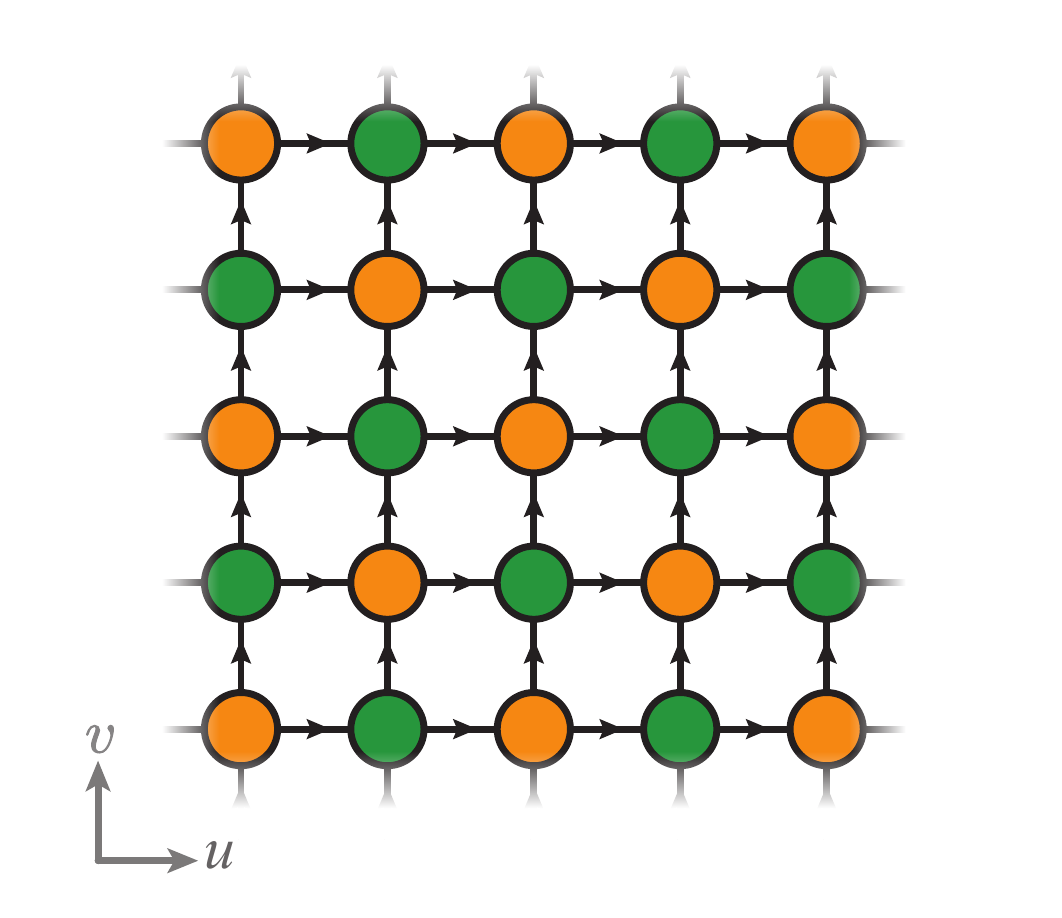} \put (0,80) {\footnotesize{(b)}} \end{overpic}
\caption{(a) For the four-leg tensor with Majorana bond number $\chi = 1$, the four Majorana modes associated with this tensor is labeled according to this figure. (b) The random GTN can be driven into an area-law entanglement phase upon staggering of the tensors on the two sub-lattices. The sublattice A is colored green while the sublattice B is colored orange. }
\label{fig:staggered}
\end{figure}

With Majorana bond number $\chi = 1$, each four-leg tensor of the square-lattice GTN has four Majorana modes associated with it, one for each leg as shown in Fig.~\ref{fig:staggered}~(a).
Each four-leg tensor is described by a $4\times 4$ covariance matrix $\Gamma_{ij} \equiv \langle \frac{\i}{2} [\hgamma_i, \hgamma_j]\rangle$ with $i,j=1,2,3,4$. For fixed fermion parity, this covariance matrix is, as discussed in Sec. \ref{sec:gtnA}, an element
of the coset space ${\rm SO(}4)/{\rm U}(2)\sim S^2$ (2-sphere), and
it can be parameterized by a real unit vector $\vec{n} = (n_1,n_2,n_3)$ in the following way:
\begin{align}
     \Gamma(\vec{n}) & = n_1 \, \i\sigma^{zy} +n_2 \, \i\sigma^{y0}+n_3 \, \i\sigma^{xy}, \nonumber \\
    & = 
    \left(
    \begin{array}{cccc}
        0 & n_1  & n_2  & n_3  \\
        -n_1 & 0 & -n_3 & n_2 \\
        -n_2 & n_3 & 0 & -n_1 \\
       -n_3 & -n_2 & n_1 & 0 \\
    \end{array}
    \right)
    \label{eqn:4leg_chi1_tensor}
\end{align}
where 
$n_1^2+n_2^2+n_3^2=1$, and $\sigma^{ab} \equiv \sigma^a \otimes \sigma^b$, where
$a,b=0,x,y,z$.
The Haar-random ensemble of $\Gamma(\vec{n})$ is given by the uniform distribution of $\vec{n}$ on a two-dimensional unit sphere.\footnote{We note in passing that the loop models discussed in 
Ref. \onlinecite{nahum2020entanglement}, Ref. \onlinecite{lang2020entanglement}, and Ref. \onlinecite{Sang2021LoopModel}
are highly special (fine-tuned) cases of this, corresponding to  covariance
matrices either (i): with $\vec{n} = (n_1,n_2,n_3) \in \{(1,0,0), (0,1,0)\}$, or
(ii): with $\vec{n} \in \{ (1,0,0), (0,1,0), (0,0,1)\}$, with corresponding probabilities.}

Now, we consider a random ensemble for these tensors, where, in order to introduce a staggered pattern on the square lattice as is shown in Fig.~\ref{fig:staggered}~(b), the tensors on the sub-lattice $A$ (green sites) and those on the sub-lattice $B$ (orange sites) are chosen from a different random distribution. 
A four-leg tensor on the sub-lattice $A$ ($B$) is generated by $\Gamma(\vec{n}_{A(B)}$) where
\begin{eqnarray}
\vec{n}_A &=& (\cos\theta,\sin\theta\cos\varphi,\sin\theta\sin\varphi) \\
\vec{n}_B &=& (\sin\theta\sin\varphi,\cos\theta,\sin\theta\cos\varphi).
\end{eqnarray}
Here, $\theta$ and $\varphi$ are random variables chosen independently for each four-leg tensor. The probability distribution for $\varphi$ is taken to be uniform in the interval $[0,2\pi\sigma)$, while $\theta=\arccos s$ with $s$ being a uniform random variable in $[1-2\sigma,1]$. 
The parameter $\sigma$, which controls the disorder strength, can be tuned from 0 to 1. Note that the random ensemble of GTN considered here corresponds to a Chalker-Coddington network model whose disorder realization is microscopically quite different from the disorder considered in Ref. \onlinecite{FulgaAkhmerovBeenakker-DIII-2012}. In the limit where $\sigma=0$, the GTN has a staggered pattern and becomes free of randomness. One can readily show analytically and numerically that the correlation function $C(r)$ defined in Eq.~\eqref{eqn:Cd} exhibits short-ranged exponential decay as opposed to power-law decay in this limit. This corresponds to a gapped phase of non-interacting fermions in two spatial dimensions. This gapped phase is expected to be stable against a finite amount of (quenched) disorder, i.e. it is expected to be stable for small $\sigma$. In the language of the random quantum circuit, this gapped phase of disordered fermions should be identified with an area-law entanglement phase (as opposed to the critical entanglement phase), a statement that will be confirmed by the numerical simulations presented later. 
For $\sigma=1$ the ensemble reproduces the Haar-random ensemble that we've introduced in Sec.~\ref{sec:numerics}. Therefore, we expect the same critical entanglement phase for large values of $\sigma$ as the one found in the Haar-random GTN in Sec.~\ref{sec:numerics}.

\begin{figure}
  \centering
  \begin{overpic}[width=0.8\columnwidth]{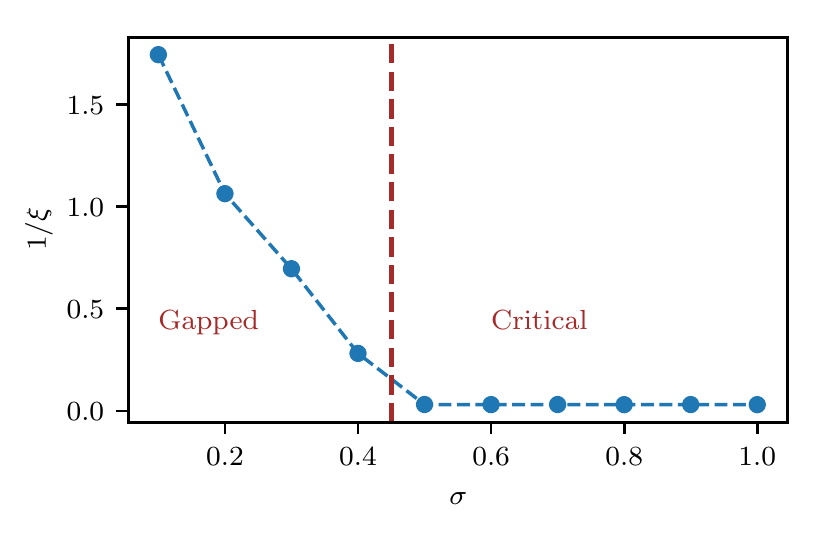} \put (-5,60) {\footnotesize{(a)}} \end{overpic} \\
  \begin{overpic}[width=0.8\columnwidth]{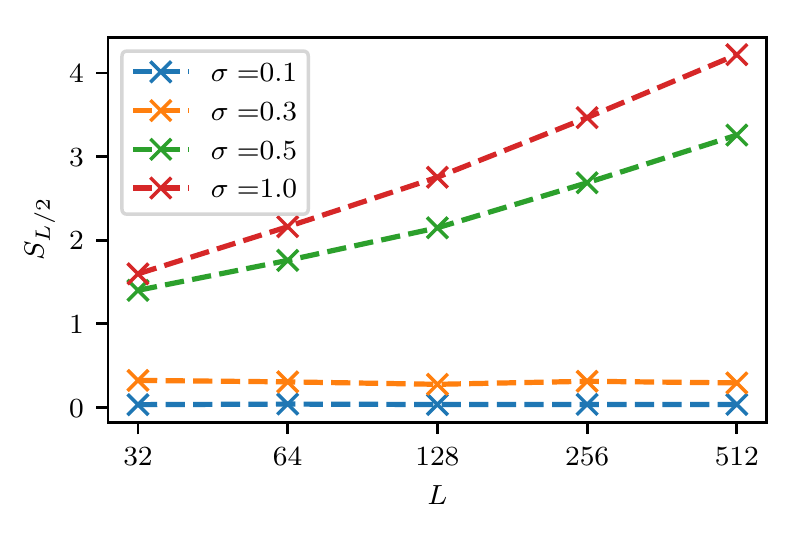} \put (-5,60) {\footnotesize{(b)}} \end{overpic}
  \caption{(a) Inverse correlation length $1/\xi$ extracted from the squared two-point correlation function in a staggered GTN with Majorana bond number $\chi=1$, as function of the disorder strength dictated by $\sigma$ (see main text).
  For $\sigma \gtrsim 0.45$ the correlation length diverges signaling a phase transition into a critical entanglement phase. (b) Scaling of the entanglement entropy of half of the system with system size for different values of the disorder strength $\sigma$.
    \label{fig:phase_transition}
  }
\end{figure}

We follow the same protocol and geometry as in Sec.~\ref{sec:numerics} and conduct numerical simulations of the random square-lattice pure-state GTN with the randomness (or disorder strength) parameterized by $\sigma$ as defined above. For the squared two-point correlation function $C(r)$ defined in Eq.~\eqref{eqn:Cd}, we fit it to the form $\exp\left(-r/\xi\right)$ and extract the correlation length $\xi$ for every choice of $\sigma$. The correlation length $\xi$ is expected to be finite in the area-law entanglement phase which corresponds to the localized (gapped) phase of disordered fermions in two spatial dimensions, and is expected to diverge as one approaches the critical entanglement phase from the area-law side. Indeed, numerical results shown in Fig.~\ref{fig:phase_transition}~(a) confirm our expectations. We also calculate numerically the half-system entanglement entropy $S_{L/2}$ (as defined in Sec.~\ref{sec:numerics}) in the ``large circuit depth" limit with $v\rightarrow \infty$. The numerical results presented in Fig.~\ref{fig:phase_transition}~(b) clearly show that when the correlation length $\xi$ is finite, $S_{L/2}$ is of order $1$, i.e. follows an area law. In contrast, $S_{L/2}$ follows a $\log L$ behavior when the correlation length diverges. The transition between the area-law entanglement phase and the critical entanglement phase occurs at $\sigma \gtrsim 0.45$. Even though our random GTN model studied here corresponds to a Chalker-Coddington network model that is microscopically different from the model studied in Ref. ~\onlinecite{FulgaAkhmerovBeenakker-DIII-2012}, our phase diagram is consistent with the phase diagram for symmetry class DIII in two spatial dimensions described in 
Ref.~\onlinecite{FulgaAkhmerovBeenakker-DIII-2012}.

\section{Random GTNs, NGCs, and Unitary Disordered Fermions in All Symmetries Classes}
\label{sec:nonunitary-to-unitary}

We have shown that any lattice pure-state GTN (including the Haar-random ensemble discussed above)
or non-unitary Gaussian circuit (NGC) can be mapped, in a fixed realization of disorder, to
a unitary Chalker-Coddington network model in symmetry class DIII residing on the same lattice. In this section, we will discuss how to obtain random ensembles of pure-state lattice GTNs that map to unitary disordered Chalker-Coddington network models in all the symmetry classes of the Altland-Zirnbauer ten-fold way symmetry classification.
The key to obtaining these other symmetry classes is to employ the idea of Clifford Algebra extensions \cite{Kitaev_2009} and consider additional constraints on the random ensemble of the covariance matrix $\Gamma$ that represents each Gaussian tensor in the GTN. The mapping introduced in Sec.~\ref{sec:transfer-matrix} still maps every realization of the GTN 
into a Chalker-Coddington
network model 
with TR symmetry, PH symmetry, and chiral symmetry, in every realization of disorder.
When the extra constraints on the random ensemble of Gaussian tensors of the GTN are properly chosen, the resulting disordered Chalker-Coddington network model can 
reside in any desired
symmetry classes in the Altland-Zirnbauer ten-fold way classification. In the following, we will first provide the construction of the complex symmetry classes AIII and A. In the discussion of symmetry class AIII, we will also present an embedding of symmetry class BDI into symmetry class AIII.
Following the discussion of the complex symmetry classes, we will also provide
a systematic construction of each of the eight real symmetry classes, namely symmetry classes DIII, AII, CII, C, CI, AI, BDI and D. Moreover, this construction makes connections between the entanglement criticality in GTNs/NGCs and criticality in unitary systems of disordered non-interacting fermions in all of the ten symmetry classes.

\subsection{Complex Symmetry Classes AIII and A}
\subsubsection{Symmetry Class AIII}
\label{sec:class_AIII}

To construct models in symmetry class AIII, we impose an additional  U$(1)$ symmetry constraint for each of the Gaussian tensors. This requires us to consider tensor networks of even Majorana bond number $\chi = 2n$. We will require that each 
tensor $\Gamma$ obeys the condition
\begin{align}
    [\Gamma, Q] = 0.
    \label{eqn:charge_conservation_AIII}
\end{align}
for a charge operator $Q$ that is given by:
\begin{align}
\label{LabelEq-DEF-Of-Q-in-AIII}
    Q \equiv \left( \begin{array}{cc}
        0 & -\i \\
        \i & 0
    \end{array}
    \right) \otimes \mathds{1}.
\end{align}
Here, we assume as usual
that the Majorana modes on $\Gamma$ are grouped together in legs, such that the $\begin{psmallmatrix}
 0 & -\i \\
\i & 0
\end{psmallmatrix}$ part of the operator $Q$ only acts within the Majorana modes on the same leg of the tensor and does not mix between the legs of the tensor.

For the whole GTN to respect the U(1) symmetry,
we also require that the U(1) symmetry is compatible with the tensor contractions, i.e. that when two U(1) symmetric tensors $\Gamma$ and $\Gamma'$ are contracted, the resulting tensor also satisfies the U(1) symmetry condition Eq.~\eqref{eqn:charge_conservation_AIII}. Remembering that the contraction between two tensors can be viewed as the projection onto a maximally entangled state of the Majorana modes residing on the contracted legs, the compatibility between the U(1) symmetry and the tensor contraction can be guaranteed by requiring that these maximally entangled states are also U(1) symmetric. When the whole pure-state GTN respects the U(1) symmetry, we can view the GTN as a charge-conserving
Gaussian tensor network based on complex fermions. As will be explained later, when the U(1) charge of each tensor is at half-filling, such a tensor network can be further interpreted as a charge-conserving
(non-unitary) quantum circuit acting on non-interacting complex fermions.

We are interested in the behavior of the ``maximally random" U(1)-symmetric lattice GTN and its corresponding random circuit.
To obtain this ensemble, consider the fact
that in a square-lattice GTN, a four-leg pure-state Gaussian tensor $\Gamma$ with Majorana bond number $\chi$ (and fixed fermion parity) can always be viewed as a point in the symmetric space $\frac{\SO(4\chi)}{\U(2\chi)}$ prior to imposing the condition of the $\U(1)$ symmetry.
With the extra U(1) symmetry constraint of Eq.~\eqref{eqn:charge_conservation_AIII}, the total space of the Gaussian tensor $\Gamma$ should be identified as the symmetric space $\frac{\U(2\chi)}{\U(\chi+q) \times \U(\chi-q)}$, where $q$ is the total U(1) charge of the Gaussian state $|\Gamma\rangle$ measured with respect to half filling. The appearance of the
symmetric space $\frac{\U(2\chi)}{\U(\chi+q) \times \U(\chi-q)}$ can be seen by noting that the U(1) symmetry constraint Eq.~\eqref{eqn:charge_conservation_AIII} ensures that there exists a complex fermion basis where the charge operator $Q$ takes the form $\begin{pmatrix}
\mathds{1}_{2\chi} & 0 \\ 0 & -\mathds{1}_{2\chi}
\end{pmatrix}$ and the covariance matrix takes the form $\begin{pmatrix}
\i G & 0 \\  0 & -\i G^*
\end{pmatrix}$ with $G$ a $2\chi\times 2\chi$ Hermitian 
matrices such that $G^2 = \mathds{1}_{2\chi}$. Physically, the matrix $\frac{1}{2}(G+ \mathds{1}_{2\chi})$ represents the two-point functions of complex fermions in the
U(1) symmetric Gaussian state $\ket{\Gamma}$. The space of $2\chi\times 2\chi$ Hermitian matrix $G$ with $\chi+q$ eigenvalues $+1$ and $\chi-q$ eigenvalues $-1$ is given by the symmetric space $\frac{\U(2\chi)}{\U(\chi+q) \times \U(\chi-q)}$.

For an isolated Gaussian tensor $\Gamma$, the total charge $q$ can, in principle, take any integer value between $-\chi$ and $\chi$. However, we require $q=0$ for each tensor in the GTN so that the GTN can be interpreted as a Gaussian quantum circuit that conserves U(1) charge. The charge conservation of the corresponding quantum circuit can be more conveniently understood in the language of the GTN.
Remember that the GTN is constructed by first forming a tensor-product of the Gaussian states given by each tensor in the GTN, and then by projecting the result of the tensor product onto the maximally-entangled-pair states on all the contracted legs. In the U$(1)$-conserving GTN, the U(1) charges of the maximally-entangled-pair states being projected onto are all at half-filling. Therefore, we need to require the U(1) charge of each Gaussian state given by each tensor to be also at half-filling, i.e. $q=0$, so that the U(1) charge of the state living on the boundary legs of the GTN, namely the total charge of state produced by the contraction of the GTN, is independent of the size of the GTN. Having set $q=0$ for each Gaussian tensor, the maximally random ensemble of pure-state square-lattice GTNs with U(1) symmetry
is given by choosing every four-leg Gaussian tensor in the GTN 
independently and randomly as a point in the symmetric space $\frac{\U(2\chi)}{\U(\chi) \times \U(\chi)}$ with the uniform probability measure on this symmetric space. 

Following the discussion in Sec.~\ref{sec:transfer-matrix}, we map each realization of the U(1) symmetric square-lattice pure-state GTN to a Chalker-Coddington network model on the square lattice. Based on Eq.~\eqref{eqn:1d_scacttering_Smatrix}, the Gaussian tensor $\Gamma$ on each site of the GTN should be identified as the scattering $S$-matrix of the scattering process occurring on the corresponding site in the Chalker-Coddington model. The classification given in Ref.~\onlinecite{SchnyderRyuFurusakiLudwig} tells us
that a Chalker-Coddington network model with its scattering $S$-matrix on each site residing in the symmetric space $\frac{\U(2\chi)}{\U(\chi) \times \U(\chi)}$ belongs to symmetry class AIII. 

To further confirm this symmetry class identification, we study the transfer matrix of 
each tensor in the U(1) symmetric square-lattice GTN. Similar to Sec.~\ref{sec:TransferM_Sq}, we treat each four-leg Gaussian tensor with each leg having Majorana bond number $\chi$ as a two-leg tensor with each leg having Majorana bond number $2\chi$. In the basis associated with the two legs ($L$ and $R$) of the Gaussian tensor $\Gamma$ where the covariance matrix $\Gamma$ is written as $\begin{psmallmatrix}
\Gamma_{LL} & \Gamma_{LR} \\
\Gamma_{RL} & \Gamma_{RR}
\end{psmallmatrix}$, the U(1) charge operator $Q$ can be chosen to take the form $ \begin{psmallmatrix}
\sigma^y \otimes \mathds{1}_{\chi} & 0 \\
0 & \sigma^y \otimes \mathds{1}_{\chi}
\end{psmallmatrix}$. As discussed in Sec.~\ref{sec:TransferM_Sq}, the full transfer matrix $\ft[\Gamma] = \begin{psmallmatrix}
\ft_p[\Gamma] & 0 \\ 0 & \ft_h[\Gamma]
\end{psmallmatrix}$ governs the scattering process at the site in the Chalker-Coddington network model given by the GTN. The U(1) symmetry of the Gaussian tensor $\Gamma$ leads to the condition that 
\begin{align}
   \big[\, \ft[\Gamma],\, \tilde{Q} \,\big] = 0,
   \label{eqn:Class_AIII_TransferM}
\end{align}
where $\tilde{Q} = \begin{psmallmatrix}
\sigma^y \otimes \mathds{1}_{\chi} & 0 \\
0 & \sigma^y \otimes \mathds{1}_{\chi}
\end{psmallmatrix} $. This condition should be interpreted as the U(1) charge conservation condition on the scattering problem in the Chalker-Coddington network model. Here, the charge operator $\tilde{Q}$ for the Chalker-Coddington model derives
from the charge operator $Q$ in Eq.~\eqref{LabelEq-DEF-Of-Q-in-AIII}
that imposes the constraint on the Gaussian tensor $\Gamma$. But $\tilde{Q}$ and $Q$ should not be identified as they act on different vector spaces. The constraint Eq.~\eqref{eqn:Class_AIII_TransferM} enables us to write 
\begin{eqnarray}
\ft[\Gamma] &=& W \left(
    \begin{array}{cccc}
    \ft_{p+}[\Gamma] & & & \\
     & \ft_{p-}[\Gamma] & & \\
     & & \ft_{h-}[\Gamma] & \\
     & &  & \ft_{h+}[\Gamma] \\
    \end{array}
    \right) W^\dag 
    \label{eqn:class_AIII_transM_full}
    \\
W &=& \frac{1}{\sqrt{2}} \begin{pmatrix}
\mathds{1}_\chi & \mathds{1}_\chi  && \\
\i\mathds{1}_\chi & -\i\mathds{1}_\chi && \\
&&\mathds{1}_\chi & \mathds{1}_\chi  \\
&& -\i\mathds{1}_\chi & \i\mathds{1}_\chi
\end{pmatrix},
\end{eqnarray}
where $W$ provides the basis rotation that diagonalizes the charge operator $\tilde{Q}$. The first two diagonal blocks in this expression correspond to the P sector of the full transfer matrix while the last two diagonal blocks correspond to the H sector. The subscripts $\pm$ indicate the charge-$\tilde{Q}$ eigenvalues $\pm 1$ of the associated blocks. On top of the charge conservation, the condition that the original P-sector and H-sector transfer matrices are complex conjugates of each other and belong to the complexified special orthogonal group $\SO(2\chi)_\mathbb{C}$ enforces that 
\begin{align}
    \ft_{p+}[\Gamma] = \left(\ft_{p-}[\Gamma]^{\T}\right)^{-1} = \ft_{h-}[\Gamma]^* = \left(\ft_{h+}[\Gamma]^{\dag}\right)^{-1},
    \label{eqn:class_AIII_transM_full_block_relation}
\end{align}
where $\ft_{p+}[\Gamma]$ belongs to the complex general linear group $\GL(\chi, \mathbb{C})$. Since the relation above implies that the full transfer matrix $\ft[\Gamma]$ with U(1) conservation is fully parameterized by $\ft_{p+}[\Gamma]$, the full transfer matrix $\ft[\Gamma]$ also corresponds to an element of $\GL(\chi, \mathbb{C})$. Moreover, it is easy to show that any element of $\GL(\chi, \mathbb{C})$ has a corresponding U(1) symmetric Gaussian tensor $\Gamma$ (via the form of full transfer matrix in Eq. \eqref{eqn:class_AIII_transM_full} and the relations in Eq. \eqref{eqn:class_AIII_transM_full_block_relation}). The result that the group of U(1)-symmetric full transfer matrices
$\ft[\Gamma]$ is given by $\GL(\chi,\mathbb{C})$ is consistent with the group of transfer matrices in symmetry class AIII as classified in Refs.~\onlinecite{Ludwig2013,SchnyderRyuFurusakiLudwig}. Therefore, we conclude that the U(1) symmetric pure-state square-lattice GTN can be mapped to a unitary Chalker-Coddington square-lattice network model in symmetry class AIII.

Note that under TR symmetry, the charge operator $\tilde{Q}$ of our ``GTN-induced" Chalker-Coddington network model transforms as
\begin{align}
  {\rm TR~symmetry~action}:~  \tilde{Q} \rightarrow \Theta_{tr}^\dag \tilde{Q}^*\Theta_{tr} = - \tilde{Q},
\end{align}
which implies the U(1) symmetry action $\exp\{ \i \alpha {\tilde Q} \}$ generated by $\tilde{Q}$ commutes with the TR symmetry action. Therefore, the TR symmetry defined for symmetry class DIII in fact plays the role~\cite{FosterLudwigHubbardPRB2008,SchnyderRyuFurusakiLudwigPRB2008,ChongWang-Senthil-PRB2014,LudwigNobelSymposium2015}
of the chiral symmetry in symmetry class AIII.

Having mapped the U(1)-symmetric pure-state GTN, 
in every realization of disorder,
to unitary Chalker-Coddington network models in symmetry class AIII, the entanglement phases of the random U(1)-conserving pure-state GTN/NGC can be identified with phases of disordered non-interacting fermions in two spatial dimensions and in symmetry class AIII. In particular, 
it is known~\cite{GadeWegnerNPB360-1991-213,GadeWegnerNPB398-1993-499,GLL-NPB583-2000-475,BocquetChalker-NetworkChiralClasses-PRB67-054204,RyuMudryLudwigFurusaki-GlobalPhaseDiagram-PRB85-2012-235115}
that this symmetry class exhibits in two spatial dimensions
a line of critical fixed points parametrized by the dimensionless conductance (a measure of the strength of disorder)
varying continuously along the line, where each point on the line is described by a distinct interacting conformal field theory. 
Being free of symmetry-allowed relevant or marginally-relevant perturbations, each one of the critical fixed points on this line will again correspond to a critical entanglement phase in the U(1)-conserving
GTN/NGC, which will exhibit properties such as a logarithmic scaling of the half-system entanglement entropy $S_{L/2}$ as defined in Sec.~\ref{sec:numerics-entanglement} based on general reasoning provided in Refs. \onlinecite{vasseur2019entanglement,jian2020measurement}. In contrast to
symmetry class DIII in two dimensions, the fixed points along this AIII line possess in general no symmetry-allowed marginally irrelevant operators (the leading irrelevant operator
has finite scaling dimension). The analogue in symmetry class AIII of second disorder moment $C(r)$ of the fermion two-point correlation function 
defined as in Eq.~\eqref{eqn:Cd}
is free of logarithmic corrections and is expected to decay as a pure power-law $1/r^2$, owing to a relationship of this quantity at an absorbing
boundary (see e.g. Appendix~\ref{app:CC_intro}) with the point-contact conductance~\cite{BettelheimGruzbergLudwig-ConformalRestriction-PRB86-2012}.

Generalizing the discussion of the GTN with global U(1) symmetry
on the square lattice to lattices in higher dimensions, we can identify entanglement phases in $D$-dimensional U(1)-symmetric lattice GTNs as phases of symmetry-class-AIII unitary disordered fermions in $D$ spatial dimensions. In particular, for all spatial dimensions $D\geq3$, it is 
known 
that a stable disordered metallic phase occurs in 
this~\footnote{See, e.g.,
Refs.~\onlinecite{GadeWegnerNPB360-1991-213,GadeWegnerNPB398-1993-499,FosterLudwigHubbardPRB2008}.}
as in all ten symmetry classes~\cite{efetovBook1996supersymmetry,WegnerBetaFunctions-NPB316-1989-663}.

Returning to the special case of the square-lattice GTN with Majorana bond number $\chi$ in $D=2$ dimensions, the U(1) symmetric condition is enforced by requiring each Gaussian tensor $\Gamma= \begin{psmallmatrix}
\Gamma_{LL} & \Gamma_{LR} \\
\Gamma_{RL} & \Gamma_{RR}
\end{psmallmatrix}$
to commute with a U(1) charge operator $Q$ which can be chosen to the take form $ \begin{psmallmatrix}
\sigma^y \otimes \mathds{1}_{\chi} & 0 \\
0 & \sigma^y \otimes \mathds{1}_{\chi}
\end{psmallmatrix}$. In the same basis, if we further require $\Gamma$ to anticommute with the operator $K = \begin{psmallmatrix}
\sigma^z \otimes \mathds{1}_{\chi} & 0 \\
0 & -\sigma^z \otimes \mathds{1}_{\chi}
\end{psmallmatrix}$, the resulting full transfer matrix $\ft[\Gamma]$ resides in the group $\GL(\chi, \mathbb{R})$ which is a subgroup of $\GL(\chi, \mathbb{C})$. This statement can be obtained by directly showing that the matrices $\ft_{p\pm}[\Gamma]$ and $\ft_{h\pm}[\Gamma]$ introduced in Eq.~\eqref{eqn:class_AIII_transM_full} are real once the extra condition $\{\Gamma, K\} = 0$ is imposed (in addition to the U(1) symmetry condition $\{\Gamma, Q\}=0$). Notice that $\GL(\chi, \mathbb{R})$ matches exactly the group of transfer matrices of symmetry class BDI ~\cite{Ludwig2013,SchnyderRyuFurusakiLudwig}. Hence, the U(1) symmetric GTN with an extra condition $\{\Gamma, K\} = 0$ for each Gaussian tensor corresponds to a Chalker-Coddington network model in symmetry class BDI. Via the GTN/NGC correspondence, such a GTN also corresponds to a U(1)-conserving NGC such that there exists a complex fermion basis in which all the gates in the circuit are purely real. Based on these correspondences, the entanglement phases in such GTN and its corresponding NGC can be identified with phases of disordered non-interacting fermions in symmetry class BDI. Focusing on $D=2$, similar to symmetry class AIII, disordered non-interacting fermions in symmetry class BDI also exhibit a line of critical fixed points that are free of symmetry-allowed relevant or marginally irrelevant operators.~\cite{GLL-NPB583-2000-475,RyuMudryLudwigFurusaki-GlobalPhaseDiagram-PRB85-2012-235115}
Therefore, 
this line of critical fixed points
corresponds to a critical entanglement phase for the U(1)-conserving GTN/NGC with the extra constraint $\{\Gamma, K\} = 0$. The conformal field theory that describes such critical fixed points in 
symmetry class BDI predicts that the corresponding critical entanglement phase should exhibit logarithmic scaling of the half-system entanglement entropy $S_{L/2}$ as defined in Sec.~\ref{sec:numerics-entanglement} and a pure $1/r^2$ scaling (free of logarithmic corrections) of the second disorder moment $C(r)$ of the fermion two-point correlation function 
defined as in Eq.~\eqref{eqn:Cd}.

Interestingly, recent work~\cite{chen2020emergent} presents a numerical study of a particular microscopic model of a random non-unitary quantum circuit with a global U(1) symmetry acting on non-interacting fermions. A critical phase is observed in which the 2nd moment of the fermion two-point function decays with a $1/r^2$ power-law behavior. We notice that the circuits involved in this particular microscopic model are not only U(1)-symmetric but also real (after a change of basis). While at the microscopic level, the random ensemble studied in Ref.~\cite{chen2020emergent} looks different from the random ensemble of GTN/NGCs with a U(1) conservation law and the extra condition $\{\Gamma, K\} = 0$, the critical properties are expected to be the same since they are governed by the universality class, which depends only the symmetry class, which in this case is BDI.

\subsubsection{Symmetry Class A}
\label{sec:class_A}

The original Chalker-Coddington network model was first introduced in Ref.~\onlinecite{Chalker1988} to tackle the two-dimensional integer quantum Hall plateau transition (in the absence of interactions). In this system, where the magnetic field breaks time-reversal symmetry, only charge is preserved and hence, it belongs to symmetry class A. In the following, we will provide a construction of lattice GTN/NGCs whose corresponding Chalker-Coddington network models reside in symmetry class A.

To construct such a lattice GTN, we need to consider another constraint on each Gaussian tensor $\Gamma$ in addition to the U(1) symmetry constraint Eq.~\eqref{eqn:charge_conservation_AIII} discussed in Sec.~\ref{sec:class_AIII}. We introduce another operator $\Lambda$ with real matrix elements (in the same basis as that of the covariance matrix $\Gamma$) such that
\begin{align}
    \Lambda^2 = \mathds{1} ,~~\Lambda^\T = \Lambda,~~[\Lambda,\i Q] = 0.
\end{align}
Note that we can view both $\Lambda$ and $\i Q$ as operators in a real matrix algebra, the latter operator satisfying
$\left(\i Q\right)^2 = -\mathds{1}$ and $\left(\i Q\right)^\T = -\i Q$.
For a pure-state GTN corresponding to the symmetry-class-A unitary Chalker-Coddington network model, we need to require each pure-state Gaussian tensor $\Gamma$ in the GTN to satisfy the following conditions:
\begin{align}
    [\i Q, \Gamma] = 0, ~~\{\Lambda, \Gamma\} = 0,
    \label{eqn:class_A_Gamma_Condition}
\end{align}
where the first condition is exactly the same as the U(1) symmetry condition discussed in Sec.~\ref{sec:class_AIII}. We also require that the operators $\i Q$ and $\Lambda$ do not mix or permute Majorana modes on different legs of the tensor $\Gamma$ so that the lattice geometry of the GTN will not interfere with the constraints Eq.~\eqref{eqn:class_A_Gamma_Condition} on each individual tensor in the GTN.

Let's discuss the square-lattice pure-state GTN as an example illustrating this. Let $\chi$ be the Majorana bond number of the GTN. The space of all $4\chi\times 4\chi$ pure-state covariance 
matrices $\Gamma$ which satisfy
the conditions in Eq.~\eqref{eqn:class_A_Gamma_Condition} can be identified with the symmetric space $\U(\chi)$:
We can choose a basis where $\i Q$ takes the form $\i\sigma^{0y}\otimes \mathds{1}_{\chi}$ and $\Lambda$ takes the form $\sigma^{z0}\otimes \mathds{1}_{\chi}$. It can then be shown that the covariance matrices $\Gamma$ satisfying the constraint Eq.~\eqref{eqn:class_A_Gamma_Condition} are in one-one correspondence with the $\chi \times \chi$ unitary matrices. We will consider the random ensemble of GTNs with the constraints Eq.~\eqref{eqn:class_A_Gamma_Condition}. In particular, it is natural to define the maximally random ensemble by having every Gaussian tensor $\Gamma$ in the GTN independently chosen as a random point in the symmetric space $\U(\chi)$ with uniform probability measure, the Haar measure.

As we've discussed in the context of symmetry class DIII, the symmetric space of the Gaussian tensor at each site
of the GTN should be identified with
the space of scattering $S$-matrices at the corresponding site of the Chalker-Coddington network model obtained from the GTN. The appearance of the symmetric space $\U(\chi)$ as the space of covariance matrices then confirms that the symmetry class of this type of Chalker-Coddington network model should be identified as symmetry class A~\cite{SchnyderRyuFurusakiLudwig}. 

In the square-lattice GTN, we can also study the space
formed by the full transfer matrix $\ft[\Gamma]$ of the four-leg Gaussian tensor $\Gamma$ constrained by Eq.~\eqref{eqn:class_A_Gamma_Condition}. As we did before, we can think of each four-leg tensor $\Gamma$ as a two-leg tensor (in a quasi-one-dimensional geometry) with each leg having Majorana bond number $2\chi$.  In the basis associated with the two legs ($L$ and $R$) of the Gaussian tensor $\Gamma$ where the covariance matrix $\Gamma$ reads $\begin{psmallmatrix}
\Gamma_{LL} & \Gamma_{LR} \\
\Gamma_{RL} & \Gamma_{RR}
\end{psmallmatrix}$, the U(1) charge operator $\i Q$ can be chosen to take the form $ \begin{psmallmatrix}
\i\sigma^{0y} \otimes \mathds{1}_{\chi/2} & 0 \\
0 & \i\sigma^{0y} \otimes \mathds{1}_{\chi/2}
\end{psmallmatrix}$ and the operator $\Lambda$ can be chosen to take the form $ \begin{psmallmatrix}
\sigma^{z0} \otimes \mathds{1}_{\chi/2} & 0 \\
0 & -\sigma^{z0} \otimes \mathds{1}_{\chi/2}
\end{psmallmatrix}$. The relative sign in lower right block of  $\Lambda$ as
compared to its upper left block is to ensure that the tensor contractions (along the quasi-one-dimensional geometry) are compatible with the constraints Eq.~\eqref{eqn:class_A_Gamma_Condition}, namely that
the contraction of two tensors $\Gamma$ and $\Gamma'$ satisfying Eq.~\eqref{eqn:class_A_Gamma_Condition} yields a third tensor that also satisfies the same conditions. The constraints on the Gaussian tensor $\Gamma$ can be translated into the following conditions on
the full transfer matrix $\ft[\Gamma] = \begin{psmallmatrix}
\ft_p[\Gamma] & 0 \\ 0 & \ft_h[\Gamma]
\end{psmallmatrix}$:  
\begin{align}
   \big[\, \ft[\Gamma],\, \tilde{Q} \,\big] = 0,~~~  \ft[\Gamma]^\dag \cdot \mathcal{J}_\Lambda \cdot \ft[\Gamma] = \mathcal{J}_\Lambda,
   \label{eqn:Class_A_TransferM}
\end{align}
where $\tilde{Q} = \begin{psmallmatrix}
\sigma^{0y} \otimes \mathds{1}_{\chi/2} & 0 \\
0 & \sigma^{0y} \otimes \mathds{1}_{\chi/2}
\end{psmallmatrix} $ and $\mathcal{J}_\Lambda = \begin{psmallmatrix}
\sigma^{z0} \otimes \mathds{1}_{\chi/2} & 0 \\
0 & \sigma^{z0} \otimes \mathds{1}_{\chi/2}
\end{psmallmatrix} $. The conditions above for the full transfer matrix $\ft[\Gamma]$ imply that the corresponding Chalker-Coddington network model conserves the U(1) charge $\tilde{Q}$ and also a current defined by $\mathcal{J}_\Lambda$ (in addition to the probability current $\Jp$,
Eq.~\ref{eqn:DIII_conserved_current}). 
The set of full transfer matrices $\ft[\Gamma]$ satisfying the conditions Eq.~\eqref{eqn:Class_A_TransferM} forms the group $\U\left(\frac{\chi}{2},\frac{\chi}{2}\right)$.
This result is in agreement with the group of transfer matrices in 
symmetry class A, which is classified in Refs.~\onlinecite{SchnyderRyuFurusakiLudwig,Ludwig2013}. We note that for the case of $\chi=2$, the group $\U\left(1,1\right)$ formed by the full transfer matrices is exactly the group of transfer matrices appearing in the symmetry-class-A Chalker-Coddington model introduced in Ref. \onlinecite{Chalker1988}. In fact, using the charge operator $\tilde{Q}$ and the two conserved currents $\Jp$ and $\mathcal{J}_{\Lambda}$, we can show that the Chalker-Coddington network model obtained from the GTN under the constraint Eq.~\eqref{eqn:class_A_Gamma_Condition} consists of four decoupled copies of Chalker-Coddington network models with each copy only conserving the U(1) charge symmetry. The four decoupled copies (or `layers') are related to each other by the action of time-reversal
(TR), particle-hole (PH) and chiral symmetries, previously introduced in the context of symmetry class DIII.
Therefore, we can conclude the random ensemble of square-lattice pure-state GTNs with each Gaussian tensor constrained by Eq.~\eqref{eqn:class_A_Gamma_Condition} can be mapped to the disordered unitary symmetry-class-A Chalker-Coddington network model on the square lattice.

Our mapping between GTN/NGCs and the corresponding Chalker-Coddington network models in symmetry class A thus relates the two-dimensional integer
quantum Hall plateau transition, which is a conformal critical point,
to an entanglement critical point in a GTN/NGC. Interestingly, 
since the integer quantum Hall plateau transition is 
a transition between topologically distinct gapped phases, the corresponding transition in the GTN/NGCs is thus also between distinct area-law phases.

Again, the mapping from pure-state lattice GTNs under the constraint Eq.~\eqref{eqn:class_A_Gamma_Condition} to Chalker-Coddington network models in symmetry class A can be generalized to any dimensions. As mentioned earlier in Sec. \ref{sec:class_AIII}, for all spatial dimensions $D\geq3$, it is known~\cite{WegnerBetaFunctions-NPB316-1989-663}
that a stable disordered metallic phase occurs in all ten symmetry classes.
Therefore, GTN in symmetry class A and in dimensions $D\geq 3$ also naturally admits a critical entanglement phase that corresponds to such a disordered metallic phase.

\subsection{Real Symmetry Classes}

Disordered non-interacting fermion systems in the real symmetry classes are known to exhibit rich behavior in two spatial dimensions ($D=2$). For example, the symmetry classes BDI and CII behave similarly to AIII, i.e. they exhibit lines of critical fixed points 
\footnote{See e.g., Ref.~\onlinecite{RyuMudryLudwigFurusaki-GlobalPhaseDiagram-PRB85-2012-235115}.
for a summary.}. Class C is similar to class A, i.e. there is a critical point~\cite{GruzbergLudwigRead-SpinQuantumHall-PRL82-1999-4524,SchnyderRyuFurusakiLudwigPRB2008}. Finally, class D is known to exhibit stable critical phases as well as quantum critical points~\cite{chalker2001thermal}.

\subsubsection{General Constructions}
\label{sec:real_classes_general_construction}

For a given symmetry class, the scattering $S$-matrices in the Chalker-Coddington network model should belong to the symmetric space given by the symmetry class. For all eight real symmetry classes, the corresponding symmetric space is summarized in Table~\ref{tab:symmetric_space}.
As we discussed in Sec.~\ref{sec:scattering_single_tensor}, the $S$-matrices in a Chalker-Coddington network model obtained from a GTN are exactly given by the covariance matrices of the Gaussian tensors in the GTN. Therefore, to realize a Chalker-Coddington network model in a specific symmetry class using the GTN, one simply needs to find the correct set of constraints on the individual Gaussian tensor $\Gamma$ in the GTN such that the space of permissible pure-state Gaussian tensors matches the symmetric space of the desired symmetry class. 

A systematic way to find a right set of constraints for all the eight real symmetry classes is to follow the so-called ``Clifford algebra extension problem"~\cite{Kitaev_2009}. As shown in Table~\ref{tab:symmetric_space}, the symmetric space $R_p$ of every real
symmetry class is labeled by an integer $p$ modulo 8. For a given symmetry class, a procedure to find an appropriate set of constraints for the Gaussian tensor in the GTN is as follows. We first find a non-negative integer $q$ such that $R_p$ with $p \equiv q+2 \mod 8$ matches the desired symmetric space,
i.e. we build all real symmetry classes upon DIII, which has $p=2$, i.e. $q=0$, is this notation. (Compare Table~\ref{tab:symmetric_space}.)
Then, we write down $q$ real skew-symmetric matrices $\Lambda_{1,2,...,q}$ such that 
\begin{align}
    & \Lambda_{i}^2 = - \mathds{1},~~~~~~\Lambda_{i}^\T = - \Lambda_{i}~~~~~~{\rm for}~i=1,2,...,q, \nonumber \\ 
    & \Lambda_{i}\Lambda_{j} = -\Lambda_{j}\Lambda_{i}~~~~~~~~~~~~~~~~~~ ~{\rm for}~i\neq j.
\end{align}
The matrices $\Lambda_{1,2,...,q}$ generate the real Clifford algebra with $q$ negative generators $\CLR{0,q}$. For each Gaussian tensor $\Gamma$ in the GTN, we require that 
\begin{align}
\Lambda_{i}\, \Gamma = -\Gamma \, \Lambda_{i}
\label{eqn:all_real_symmetry_class_constraints}
\end{align}
for $i=1,2,...,q$. Since $\Gamma$ is also a real skew-symmetric matrix such that $\Gamma^2=-\mathds{1}$, the set of matrices $\Lambda_{1,2,...,q}$ together with $\Gamma$
all together generate the real Clifford algebra with $q+1$ negative generators $\CLR{0,q+1}$. Therefore, we can 
identify the space of permissible  matrices $\Gamma$ as the space of extensions of the real Clifford algebra $\CLR{0,q}$ (with all $q$ negative generators $\Lambda_{1,2,...,q}$ fixed) to the real Clifford algebra $\CLR{0,q+1}$. It is known that the space of such Clifford algebra extensions is given by symmetry space
$R_p$ with $p= q+2 \mod 8$. (See Table~\ref{tab:symmetric_space}.)
For the Clifford algebra extension problems, the corresponding symmetric spaces $R_p$ are also referred to the classifying spaces. Therefore, when a random GTN consists of only Gaussian tensors obeying the constraints given above, its corresponding Chalker-Coddington network models belong to the desired symmetry class. Moreover, each of the symmetric spaces admits a natural uniform measure~\footnote{They  are generalizations of spheres; see e.g. corresponding comments in  Ref.~\onlinecite{RyuSchnyderFurusakiLudwig-NewJPhysics12-2010-065010}}. Therefore, we can always define the maximally random ensemble of GTNs by having each of its Gaussian tensors drawn from the specified symmetric space according to its uniform probability measure.

In the special case with $q=0$, no constraint is imposed on the Gaussian tensor $\Gamma$. 
In this case, we recover the random pure-state GTN studied in Sec.~\ref{sec:numerics} and in Sec.~\ref{sec:transfer-matrix}. Indeed, for $q=0$, the symmetric space $R_2 = \frac{\SO(2N)}{\U(N)}$ is the space of pure-state covariance matrices with no additional constraint. Also, the symmetry class associated with the symmetric space $R_2$ is 
symmetry class DIII, which is consistent with the discussion in Sec.~\ref{sec:transfer-matrix}.

For a random ensemble of $D$-dimensional GTNs with each of its Gaussian tensors obeying Eq.~\eqref{eqn:all_real_symmetry_class_constraints}, entanglement phases in this random ensemble can be identified with the phases of unitary systems of disordered non-interacting fermions in $D$ spatial dimensions 
in the symmetry class whose symmetric space is given by $R_{q+2}$. However, we would like to comment that the procedure provided above is not the only way to realize GTNs that are mapped to Chalker-Coddington network models 
in the desired symmetry class. For example, an alternative procedure is provided by the Clifford algebra extension problem with positive generators instead of those with negative generators 
discussed above (see App.~\ref{app:clifford_extension} for more details). Also, a different construction for symmetry class BDI as emerging from symmetry class AIII was already discussed in Sec.~\ref{sec:class_AIII}.

\begin{table*}[t]
\centering
\begin{tabular}{||c c c ||} 
 \hline
 ~~Symmetry Class~~  & ~~Symmetric Space $R_p$~~ & ~~$p$ mod 8~~   \\ 
 \hline\hline
BDI & $\SO(N+N')/(\SO(N) \times \SO(N'))$ &  0 \\
D & $\SO(N)$ & 1  \\ 
DIII & $\SO(2N)/\U(N)$ & 2  \\
AII & $\U(2N)/{\rm Sp}(N)$ & 3  \\
 CII & ${\rm Sp}(N+N')/({\rm Sp}(N) \times {\rm Sp}(N'))$ & 4  \\
 C & ${\rm Sp}(N)$ & 5  \\ 
 CI & ${\rm Sp}(N)/\U(N)$ & 6 \\
 AI & $\U(N)/\rO(N)$ & 7 \\
 \hline
\end{tabular}
\caption{Table~\cite{SchnyderRyuFurusakiLudwig,SchnyderRyuFurusakiLudwig,RyuSchnyderFurusakiLudwig-NewJPhysics12-2010-065010,LudwigNobelSymposium2015}
of all eight real symmetry classes of $S$-matrices in the current context (which are referred to as ``time-evolution operators" in those references) 
in the Cartan/Altland-Zirnbauer classification,
and their associated symmetric spaces $R_p$ (which are also known mathematically
as classifying spaces). Each symmetry class and its symmetric space is labeled by an integer $p \mod 8$.  }
\label{tab:symmetric_space}
\end{table*}

\subsubsection{An alternative construction for symmetry class D}
\label{sec:class_D}
In Sec.~\ref{sec:real_classes_general_construction}, we've discussed a systematic construction of pure-state GTNs whose corresponding Chalker-Coddington models belong to any of the
eight real symmetry classes. However, in some cases, this construction may require 
an unnecessarily
large number of Majorana modes in each Gaussian tensor of the GTN. For example, when applied to the case of symmetry class D, the construction given in Sec.~\ref{sec:real_classes_general_construction} requires a minimum of $q=7$ negative Clifford algebra generators $\Lambda_{1,2,..,7}$. The minimal matrix dimensions to accommodate the algebra of the operators $\Lambda_{1,2,..,7}$ and the covariance matrix $\Gamma$ is 16 before we take into account the ``leg structure" (or geometry) of the Gaussian tensor $\Gamma$ in the GTN. Therefore, the construction given in Sec.~\ref{sec:real_classes_general_construction} requires at least 16 Majorana modes in each Gaussian tensor in order to realize the GTN that corresponds to a symmetry-class-D Chalker-Coddington network model. In the following, we provide an alternative and minimal construction for 
symmetry class D which is applicable to the square-lattice GTN with Majorana bond number $\chi =1$.

For a generic square-lattice GTN with Majorana bond number $\chi =1$ (which, as discussed, is in symmetry class DIII), each four-leg Gaussian tensor can be described by a $4\times 4$ covariance matrix $\Gamma(\vec{n})$ that can be parameterized following Eq.~\eqref{eqn:4leg_chi1_tensor}. Here, we again adopt the ordering of Majorana modes on a single four-leg tensor as shown in Fig.~\ref{fig:staggered}~(a). In order to realize the GTN that corresponds to a symmetry-class-D Chalker-Coddington network model, we further require that 
\begin{align}
\Gamma(\vec{n}) \cdot \Lambda' =  -\Lambda' \cdot \Gamma(\vec{n}),
\label{eqn:class_D_constraint}
\end{align}
where $\Lambda' = \begin{psmallmatrix}
\sigma^z & \\
& -\sigma^z
\end{psmallmatrix}$. 
Note that ${\Lambda'}^2 = \mathds{1}$ and ${\Lambda'}^\T = \Lambda'$. Hence, $\Lambda'$ and $\Lambda'\Gamma$ together generate the real Clifford algebra $\CLR{2,0}$ with two positive generators. The constraint Eq.~\eqref{eqn:class_D_constraint} on $\Gamma(\vec{n})$ can be satisfied by choosing $\vec{n} = (\cos\theta, \sin \theta, 0)$. In the following, we will denote $\Gamma(\vec{n} = (\cos\theta, \sin \theta, 0) ) $ as $\Gamma(\theta)$ for simplicity. Now, we view the four-leg tensor shown in Fig.~\ref{fig:staggered}~(a) as a two-leg tensor (with the Majorana modes $\hgamma_{1,2}$ residing on the leg $L$ and $\hgamma_{3,4}$ residing on the leg $R$), which enables us to discuss the full transfer matrix $\ft[\Gamma(\theta)]$ of the tensor $\Gamma(\theta)$. In addition to all the common properties shared by all full transfer matrices, the constraint Eq.~\eqref{eqn:class_D_constraint} leads to an extra current conservation relation for the full transfer matrix $\ft[\Gamma(\theta)]$:
\begin{align}
  \ft[\Gamma(\theta)]^\dag \cdot \mathcal{J}_{\Lambda'} \cdot \ft[\Gamma(\theta)] = \mathcal{J}_{\Lambda'},
   \label{eqn:Class_D_current_conservation}
\end{align}
where the extra current operator is given by $\mathcal{J}_{\Lambda'} = \begin{psmallmatrix}
\sigma^{z} & 0 \\
0 & \sigma^{z} 
\end{psmallmatrix} $. With this extra current conservation condition, the full transfer matrix $\ft[\Gamma(\theta)]$ can be identified as an element of the group $\rO(1,1)$. In fact, according to the classification given in Refs.~\onlinecite{SchnyderRyuFurusakiLudwig,Ludwig2013}, the group $\rO(1,1)$
indeed matches the group of transfer matrices of Chalker-Coddington network models in symmetry class D. 

In the Chalker-Coddington network model that corresponds to the square-lattice GTN with a Gaussian tensor of the form $\Gamma(\theta)$ on each site, the scattering process at each site, which is governed by the full transfer matrix $\ft[\Gamma(\theta)]$, conserves two types of currents $\Jp$ and $\mathcal{J}_{\Lambda'}$. We can choose a basis where both currents are diagonal matrices. In this basis, the $4\times 4$ full transfer matrix $\ft[\Gamma(\theta)]$ decouples into two $2\times 2$ blocks. That is to say that the Chalker-Coddington network model studied here in fact consists of two decoupled layers of Chalker-Coddington network models. In fact, each layer can be identified as a symmetry-class-D Chalker-Coddington network model (with one mode per edge) on the square lattice. The two layers of symmetry-class-D Chalker-Coddington network model are interchanged under the chiral symmetry action given $\Sigma_c$.

In fact, this alternative construction for symmetry class D is a special case of another systematic construction of GTNs that correspond to Chalker-Coddington network models in any of the eight real symmetry class and in any dimensions. This systematic construction, which is an alternative to the construction given in Sec.~\ref{sec:real_classes_general_construction}, is summarized in App.~\ref{app:clifford_extension}.

\section{Three-dimensional tensor networks}
\label{sec:3d}

\begin{figure}
  \includegraphics[width = 0.45\textwidth]{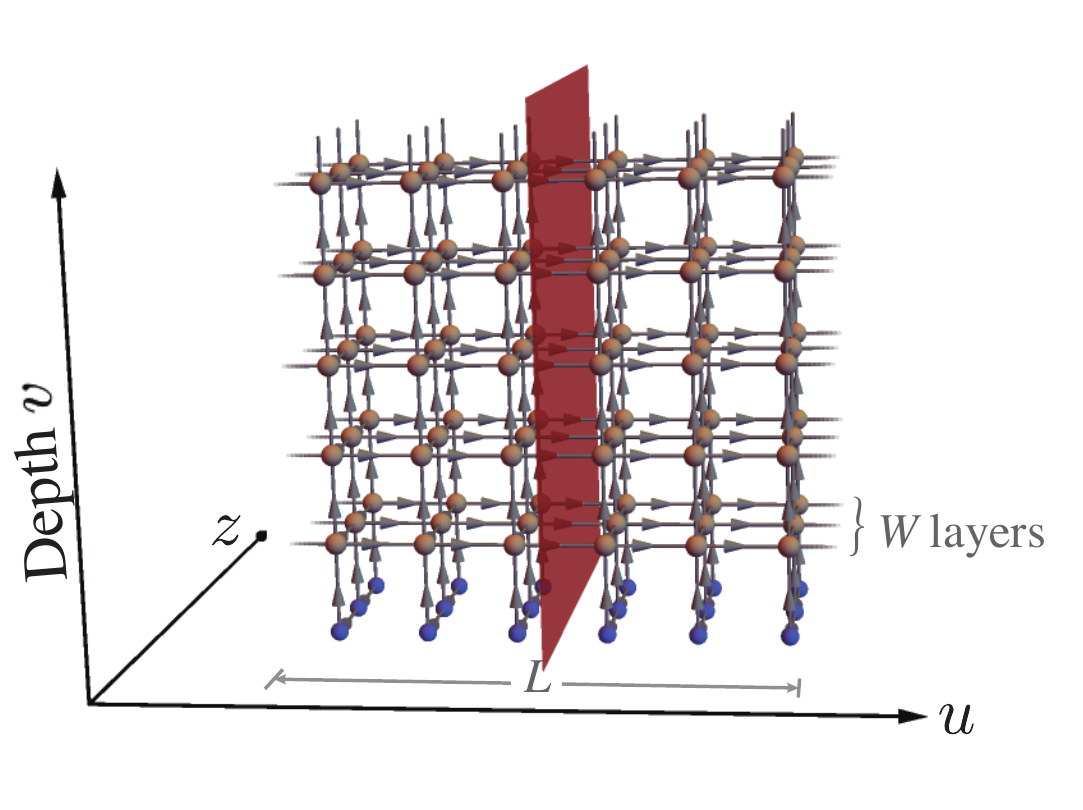}
  \caption{Illustration of a quasi-three-dimensional tensor network, which can be understood as a stack of $W$ (here $W=3$) layers of the two-dimensional network connected in the $z$-direction. We apply periodic boundary conditions in the $u$-direction.
  \label{fig:3d_network}}
\end{figure}

As further generalization of our construction, we consider the contraction of a three-dimensional random Gaussian tensor network, as sketched in Fig.~\ref{fig:3d_network}. In this construction, the square lattice of the two-dimensional tensor network is replaced by a cubic lattice, and the state on the boundary is now defined on a two-dimensional strip rather a one-dimensional chain. Each tensor in this cubic-lattice GTN is given by an independent Haar-random Gaussian pure state.

In the following, we will focus on the entanglement properties of this construction. For concreteness of this discussion, we will consider a tensor network of length $L$ and thickness $W$ and some depth $v \gg W,L$, as shown in Fig.~\ref{fig:3d_network}. The boundary state (residing on the top boundary of the GTN) is thus defined on a square lattice of dimensions $L \times W$ with $\chi$ Majorana modes on each site. We choose periodic boundary conditions in the $u$-direction, and open boundary conditions in the $z$-direction. We've also chosen a simple product state (graphically represented by the blue dots in Fig. \ref{fig:3d_network}) at depth $v=0$ as the initial state. With these choices, the three-dimensional case can be understood as a stack of $W$ connected layers of 
two-dimensional random Gaussian tensor networks in the $u$-$v$ plane. The entanglement cut we consider is also shown in Fig.~\ref{fig:3d_network}: the $L \times W$ system where the boundary state resides is cut into two halves of size $L/2 \times W$ and the length of the cut itself is $W$.

In this geometry, an area-law scaling of the entanglement entropy would correspond to
\begin{equation}
S \sim W.
\end{equation}
This scaling would be expected for the ground state of a two-dimensional (non-random) fermionic Hamiltonian with either a gapped spectrum or a single gapless (Dirac) point~\cite{Eisert2010,wolf2006}, as well as certain classes of critical systems~\cite{fradkin2006entanglement,hsu2009universal}.
If, on the other hand, the ground state is described by a finite Fermi surface, one expects a logarithmic violation of the area law, thus leading to a scaling of the form
\begin{equation} \label{eqn:area-law-violation}
S \sim W \log (L/L_0).
\end{equation}
Note that this is the same scaling that would also be expected for the ground state of a stack of $W$ decoupled one-dimensional systems each with a gapless Hamiltonian. 

To gain some intuition into the three-dimensional GTN, we can choose the Majorana bond number in the $z$-direction, i.e. connecting layers, independently from the other directions; we will use $\chi_z$ for the Majorana bond number along the $z$-direction and $\chi$ for the two directions within the $u$-$v$ plane. 
Consider first the case $\chi_z=0$. This corresponds exactly to decoupled layers of the previously described two-dimensional random Gaussian tensor networks, and the entanglement entropy is thus the sum of the contributions from each individual layer. We know that the entropy scaling of one layer is $S = \zeta_1(\chi) \log (L/L_0)$ with $\zeta_1(\chi)$ as shown in Fig.~\ref{fig:entropyscaling}. A stack of $W$ independent layers will thus have entropy
\begin{equation}
\chi_z = 0: \quad S = W \cdot \zeta_1(\chi) \log (L/L_0)
\end{equation}
where $\zeta_1(\chi)$ is the scaling prefactor of a single layer with Majorana bond number $\chi$.

Now consider the case of $\chi_z \gg \chi$, i.e. where the legs along the $z$-direction carry much more entanglement than the other ones in the $u$-$v$ plane. Heuristically, this is similar to replacing one column of $W$ tensors (along the $z$-direction) by a single tensor, which is equivalent to a single layer with Majorana bond number $W \chi$. This situation will produce the entanglement scaling $S \sim \zeta_1(W\chi) \log L$, where $\zeta_1(W \chi)$ is the scaling prefactor for a single layer with bond dimension $W \chi$. As we have shown in Sec.~\ref{sec:numerics-entanglement}, $\zeta_1$ scales linearly with the bond dimension, and therefore $\zeta_1(W\chi) \sim W \zeta_1(\chi)$. We thus conclude that for the $D=3$-dimensional network with $\chi_z \gg \chi$,
\begin{eqnarray}
\chi_z \gg \chi: \quad S &=& \zeta_1(W\chi) \log (L/L_0) \\ &\sim& W \cdot \zeta_1(\chi) \log (L/L_0) \nonumber
\end{eqnarray}

\begin{figure*}
\begin{overpic}[width=3.5in]{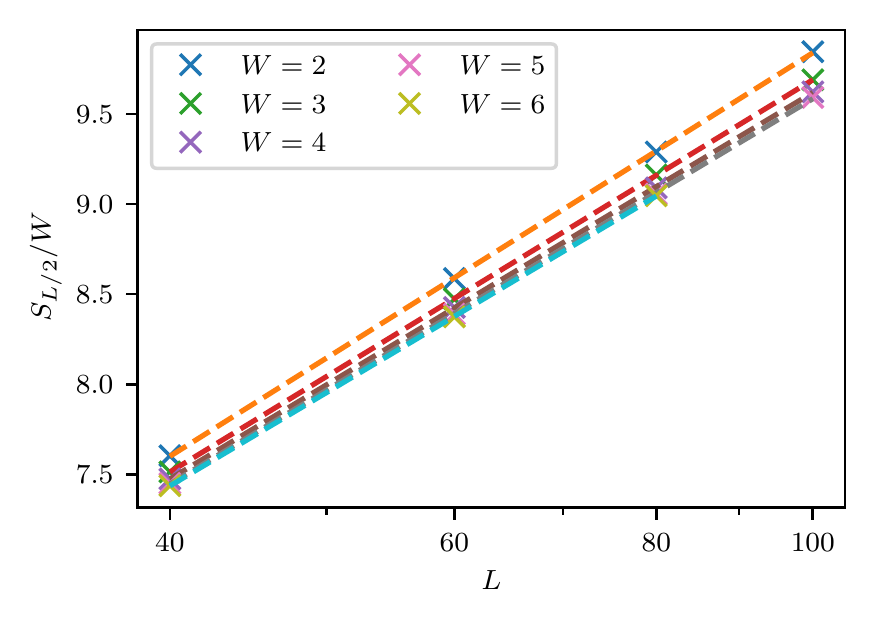} \put (0,70) {(a)} \end{overpic}
\begin{overpic}[width=3.5in]{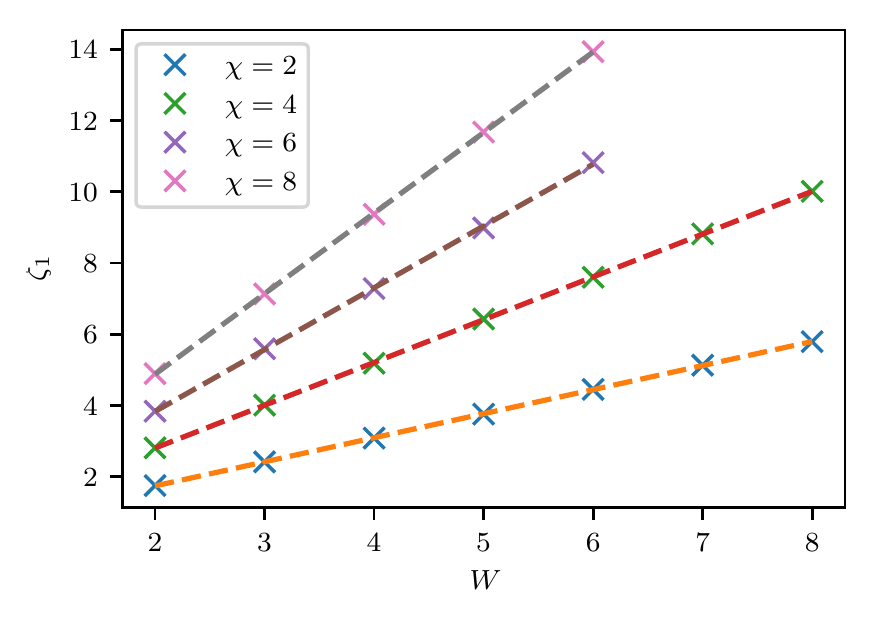} \put (0,70) {(b)} \end{overpic}
\caption{Entropy scaling in three-dimensional systems. The left panel (a) shows the scaling for fixed bond number $\chi=8$ (where we use the same bond number in all three directions) with the length $L$ of the system for a variety of $W$. Note that the $y$ axis is entropy divided by thickness $W$ for better readability. In all cases, a logarithmic scaling with $L$ is observed. The right panel (b) shows the prefactor of this logarithmic scaling as a function of thickness $W$ and for various bond numbers $\chi$. Together, these results clearly establish $W \log L$ scaling of the entropy. This is using periodic boundary condition in the $L$ and open boundary condition in the $W$ direction, but results for periodic boundary condition in both directions are qualitatively similar. \label{fig:3ddata}}
\end{figure*}

Since we find similar scaling with $W$ and with $L$
in the limit of both small and large $\chi_z$, it is natural to expect that it holds also for the isotropic case with $\chi_z = \chi$.
This is substantiated by our numerical results.
Fig.~\ref{fig:3ddata}~(a) shows the scaling of the von Neumann entanglement entropy $S_{L/2}$ for the cut shown in Fig.~\ref{fig:3d_network} as a function of $L$ (on a logarithmic scale) for different choices of $W$ from $W=2$ to $W=6$ and fixed Majorana bond number $\chi=\chi_z=8$. This confirms the suspected scaling of the entropy proportional to $\log (L/L_0)$.
Fig.~\ref{fig:3ddata}~(b) shows the extracted prefactor $\zeta_1$ as function of both $W$ and $\chi$. Most importantly, we find that the prefactor of the logarithm scales linearly with $W$, i.e. consistent with the logarithmic violation of the area law shown of Eq.~\eqref{eqn:area-law-violation}.

\section{Outlook}
\label{sec:outlook}

In this work, we have established 
correspondences, as shown in Fig.~\ref{fig:correspondences}, among non-unitary Gaussian circuits (NGCs), pure-state Gaussian tensor networks (GTNs) and unitary non-interacting fermion systems subject to static Hermitian Hamiltonians/undergoing static unitary time evolution.
These correspondences enable the identification of entanglement phases and criticality in random NGCs and in random pure-state GTNs with their counterparts in disordered Hamiltonian systems of non-interacting fermions. One natural direction to consider is
the effect of interactions in such tensor networks and in non-unitary circuits. More specifically, one can consider 
deformations of a GTN into a more general fermionic tensor network where the quantum state associated with each tensor is no longer given by Gaussian states. In the language of quantum circuits, such deformations turn a NGC into a more generic non-unitary quantum circuit that can no longer be fully described by its action on single fermionic operators. One interesting question worthy of future investigation concerns the stability of entanglement phases and entanglement criticality obtained in GTNs/NGCs to such deformations. Moreover, none of the pure-state GTNs and NGCs we've investigated exhibit a volume-law entanglement phase (in line with Ref. \onlinecite{fidkowskiHaahHastingsHowDynamicalMemoriesForget-arXiv2008.10611}), while, as exemplified by Ref.~\cite{vasseur2019entanglement,
li2018quantum,skinner2019measurement,li2019measurement,sang2020measurement,Lavasani2020measurement}, 
the volume-law entanglement phase certainly exists in non-Gaussian/interacting random tensor networks and in generic random non-unitary circuits. One may ask how a volume-law entanglement phase can emerge when the pure-state random GTNs and NGCs are deformed by interactions.

Another possible avenue that could be explored are mixed-state GTNs. Unlike the pure-state GTNs, a mixed-state GTN does not naturally correspond to a NGC. However, as discussed in App.~\ref{app:TranferM_mGTN_Derivation}, a mixed-state GTN still admits a transfer matrix description and can be mapped to a Chalker-Coddington network model residing inside symmetry class D (if no further constraint is imposed). While it may be most natural to expect a volume-law scaling of the entanglement entropy in the mixed-state GTNs, it remains a question whether there can be different entanglement phases in random mixed-state GTNs that are distinguishable under other measures of entanglement and that correspond to different phases of unitary disordered fermions in symmetry class D.

Finally, as pointed out in Sec.~\ref{sec:UnitaryTimeEvolutionAndMeasurements}, there is a subtle yet important distinction between 
physical systems whose non-unitarity is induced by measurement, and systems evolving
under generic random non-unitary circuit dynamics. While both can be described by ensembles of non-unitary circuits, the former further requires two extra conditions. The first condition is that the ensemble of non-unitary circuits generates a positive operator-valued measure. The second condition is that the probability for each circuit to appear needs to follow Born's rule (as explained in Sec.~\ref{sec:UnitaryTimeEvolutionAndMeasurements}), whereas for circuits where non-unitarity does not arise from measurements, the probability measure can be chosen freely.
It remains an open problem to elucidate the exact relation between the two types of systems. 
In particular, the general question about the relationship between the universality classes describing the
entanglement criticality in these two types of systems
deserves further investigation.
For the specific case of the Haar-random ensemble of pure-state GTNs introduced in Sec.~\ref{sec:numerics}, we have shown in App.~\ref{app:random_GTN_vs_MeasurementProblem} that, when viewed as 
a random ensemble of NGCs, the ensemble generates a positive operator-valued measure.
However, in the present study we've assumed that each realization of the random GTN/NGC appears with equal probability.  While this is clearly different from the Born-rule probability that one needs to use when the same NGC ensemble is used to describe the evolution under generalized (non-projective) measurements, how sensitive different universal behavior shown in different parameter regimes of Fig. \ref{fig:phase_transition} (a) is to this difference in the choice of probability requires further study.

\acknowledgements
 The authors thank Romain Vasseur for inspiring discussions at
the early stage of this work and Hassan Shapourian for helpful discussions on the numerical simulations of the transition from the critical entanglement phase to the area-law entanglement phase. C.-M. J. thanks Xiao-Liang Qi for helpful discussions.
C.M.J. and A.W.W.L. thank R. Vasseur and Y-Z You, A.W.W.L. thanks Y. Li, X. Chen and M. P. A. Fisher 
as well as A. Zabalo, M. J. Gullans, J. H. Wilson, R. Vasseur, S. Gopalakrishnan, J. H. Pixley, and D. A.  Huse
for collaboration on related topics. This research is supported in part by a faculty
startup grant at Cornell University (C.-M.J.).
This research is funded in part by the Gordon and Betty Moore Foundation through Grant No. GBMF8690 to UCSB to support the work of A. K. Use was made of the computational facilities administered by the Center for Scientific Computing at the CNSI and MRL (an NSF MRSEC; DMR-1720256) and purchased through NSF CNS1725797. This work was supported in part by the NSF under Grant No. DMR-1309667 (A.W.W.L.).

\appendix

\section{Non-interacting systems of fermions subjected to static disorder - Brief review}
\label{app:CC_intro}

\subsection{General considerations and description in terms of Chalker-Coddington models}

We begin by considering a system of non-interacting fermions described
by a Hermitian Hamiltonian in $D=2$ spatial dimensions subject to disorder which is static, i.e., time-independent. (We will also briefly comment on the case of $D=3$ spatial dimensions.)
In every realization of disorder, the Hamiltonian generates a unitary time evolution.  Upon Fourier transforming from time to energy, only fermions at the same energy couple to each other, 
owing to the absence of interactions. The (2+1)-dimensional fermion system at any fixed energy $E$ is described by a $D=2$-dimensional statistical mechanics system~\cite{LudwigFisherShankarGrinsteinPRB1994}. 
Specifically, in terms of the 1st-quantized Hamiltonian ${\cal H}$ of the (2+1)-dimensional fermion system, the  retarded (advanced) fermion 2-point function at energy  $E$ is expressed as
\begin{eqnarray}
\nonumber
&&
G^{\pm}_{a_1, a_2}({\vec r}_1, {\vec r}_2; E\pm \i \epsilon)
\\ \nonumber
&=& \langle {\vec r_1}|
\left [ \left (  \mp \i E + \epsilon \pm \i {\cal H} \right)^{-1}
\right ]_{a_1, a_2}| {\vec r}_2 \rangle
\\  \label{Eq-GreensFctContinuum-new}
&=& \langle {\hat c}_{\pm, a_1}({\vec r}_1) {\hat c}^\dagger_{\pm, a_2}({\vec r}_2)\rangle_{E\pm \i \epsilon}, \ \  \
(\epsilon \to 0^+),
\end{eqnarray}
where
additional indices $a_1,a_2$ may possibly appear  (as indicated) to 
characterize
additional quantum numbers, when needed.
(The last equality corresponds to the 2nd quantized formulation, 
${\hat c}_{\pm, a}({\vec r})$ and ${\hat c}^\dagger_{\pm, b}({\vec r})$ denoting canonical fermion creation and annihilation operators.)
From now on, we assume that the system exhibits critical behavior at energy $E=0$ and we will often omit the energy $E$ from our expressions. In all symmetry classes that possess
either particle-hole (charge-conjugation) or chiral symmetry, which are the classes we are most interested in here in this work, the single particle
Hamiltonian ${\cal H}$ changes sign under these operations, and therefore $E=0$
is a special value of energy in those cases. At this energy the system is known
to be critical. In the other symmetry classes it is typically possible to choose a generic value of energy. Because the (2+1)-dimensional fermion system is non-interacting, all observables can be expressed in terms of the 2-point function. We will now rewrite this
2-point function in the language of the so-called Chalker-Coddington network model~\cite{Chalker1988}.

For this purpose, it is convenient to consider $D=2$-dimensional position space being discretized on a lattice which we choose here to be a square lattice
(the details of the lattice are unimportant), and we also choose an evolution of the $D=2$-dimensional system in {\it discrete} time
steps. In this way, one arrives at the Chalker-Coddington formulation of the system. Here, these discrete time
steps are those of the time $\tH$ associated with the Hamiltonian $\mathcal{H}$ of the $D=2$-dimensional fermion system and should not be confused with the coordinate $t$ of the square-lattice GTN, which is also the circuit time for the NGC, corresponding to the Chalker-Coddington model. Because time steps for the Hamiltonian $\mathcal{H}$ are discrete,
the fermion 2-point function is now expressed in the form 
\begin{equation}
\label{Eq-GreensChalkerCoddington-new}
\mathbf{G}^{\pm}({I}_1, {I}_2; E\pm \i \epsilon) =
\langle I_1|\left (1- e^{\pm \i E-\epsilon} \  U^{\pm}
\right )^{-1}
|I_2\rangle.
\end{equation}
Here
$I_j$ denotes a position ${\vec r}_j$  on the square lattice (chosen to be a center of a link) as well as,
if needed, an additional quantum number $a_j$ of the fermion at that lattice position [i.e. $I_j = ({\vec r}_j, a_j)$].
As mentioned above, in the cases of interest, the energy is usually set to zero, i.e. $E=0$. The matrix $U^{\pm}\equiv U^{\pm 1}$ (where $U=U^{+1}$) 
in Eq.~\eqref{Eq-GreensChalkerCoddington-new}
is unitary, and can be thought of  as arising from a time-evolution by a small time-step $\delta \tH$ with the first quantized Hamiltonian ${\cal H}$, i.e. writing $U^{\pm}=\exp \{ \mp \i \delta \tH \ {\cal H}\}$, thereby recovering Eq.~(\ref{Eq-GreensFctContinuum-new}) from Eq.~(\ref{Eq-GreensChalkerCoddington-new}) in the limit of small $\delta \tH$. Here, 
$\tH$ denotes the time associated with the Hamiltonian $\mathcal{H}$ as introduced above and we reiterate that $\tH$ is to be distinguished from the $t$-coordinate of the GTN. In fact, the coordinates $(x,t)$ or $(u,v)$ introduced in the main text for the square-lattice GTN should be viewed as the spatial coordinates of the corresponding Chalker-Coddington model.

Just as we  can, in continuous time $\tH$, represent the resolvent appearing in Eq.~(\ref{Eq-GreensFctContinuum-new}) as the Laplace transformation of the (retarded) time-evolution operator, 
\begin{eqnarray}
\label{Label-Eq-LaplaceTransform-new}
\left (  \mp \i E + \epsilon \pm \i {\cal H} \right)^{-1}=
\int_{0}^{\infty} d \tH \exp\{- \tH [ \mp \i E + \epsilon \pm \i {\cal H}]\},
\qquad \ 
\end{eqnarray}
the expression
Eq.~(\ref{Eq-GreensChalkerCoddington-new}) appearing for discrete time is the discrete Laplace transform of the discrete time-evolution operator, 
\begin{eqnarray}
\label{label-Eq-DiscreteLaplaceTranform-new}
&&\left (1- e^{\pm  \i E -\epsilon} U^{\pm} \right )^{-1}=\sum_{n=0}^{\infty} \ \exp (- n [\mp \i E + \epsilon ]) \ \left( U^{\pm 1} \right)^n.   \ \ \qquad 
\end{eqnarray}
In the discrete formulation, the quantum state of the system at (discrete) time $\tH$ is described by
a wave function $\psi_I(t)$,
\begin{eqnarray}
&&|\psi(\tH)\rangle
= \sum_{I} \psi_I(\tH) \ |I\rangle
\\ \label{Label-Eq-QuantumStateNetwork}
&=& \sum^{\rm lattice \ links}_{j} \ \ 
\sum^{{\rm at \ link} \ {\vec r}_j}_{a_j}
\  \psi_{{\vec r}_j, a_j}(\tH)  \ |{\vec r}_j, a_j\rangle.
\end{eqnarray}

The matrix $U^{\pm }$ (having row and column indices $I_i$ and $I_j$) describes the unitary (forward/backward)  time-evolution of the 
quantum state of the network
by one discrete time-step:
\begin{eqnarray}
|\psi(\tH+1)\rangle = \sum_{I} \psi_I(\tH+1) \ |I\rangle,
\end{eqnarray}
where
\begin{eqnarray}
\label{Eq-TimeEvolutionNetworkWaveFunction-new}
\psi_I(\tH+1) = \sum_J \ U_{I,J}  \ \psi_J(\tH).
\end{eqnarray}

In one time-step the wave function evolves from link ${\vec r}_j$ to an adjacent link ${\vec r}_i$ (${\vec r}_i$ and ${\vec r}_j$ are two links attached 
to the {\it same}  lattice point of the
square lattice, usually referred to a ``node'' in this context; all other matrix elements vanish owing to the locality of the evolution) or, 
depending on the case (or symmetry class),
`reflects back' to the same link ${\vec r}_j$.
Thus, in
one time-step there is probability flux moving from one link to an adjacent link through a node
(or possibly `reflecting back').
For this reason the
matrix $U_{I,J}$  simply encodes the information of a ``scattering matrix'' or ``$S$-matrix'' for scattering of
probability flux at a node.

The information contained at each node in a ``scattering matrix'' can be recast in the familiar way in the language of a ``transfer matrix''
at the same node. This process is reviewed explicitly in Sec.~\ref{sec:GTN_CC_mapping} (where the square lattice is rotated by 45 degrees).
Since we are interested in situations where the original non-interacting fermion problem is subject to
static (quenched) disorder in $D=2$-dimensional position space, these $S$-matrices, and consequently also the transfer matrices
defined at each node,  will in general, in each realization of disorder, differ from node to node.

Finally, we come back to  higher spatial dimensions $D$: While the
above discussion was cast in the language of the correspondence of
Hamiltonians of non-interacting fermions in $D=2$ spatial dimensions subject to static disorder,  with
$D=2$ dimensional Chalker-Coddington models
upon discretization of space- and time-coordinates (where the  time-coordinate refers to the coordinate $\tH$), the same procedure carries through in higher dimensions. Thus, a non-interacting fermion system in $D=3$ spatial dimensions whose unitary time-evolution is 
governed by a Hamiltonian subject to static disorder in $D=3$ dimensional space, corresponds to a $d=2$-dimensional
quantum circuit subject to disorder in both space and time via a $D=3$-dimensional Chalker-Coddington model. (A relatively
recent detailed discussion of Chalker-Coddington models in $D=3$ appeared in Refs.~\onlinecite{OrtunoSomozaChalker-3DClassCNetwork-PRL2009,SonRaghu3DChalkerCoddingtonAII-Aug-2020} for symmetry classes C and AII.)

In any spatial dimension $D$, the properties of systems of non-interacting fermions subject to static disorder are well 
studied.
In particular, in the absence of extra
conservation laws arising from unitarily implemented symmetries,  these systems correspond to the so-called ``ten-fold
way" Altland-Zirnbauer classification, the ten Altland-Zirnbauer symmetry classes which exhaustively classify the behavior of fermionic quantum systems  invariant under symmetries which arise from the most general anti-unitary symmetry operations (including amongst others, e.g.,  time-reversal). 
There are then 10 classes of such local scattering $S$-matrices, and unitary time-evolution operators
$U^{\pm}$.
In the presence of additional unitary symmetries, more classes can arise in this manner.

For the discussion of the numerical results, it is also important to take boundaries into account. In GTN/NGC dynamics, one is typically interested in the physical state at late times, i.e. after evolving for a sufficiently long time such that physical quantities have reached a steady-state. In our GTN simulations in Sec.~\ref{sec:numerics}, this corresponds to considering the state for sufficiently large network depth $v$. Hence, the corresponding Chalker-Coddington model has a spatial boundary at a
certain large value of $v$, denoted as $v_b$ (the subscript ``$b$" standing for boundary). One could also consider terminating the GTN (and its corresponding Chalker-Coddington model) at a certain large $t$-coordinate value $t=t_b$, namely a large circuit time $t$ for the corresponding NGC, instead and study the fermion correlation on the boundary at $t=t_b$. However, 
the choice of the direction of the spatial
boundary of the Chalker-Coddington model does not affect universal results. Here, we reiterate that both the $v$- and $t$-coordinates of the GTN should be understood 
as spatial coordinates
of the Chalker-Coddington model (which are to be distinguished from the discrete time $\tH$ that is introduced earlier in this appendix to relate the Chalker-Coddington model to the static Hamiltonian $\mathcal{H}$).

In the language of the  corresponding Chalker-Coddington model, the fact that the GTN/NGC is simply stopped at depth $v=v_b$, or at coordinate $t=t_b$, amounts to a  particular boundary condition on the Chalker-Coddington model. We now discuss what boundary condition this is. Physically, it is clear that from the point of view of the discrete time-evolution (with discrete time step in $\tH$) of the Chalker-Coddington model that any quantum mechanical probability flux that hits this boundary from ``inside'' the Chalker-Coddington network simply escapes to what would be depths $v$ or $t$-coordinates larger than $v=v_b$ or $t=t_b$ (i.e. ``outside'' the Chalker-Coddington model). Moreover, the actual GTN and the Chalker-Coddington model end at $v=v_b$ or $t=t_b$, and are thus not present at depths $v$ or $t$-coordinates larger than $v=v_b$ or $t=t_b$.
Therefore, no quantum mechanical probability flux will ever enter the Chalker-Coddington model through this boundary from depths $v$ or $t$-coordinates larger than $v=v_b$ or $t=t_b$,
i.e. from ``outside'' the 
Chalker-Coddington network. 
This type of  boundary condition is well known in the context of Chalker-Coddington models: It is what is called an {\it
absorbing boundary 
condition}~\cite{XiongReadStone-MesoscopicConductance-PRB56-1997-3982,BettelheimGruzbergLudwig-ConformalRestriction-PRB86-2012}. One can also show at the microscopic level of the
GTN that the absorbing boundary condition should be applied to the boundary of the corresponding Chalker-Coddington model at depth $v=v_b$ or $t$-coordinate $t=t_b$.
We remark that when
Chalker-Coddington models are used to describe and/or  compute the electrical or thermal conductance properties of systems of non-interacting fermions in $D$ spatial dimensions subject to static disorder, such absorbing
boundary conditions represent idealized  contacts of the conducting system with so-called {\it ideal leads} to which the system is
connected in order to measure and/or define corresponding conductances~\footnote{Contacts occurring in the real world are of course known to
possess contact resistances which have to be taken into account, but which are solely properties of the physical contacts themselves.
The notion of an idealized contact allows the discussion of the conduction properties of the sample, without reference to the
details of the contact which could be included into the calculation of the conductance of the system connected to a realistic contact at a later point.}.

\subsection{Field Theory description of disorder averaged observables}
\label{sec:app-field-theory}
On  length scales much longer than the mean free path arising from disorder (serving as a microscopic short-distance `cut-off' scale), 
the theoretical description of these systems is known to be very systematic and geometrical. Disorder averaged observables for any Hamiltonian with static disorder in any one of the 10 symmetry classes  possess a `hydrodynamic' description in terms of   a specific non-linear sigma model (NLSM) field theory -- one for each symmetry 
class~\cite{ZirnbauerSusySymmSpacesMJathPhys1996,WegnerBetaFunctions-NPB316-1989-663}.
 (For a more  recent discussion of this dictionary, see e.g. Refs.~\onlinecite{SchnyderRyuFurusakiLudwigPRB2008,RyuSchnyderFurusakiLudwig-NewJPhysics12-2010-065010,LudwigNobelSymposium2015}.)
The observable described by these long length-scale theories is the average of the modulus square of the retarded 2-point function, 
Eqs.~(\ref{Eq-GreensFctContinuum-new}), (\ref{Eq-GreensChalkerCoddington-new}), and 
higher disorder moments thereof.

In a nutshell, the field theory represents the disorder average of the absolute square of the 2-point function, which is the average of the product of the retarded and the advanced
2-point function, 
\begin{eqnarray}
\nonumber
&&
\overline{G_{a,a}^+({\vec r}_1, {\vec r}_2) \ 
[G_{b,b}^+({\vec r}_1, {\vec r}_2)]^*}
=\overline{G_{a,a}^+({\vec r}_1, {\vec r}_2) \ \ G_{b,b}^-({\vec r}_2, {\vec r}_1)}
\\ \nonumber
&&
=\overline{\Big\langle 
\left( {\hat c}_{+,a}({\vec r}_1)
{\hat c}^\dagger_{-,b}({\vec r}_1)\right )
\ 
\left ({\hat c}^\dagger_{+,a}({\vec r}_2)
{\hat c}_{-,b}({\vec r}_2)
 \right )
\Big\rangle} 
\\ \label{Label-Eq-Q-field-Correlator}
&&
\propto
\langle
Q^{+-}_{a, b}({\vec r}_1)
Q^{-+}_{b, a}({\vec r}_2)
\rangle.
\end{eqnarray}
Here, $Q^{+-}_{a, b}$  is a complex hermitian Hubbard-Stratonovich field [thus satisfying
$(Q^{+-}_{a, b})^*=$ $Q^{-+}_{b, a}$], whose averages are
evaluated using the action for the NLSM (parametrized by
$Q^{+-}_{a, b}$)
in the corresponding symmetry class. In this formulation,
the indices are expanded to include replica indices $\alpha, \beta \in \{ 1, ..., n\}$, i.e.
$a \to (a,\alpha)$ and $b \to (b, \beta)$.
The number $n$ of replicas is taken to zero, i.e. $n\to 0$, at the end of the calculation.

Let us now  specialize to symmetry class DIII which is discussed in the main part of the paper. Since this describes a superconductor, the 2nd quantized Hamiltonian can be written as a bilinear of Majorana fermions ${\hat \gamma}_a({\vec r})$, and the  corresponding
1st quantized Hamiltonian is anti-symmetric and purely imaginary. We can obtain a formulation in terms of complex fermions, as the one used 
in Eqs.~(\ref{Eq-GreensFctContinuum-new}) and (\ref{Label-Eq-Q-field-Correlator})
above, by introducing a second copy ${\hat \eta}_a({\vec r})$
of Majorana fermions and defining ${\hat c}_a({\vec r})\equiv$
$[{\hat \gamma}_a({\vec r}) + \i {\hat \eta}_a({\vec r})]/\sqrt{2}$.
Then, Eq.~(\ref{Eq-GreensFctContinuum-new}) represents the 2-point function in the form written. Consider now the quantity in Eq.~(\ref{Label-Eq-Q-field-Correlator})
where the two points ${\vec r}_1$ and ${\vec r}_2$ are located near the (`final' $v=v_b$ or `final' $t=t_b$) boundary of the circuit for $E=0$ (where the system is critical). 
Because ${\vec r}_1$ and ${\vec r}_2$ are near the boundary it turns out that we can now set $\epsilon \to 0$. We see from Eq.~(\ref{Eq-GreensFctContinuum-new}) that at $E=\epsilon=0$ this expression is
anti-symmetric under exchange of $({\vec r}_1, a_1)$ and
$({\vec r}_2, a_2)$. Expressing this in terms of Majorana fermions ${\hat \gamma}$ and ${\hat \eta}$ defined above, the first
2-point function
in Eq.~(\ref{Label-Eq-Q-field-Correlator})
equals
$2 \langle {\hat \gamma}_{a}({\vec r}_1)
{\hat \gamma}_{a}({\vec r}_2) \rangle$. 
We can now simply replace ${\hat c}_{-,b}$ in the same equation
by a second copy ${\hat c}_{+,b}$ of the fermion ${\hat c}_{+,a}$  yielding the same 2-point function.
Using  this in
Eq.~(\ref{Label-Eq-Q-field-Correlator}) we see that the left hand side of this equation is proportional to (i.e. four times) the second moment of the 
Majorana fermion 2-point function,
$\overline{\langle {\hat \gamma}_{a}({\vec r}_1)
{\hat \gamma}_{a}({\vec r}_2) \rangle \ 
\langle {\hat \gamma}_{b}({\vec r}_1)
{\hat \gamma}_{b}({\vec r}_2) \rangle}$,
evaluated at two points on the boundary. The indices $a \not = b$ can be taken to be replica indices.
This is the quantity evaluated numerically in Eq.~(\ref{eqn:Cd})
of Sect.~\ref{Label-SubSection-CorrelationFunctions} of the main part of the paper.
Next we discuss the right hand side of the 
Eq.~(\ref{Label-Eq-Q-field-Correlator}).

In symmetry class DIII  of interest in the main part of this paper, the NLSM field $Q^{+-}_{a,b}$
is known to be
real and an element of the (special) orthogonal group, 
$Q^{+-}_{a,b}=$
$O_{a,b}\in \SO(n)$, 
where $a,b=1, ...n$ are replica indices,
and $Q^{-+}_{b, a} = (O^{-1})_{a,b}$ the inverse group element.
Moreover, at an absorbing boundary such as the one at depth $v=v_b$ or $t$-coordinate $t=t_b$ for the Chalker-Coddington model that corresponds to a GTN that terminates at $v=v_b$ or $t=t_b$,
the NLSM field $O({\vec r})\in \SO(n)$ tends to the identity element in the group. Parametrizing this field in terms of the Lie algebra, $O({\vec r})=$
$\exp\{ \i \sum_{a<b} T_{ab}  \ \phi_{ab}({\vec r})
\}$, where $T_{ab}$ are suitably normalized matrices anti-symmetric in $a$ and
$b$, the absorbing boundary condition is a Dirichlet boundary condition on
the fields $\phi_{ab}$ (anti-symmetric in indices $a$ and $b$).
The correlation function of Majorana bilinears at the boundary will then be given by the normal derivative at the boundary of the NLSM field $O({\vec r})$,
and will hence be proportional to ${\partial \over \partial v}\phi_{ab}({\vec r})$ or
${\partial \over \partial t}\phi_{ab}({\vec r})$, where the derivatives are
taken at the `final' depth $v=v_b$ or the `final' $t$-coordinate $t=t_b$ of the Chalker-Coddington model, depending on the formulation we chose to consider.

The action of the NLSM in class DIII,
\begin{eqnarray}
\label{Label-Eq-NLSM-DIII}
S=
\int d^d {\vec r}
~ \Tr \left\{ {1\over 2g} ({\vec \nabla} O^{-1})
({\vec \nabla} O)
- \epsilon  
(O + O^{-1})
\right\}  \qquad
\end{eqnarray}
simplifies, upon rescaling in the usual manner $\phi_{ab}$ $\to \phi_{ab}/\sqrt{g}$, to leading order for small coupling constant $g$, i.e.
in the metallic phase, to a Gaussian action in the free scalar fields
$\phi_{ab}$. The 2-point function of the normal derivatives along the boundary
thus gives a power-law decaying with distance along the boundary with
exponent $=2$. This power-law acquires a significant correction from the leading irrelevant operator, which is known to be marginally irrelevant, at the metallic Gaussian fixed point of the NLSM. The functional form of the
boundary correlation function of the
Majorana fermion bilinear in the presence of this  marginally irrelevant operator in the bulk is computed
in App.~\ref{app:marginal} by computing suitable renormalization group (RG) functions, and solving the corresponding Callan-Symanzik (RG) equation for the boundary correlation function. The resulting functional
form for this  correlation function has been fit successfully in the main part of this
paper to the same function, computed numerically.

\begin{widetext}

\section{Logarithmic corrections to scaling of 2nd moment of the fermion correlation function in the symmetry-class-DIII metallic phase}
\label{app:marginal}

\subsection{Setup}
\label{LabelSubsectionSetupandResults}

We start from the long-wavelength formulation
of disordered Majorana fermions in symmetry class DIII and $D=2$ spatial dimensions, 
the NLSM in  Eq.~(\ref{Label-Eq-NLSM-DIII}),
valid on length scales large compared to the mean free path.
This is a special type of NLSM (also known as the ``Principal Chiral model'') in which the field $O({\vec r}) \in \SO(n)$ is an element of a group (in the present case the (special) orthogonal group), which we parametrize as
\begin{eqnarray}
\label{LieAlgebraParametrization-O-n}
&&
O({\vec r})
=
\exp\left\{\i \sum_A \phi_A({\vec r}) \ T_A \right\},
\end{eqnarray}
where $n$, the number of replicas,
tends to zero at the end of the calculation, a well-understood limit in the present situation.
Here,  $\phi_A({\vec r})$ are real fields, and $T_A$ are $n(n-1)/2$ matrices generating infinitesimal $\SO(n)$ rotations which 
form a basis of the Lie algebra in the defining ($n$-dimensional)
representation
suitably normalized, i.e. $\Tr (T_A T_B )= \delta_{A,B}$, as in App.~\ref{LabelSubsectionSummaryGroupTheoryFacts} below. One can choose the subscript $A$ that labels the $\SO(n)$ generator $T_A$ to be $A=(a,b)$ with $1 \leq a < b \leq n$, but this will be unimportant in the present section.

Upon inserting the parametrization 
from Eq.~(\ref{LieAlgebraParametrization-O-n}) into the action Eq.~(\ref{Label-Eq-NLSM-DIII}) appearing in the Boltzmann weight
$\exp\{- S \}$ for the resulting statistical mechanics model describing disorder averaged observables, this action can be written in a perturbative expansion
in the parameter $g$~\footnote{signifying physically the inverse longitudinal thermal conductivity divided by $k_B T$ (where $T$ is temperature), in the zero temperature limit, 
of the system of fermionic BCS quasiparticles deep inside the superconducting phase of a superconductor in symmetry class DIII.} 
about a Gaussian fixed point theory (describing the metallic fixed point in class DIII) as
\begin{eqnarray}
\label{LabelEqSmall-g-Expansion-Class-DIII}
\nonumber 
S &=&  S_0 + S_{int}
\\ \nonumber
S_0 &=& \int_{\vec r}
{1\over 8\pi}
(\partial_{\mu} \varphi_A)
(\partial_{\mu} \varphi_A)
\\ \nonumber
S_{int} &=& \lambda 
\int_{\vec r}
\left [(\partial_{\mu} \varphi_{A_1}) \varphi_{B_2}\right ]
C_{A_1 B_2 H} C_{H A_2 B_1}
\left [\varphi_{A_2} (\partial_{\mu}\varphi_{B_1})\right ] +\mathcal{O}(\lambda^2)
\label{LabelEq-NLSM-Fourth-Order-Expansion}
\end{eqnarray}
where $\int_{\vec r} = \int d^2 {\vec r}$,
$\phi_A = (\sqrt{g/ (4 \pi)}) \ \varphi_A 
$ ,
$\lambda = \kappa_0  \ g$ where $\kappa_0$ is a fixed positive rational number (whose value is immaterial), $\mathcal{O}(\lambda^2)$ denotes terms of order $\lambda^2$,
and  the totally antisymmetric and cyclically invariant coefficients $C_{A B C}$ characterize the structure constants of the Lie algebra of  the group $\SO(n)$,
\begin{eqnarray}
\label{LabelEq-LieAlgebra}
&&
[T_A, T_B]
=
\i
C_{A B C} \ T_C.
\end{eqnarray}
The action $S_0$ describes the metallic fixed point of free scalar fields in $D=2$ dimensions, an elementary two-dimensional CFT in which each scalar field
is a sum of holomorphic and anti-holomorphic fields,
\begin{eqnarray}
\label{LabelEqFreeScalarFields}
&&\varphi_A({\vec r}) = \varphi^A_L (z) + \varphi^A_R(z^*)
\end{eqnarray}
where $z=x + \i y$, $z^* = x - \i y$ when ${\vec r}= (x, y)^\T$.
Using
Eqs.~\eqref{LabelEq-NLSM-Fourth-Order-Expansion} and \eqref{LabelEqFreeScalarFields}
and $(1/4) \partial_{\mu} \partial_{\mu}=$
$  (\partial/\partial z) (\partial /\partial z^*)=$
$\partial_z \partial_{z^*}$ we arrive at the following form of the action that we will use in the sequel
\begin{eqnarray} \nonumber
S &=&  S_0 + S_{int}
\\ \nonumber
S_0 &=& \int_{\vec r}
{1\over 2\pi}
(\partial_{z} \varphi_L^A)
(\partial_{z^*} \varphi_R^A)
\\ \label{LabelEqSmall-g-Expansion-Class-DIII-complex}
S_{int} &=& \lambda 
\int_{\vec r}
\left [(\partial_{z} \varphi_L^{A_1}) \varphi_L^{B_2}\right ]
C_{A_1 B_2 H} C_{H A_2 B_1}
\left [\varphi_R^{A_2} (\partial_{z^*}\varphi_R^{B_1})\right ]+\mathcal{O}(\lambda^2).
\end{eqnarray}
We will need the correlators of the scalar fields at $\lambda=0$,
\begin{eqnarray}
\nonumber
\langle \varphi_L^A(z_1) \varphi_L^B(z_2)\rangle &=& - \delta_{AB} \ln z_{12}, \qquad  \quad
\\ \nonumber
\langle \varphi_R^A(z^*_1) \varphi_R^B(z^*_2) \rangle &=& - \delta_{AB} \ln z^*_{12}
\\ \nonumber
\langle (\partial_z\varphi_L^A)(z_1) \varphi_L^B(z_2)\rangle &=& {(-1)\delta_{AB}\over  z_{12}},
\\ \nonumber
\langle (\partial_{z^*}\varphi_L^A)(z^*_1) \varphi_L^B(z^*_2)\rangle &=& {(-1) \delta_{AB}\over  z^*_{12}},
\\ \nonumber
\langle\varphi_L^A(z_1)(\partial_z \varphi_L^B)(z_2)\rangle &=& {\delta_{AB}\over  z_{12}},
\\
\label{LabelEqFreeScalarCorrelators}
\langle \varphi_L^A(z^*_1) (\partial_{z^*}\varphi_L^B)(z^*_2)\rangle &=& {\delta_{AB}\over  z^*_{12}}.
\end{eqnarray}
No summation over repeated indices is implied here.

\subsection{Absorbing Boundary}
\label{LabelSubsection-AbsorbingBoundary}

We also need to discuss the absorbing boundary condition  which we place
at $y = \Im (z) =0$, the real axis of the complex $z$-plane,  at which (as already discussed in App.~\ref{sec:app-field-theory}) the NLSM field
from Eq.~\eqref{LieAlgebraParametrization-O-n} tends to the identity group element, $O({\vec r}) \to 1$, implying
$\phi_A({\vec r}) \to 0$. 
This hence also implies a Dirichlet boundary condition on the scalar field in Eq.~(\ref{LabelEqFreeScalarFields}),
\begin{eqnarray}
\nonumber
\varphi_A({\vec r}) = \varphi_L^A(z) + \varphi_R^A(z^*) \to 0, \ \  &{\rm as}& \ \Im z = y \to 0
\\ 
\label{LabelEqFreeScalarFields-Boundary}
{\rm or} \quad \varphi_R^A(z^*) \to (-1) \varphi_L^A(z), \ \  &{\rm as}& \ \Im z = y \to 0.
\qquad 
\end{eqnarray}
Because the scalar field $\phi_A$ vanishes at the absorbing boundary, the 2-point function of the field $O({\vec r})$ near the boundary
becomes that of the normal derivative $\left({\partial \over \partial y} \varphi_L^A \right)$ (or equivalently of the normal derivative of  $-\varphi_R^A$). 
That 2-point function 
reads in the non-interacting  fixed-point theory [Eq.~(\ref{LabelEqSmall-g-Expansion-Class-DIII-complex})
at $\lambda=0$]
\begin{eqnarray}
\Big\langle
\left({\partial \over \partial y} \varphi_L^A\right)(x_1) \ 
\left({\partial \over \partial y} \varphi_L^B\right)(x_2) \Big\rangle
=
\label{LabelEq-Boundary-2-point-function}
\Big\langle
\left[{\partial z\over \partial y} (\partial_z\varphi_L^A)\right](x_1)
\ 
\left[{\partial z \over \partial y} (\partial_z \varphi_L^B)\right](x_2) \Big\rangle
= {\delta_{A,B}\over x_{12}^2}.
\end{eqnarray}
In conclusion, the NLSM field $O({\vec r})$ near the boundary becomes a  {\it boundary operator} which we 
denote~\footnote{The subscript $s$ stands for ``surface'', which in the present context just means in general ``boundary'' - in particular a one-dimensional boundary (the real axis) of $D=2$-dimensional (bulk)  position space.}
by $\Phi_s(x)\equiv \lim_{y \to 0}\left({\partial \over \partial y} \varphi_L^A \right)$; its
2-point function at the fixed point is~\footnote{The 2-point function is independent of the choice of index $A$ due to permutation symmetry of the replica indices.}
\begin{eqnarray}
\label{BoundaryTwoPointFunction}
\langle\Phi_s(x_1)\Phi_s(x_2)\rangle = {1\over x_{12}^2}.
\end{eqnarray}

In the next subsection, we will discuss the effect of the interaction $\lambda$
on this boundary 2-point function.
For this purpose, it will be crucial to understand the behavior of the interaction term $S_{int}$ 
from Eq.~(\ref{LabelEqSmall-g-Expansion-Class-DIII-complex})
near the absorbing boundary. Here, the region of integration over ${\vec r}$ will be the  upper complex $z$-plane, and  thus the argument of $\varphi_L^A(z)$
will be in the upper complex plane. On the other hand, due to the Dirichlet boundary condition, the second line of
\eqref{LabelEqFreeScalarFields-Boundary}
implies that
$
\left (
\left [\varphi_R^{A} (\partial_{z^*}\varphi_R^{B})\right ]
-
\left [\varphi_R^{B} (\partial_{z^*}\varphi_R^{A})\right ]
\right )_{z^*}$
is the analytic continuation of the expression
$
\left (
\left [\varphi_L^{A} (\partial_{z^*}\varphi_L^{B})\right ]
-
\left [\varphi_L^{B} (\partial_{z^*}\varphi_L^{A})\right ]
\right )_{z}$
from the upper half complex plane
into the lower half complex plane~\footnote{The anti-symmetrization is implicit due to the contraction with the anti-symmetric structure constant $C_{A B C}$.}, the two expressions becoming equal to each other on the real axis. Thus, the
right-moving factor (involving $\varphi_R$)  of the interaction operator 
in $S_{int}$ is located precisely at the mirror image with respect to the real axis of the left-moving factor, and all scalar fields are left-moving ($\varphi_L$) after this analytic continuation. This  fact turns out to be crucial
for the ability to perform the integral over half-space, the upper half complex plane, in an efficient manner.

Before proceeding to the effect of the interaction $\lambda$ on the correlation function
in Eq.~(\ref{BoundaryTwoPointFunction}), we turn to the $N$th disorder moments of the square of the Majorana
correlation function appearing in Eq.~(\ref{eqn:Cd}), namely
$\overline{ \langle \i \hat{\gamma}_{p,m} \hat{\gamma}_{p+r,n} \rangle^{2N} }$, discussed at the very end of
Sec. \ref{LabelSubSubSectionCriticalEntanglementPhase}. As follows from the discussion in the two paragraphs preceeding
Eq.~(\ref{Label-Eq-NLSM-DIII}), these moments are described in NLSM language by the 2-point function of  the $n$-fold product of fields in $N$ different replicas at the same boundary point $x$,
$\lim_{y \to 0} \left [ \left({\partial \over \partial y} \varphi_L^{A_1}(x) \right) ...
\left({\partial \over \partial y} \varphi_L^{A_N}(x) \right)\right ] $. Since at the metallic fixed point,
Eq.~(\ref{LabelEqSmall-g-Expansion-Class-DIII-complex}) with
$\lambda=0$, all replica indices are decoupled, this 2-point function equals the $N$th power of the $N=1$
result from Eq.~(\ref{BoundaryTwoPointFunction}), i.e. it decays with the $2N$th power of distance. This was the result mentioned in the main text at the end of Sec.~\ref{LabelSubSubSectionCriticalEntanglementPhase}.

\subsection{Renormalization Group Calculation}
\label{LabelSubsectionRenormalizationGroupCalculation}

The purpose of this subsection is to obtain the  renormalization group (RG) anomalous dimension function
(denoted by $\gamma_s(\lambda)$ below) of the boundary operator
$\Phi_s$, which leads to the functional form of the 2-point function in the presence of the marginally irrelevant bulk perturbation
$\Phi$ as determined as the solution of the Callan-Symanzik (RG) equation for this function.

The most efficient way to perform the 1-loop RG calculation is using the Operator Product Expansion (OPE) and tracking the change of the
action $S$ upon changing  a hard short-distance 
cutoff~\cite{Cardy-LogCorrections-JPhysA19-1986-L1093,Ludwig-Potts-OPE-RG-NPB285-1987-97,LudwigCardy-c-Theorem-NPB687-1987-687,WieseLudwig-4Loop-NPB661-2003-577}. We will need to consider the renormalization of the bulk operator  
$\Phi({{\vec r}})=$
$\Phi(z, z^*)$
appearing in the interaction term
in \eqref{LabelEqSmall-g-Expansion-Class-DIII-complex}:
\begin{eqnarray}
\label{LabelEq-InteractionOperator}
&&
S_{int}=\lambda \int_{\vec r} 
\Phi({\vec r})
+ {\cal O}(\lambda^2)
\\ \nonumber
\end{eqnarray}
as well as that of the boundary operator
$\Phi_s(x)=$ ${\partial \over \partial y} \varphi_L^A(x)$ 
appearing in Eq.s~(\ref{BoundaryTwoPointFunction},\ref{LabelEq-Boundary-2-point-function}). In the current subsection, we carry out this RG calculation using the results obtained for the corresponding OPE coefficients in App.~\ref{LabelSubsection-DetailsOfOPE}
below. These OPE coefficients are
\begin{eqnarray}
\nonumber
&&
{\Phi}(z_1, z^*_1) {\Phi}(z_2, z^*_2) \sim
{-b \over z_{12} z^*_{12} }
 {\Phi}(z_2, z^*_2) + ..., \ z_1 \to z_2,
\\ 
\label{LabelEq-Result-OPE-Coefficients}
&&
{\Phi}(z, z^*) {\Phi}_s(x=0)
\sim
{- b_s\over
z z^*} {\Phi }_s(x=0) + ....,
\ z \to 0,
\qquad
\end{eqnarray}
where the ellipsis denotes subleading terms in the considered limit.
As mentioned, the numbers $b$ and $b_s$ are computed explicitly in App.~\ref{LabelSubsection-DetailsOfOPE}.

These OPE coefficients turn out to imply the following renormalization group (RG) equations for the bulk coupling constant $\lambda$, as well as for a 
coupling constant~\footnote{The renormalization of the boundary operator $\Phi_s$ by the bulk perturbation specified by the bulk operator $\Phi$ arises from the integral over bulk positions ${\vec r}=(z, z^*)$ in the upper half complex plane. Owing to the
OPE in the 2nd line in \eqref{LabelEq-Result-OPE-Coefficients}, this integral  is logarithmically divergent at short distances
coming the integration region near the position of the boundary operator $\Phi_s(x=0)$ located on the real axis at position $x=0$.
Due the  analytic continuation property of the interaction operator $\Phi(z,z^*)$ discussed in the paragraph below
\eqref{LabelEq-Boundary-2-point-function}, and since the integral over the bulk interaction operator is only over the upper-half
complex plane, the effect of this integration over the location of the bulk interaction on the RG equation for the boundary operator $\Phi_s$ is half of what it would have been had the boundary operator $\Phi_s$ been replaced by another bulk operator. This type
of mechanism
was first used~\cite{AffleckShaojin-LogCorrectionsQuantumImpurityProblems-JPhysA32-1999-7815}
in a similar (but not identical) context for the renormalization of a boundary operator due to a bulk operators in the context of one-dimensional Heisenberg quantum spin chains.}~\cite{AffleckShaojin-LogCorrectionsQuantumImpurityProblems-JPhysA32-1999-7815}
$h_s$ conjugate~\footnote{imagine a surface (boundary) ``boundary magnetic field''}
to the boundary operator ${\Phi}_s(x)$ when added to the action as  $S \to S(h_s) =$ $S -  h_s \int dx {\Phi}_s(x)$,
\begin{eqnarray}
\nonumber
{ d \lambda \over dl}&=& \beta(\lambda)
= - \pi (- b) \lambda^2 + ...
\\ \label{LabelEq-RG-Equations}
{ d h_s \over dl}
&=& 
-2 \pi (- b_s/2) \lambda  h_s + ... 
\qquad \quad
\end{eqnarray}
where $dl$ is the infinitesimal increase of the logarithm of the short distance cutoff during a RG step.
The second equation provides the anomalous dimension function
$\gamma_s(\lambda)$  of the boundary operator  ${\Phi}_s(x)$, defined by
\begin{eqnarray}
\label{LabelEq-AnomalousDimensionBoundaryOperator}
&&
\gamma_s(\lambda) 
=
(1-1) - {1\over h_s} {d h_s \over dl} 
= 2 \pi (-b_s/2) \lambda + ...,
\end{eqnarray}
where the $(1-1)$ part means that
the dimension of the boundary is $1$, and the scaling dimension of the boundary operator at $\lambda=0$ is also $1$.

The 2-point function of the boundary operator in the presence of the coupling constant $\lambda$
of the marginally irrelevant bulk operator ${\Phi}$ is then found by solving the RG equation  for this 2-point function
(the usual Callan-Symanzik
equation), yielding
\begin{eqnarray}
\langle {\Phi}_s(x_1) {\Phi}_s(x_2)\rangle
\label{LabelEq-SolutionCallanSymanzik}
={1\over (x_{21})^2}
 \ \exp\left\{ -2 \int_{\lambda(1)}^{\lambda(x_{12}/a)}
 d \lambda 
 {\gamma_s(\lambda)
 \over \beta(\lambda)
 }
 \right\}
 \ F[\lambda(x_{12}/a)],
 \qquad \quad
\end{eqnarray}
where
\begin{eqnarray}
\lambda(e^l)
=
{\lambda
\over
1 + \pi (-b)  \lambda  \ l}
\end{eqnarray} is the running coupling constant, and $F[...]$ is a function that is finite at zero argument.

In App.~\ref{LabelSubsection-DetailsOfOPE} below we obtain the results
\begin{eqnarray}
\label{LabelEq-Quote-Results-for-OPE-Coeff}
&&
b = {\cal C}^{(2)}_{adj}, \quad b_s = - {\cal C}^{(2)}_{adj},
\quad
{\rm and}  \ \ {\cal C}^{(2)}_{adj} = 2 (n-2)
\qquad \ \ 
\end{eqnarray}
where ${\cal C}^{(2)}_{adj}$ is the value of the quadratic Casimir invariant in the adjoint representation of $\SO(n)$. As a 
first basic check, we then conclude from the
first line in Eq.~(\ref{LabelEq-RG-Equations}) that the bulk coupling constant $\lambda$ is indeed marginally irrelevant (in the infrared, where $l$ increases)
in the replica limit $n \to 0$.

Finally, inserting  the values obtained for $b$ and $b_s$ into the functions $\beta$ and $\gamma_s$ appearing in the integrand of the  integral  in Eq.~(\ref{LabelEq-SolutionCallanSymanzik})
above, we obtain for the boundary 2-point function
\begin{eqnarray}
\label{LabelEqSolutionCallanSymanzik}
&&
\langle {\Phi}_s(x_1) {\Phi}_s(x_2)\rangle
= {
[1 +  4 \pi  \lambda \ln (x_{12}/a)]^2
\over
(x_{12})^2},
\end{eqnarray}
where $\lambda_0=$ $4 \pi \lambda=$ $4 \pi \kappa_0 g$.
This is the result reported in Eq.~(\ref {eqn:corr-decay})
of the main text (up to a multiplicative factor which can always be absorbed by redefining the normalization of the field $\Phi_s$).

\subsection{Summary of relevant Group Theory facts, and derivation of Eq.~(\ref{LabelEqSmall-g-Expansion-Class-DIII})}
\label{LabelSubsectionSummaryGroupTheoryFacts}

\subsubsection{Summary of Group Theory facts}
\label{LabelSubsubsction-SummaryOfGroupTheoryfacts}

In the following
we summarize some basic group theory facts (and conventions) that we will use. We normalize the generators $T_A$ of infinitesimal $\SO(n)$
rotations in the defining ($n$-dimensional) representation via
\begin{eqnarray}
\label{LabelEqKillingForm}
&&
\Tr ( T_A T_B ) = \delta_{A,B}.
\end{eqnarray}
The structure constants defined in Eq. (\ref{LabelEq-LieAlgebra})
are known to define the generators 
\begin{eqnarray}
\label{LabelEq-Adjoint-Generators}
\left( T^{adj}_A\right)_{BC}
= (-\i) C_{ABC}
\end{eqnarray}
of infinitesimal $\SO(n)$ rotations
 in the $n(n-1)/2$-dimensional  adjoint representation. Using those, the quadratic Casimir invariant in the adjoint
 representation, ${\cal C}^{(2)}_{adj}$, is expressed as
\begin{eqnarray}
\label{LabelEq-Adjoint-Casimir}
C_{ABC} C_{DBC} = {\cal C}^{(2)}_{adj} \ \delta_{AD},
\end{eqnarray}
where ${\cal C}^{(2)}_{adj}=2 (n-2)$.
We will also need some basic information about 
quartic invariants, which arise when
considering traces of four generator matrices $T_A$. 
In particular~\footnote{The following equation can be obtained, e.g.,  by repeating for the generators of the defining representation the steps presented in Appendix C.1 of Ref.~\cite{WieseLudwig-4Loop-NPB661-2003-577}
for the generators of the adjoint representation. The structure constants ${f_C}^{AB}$ in that reference correspond to $i C_{ABC}$ in the notations of the present paper.}
\begin{eqnarray}
\label{LabelEq-QuarticCasimirInvariant}
\Tr\left (T_A T_B T_C T_D\right )
=
\alpha~
\Tr\left (T_{\{A} T_B T_C T_{D\}}\right )
+ \beta~
\left [
(-\i) C_{HAB} (-\i) C_{HCD}
+ (-\i)
C_{HDA}
(-\i) C_{HBC}
\right ],
\qquad \quad
\end{eqnarray}
where $\alpha$ and $\beta$ are positive numbers  (whose specific values are not needed here) and
the curly brackets under the trace on the right hand side of the first line
denote complete symmetrization of the four indices $A, B, C, D$ -  that symmetrized trace (multiplying $\alpha$)
defines the quartic invariant of the group $\SO(n)$.

\subsubsection{Derivation of Eq.~(\ref{LabelEqSmall-g-Expansion-Class-DIII})}
\label{LabelSubsubsection-DerivationSmall-g-Expansion}

After taking the derivative of the expansion of the
parametrization 
from Eq.~(\ref{LieAlgebraParametrization-O-n}) to second order in $\phi_A$
\begin{eqnarray}
\label{LabelEq-SecondOrderExpansionGroupElement}
\partial_{\mu}
O({\vec r})
=
\i (\partial_{\mu} \phi_A) T_A
+{\i^2 \over 2!}
\left[
(\partial_\mu \phi_{A_1}) \phi_{A_2}
+ \phi_{A_1} (\partial_{\mu} \phi_{A_2})
\right] \ T_{A_1} T_{A_2} + ...
\quad
\end{eqnarray}
and similarly for $\partial_{\mu}O^{-1}({\vec r})$, we insert
the result into the action Eq.~(\ref{Label-Eq-NLSM-DIII})
yielding
\begin{multline}
\Tr [(\partial_{\mu} O^{-1})
(\partial_{\mu}O)] =
\Tr \left (
\{-\i (\partial_{\mu} \phi_A) T_A\}
\{\i (\partial_{\mu} \phi_B) T_B\}
\right)+
\\
+{1 \over 4}
\left[
(\partial_\mu \phi_{A_1}) \phi_{A_2}
+ \phi_{A_1} (\partial_{\mu} \phi_{A_2})
\right] \times
\label{LabelEq-ExpansionNLSM}
\left[
(\partial_\mu \phi_{B_1}) \phi_{B_2}
+ \phi_{B_1} (\partial_{\mu} \phi_{B_2})
\right]
 \Tr \left(
T_{A_1} T_{A_2}T_{B_1}T_{B_2}
\right )
\end{multline}
where we have used the fact that the trace of three generator matrices is anti-symmetric
for groups whose defining representation is real, such as 
$\SO(n)$ and ${\rm USp}(2n)$, in particular
\begin{eqnarray}
\label{LabelEq-NoCubicCasimir}
\Tr ( T_A T_B T_c ) \propto C_{ABC},
\end{eqnarray}
implying that all traces of three $T$-matrices appearing in the expansion vanish
(due to the symmetry of the terms with which these traces are contracted).
Finally, we use Eq.~\eqref{LabelEqKillingForm} in the first term, and
Eq.~\eqref{LabelEq-QuarticCasimirInvariant} in the second term on the right hand side of 
Eq.~\eqref{LabelEq-ExpansionNLSM}.
The first term on the right hand side of Eq.~\eqref{LabelEq-QuarticCasimirInvariant},
totally symmetric in all four indices, yields when contracted against the other terms a total derivative $(\partial_{\mu}\partial_{\mu}) \left(\phi_{A_1} \phi_{A_2} \phi_{B_1} \phi_{B_2}\right)$ which can be dropped, while the second term on the right hand side of the same equation vanishes upon contraction due to symmetry. The remaining third term on the right hand side of Eq.~(\ref{LabelEq-QuarticCasimirInvariant}) then gives the interaction term listed in  Eq.~(\ref{LabelEqSmall-g-Expansion-Class-DIII}), concluding the derivation of this equation.

\subsection{Details of the Calculations of Operator Product Expansion  (OPE) coefficients}
\label{LabelSubsection-DetailsOfOPE}

\subsubsection{Bulk-Bulk OPE}
\label{LabelSubsubsection-Bulk-Bulk-OPE}
 We begin with the bulk operator $\Phi({\vec r})$ appearing in the interaction $S_{int}$ in Eq.
(\ref{LabelEqSmall-g-Expansion-Class-DIII-complex}) which factorizes into  a product of a holomorphic  (``left-moving'')
and an anti-holomorphic (``right-moving'') part.

\vspace{.3cm}

\noindent ({\it i}): We will first discuss the OPE of two holomorphic parts of this operator at two positions $z_1$ and $z_2$:
\begin{align}
\MoveEqLeft[4]
[(\partial_{z} \varphi_L^{A_1}) \varphi_L^{B_2} ]_{z_1}C_{A_1 B_2 H}
\ \ 
 [(\partial_{z} \varphi_L^{{A'}_1}) \varphi_L^{{B'}_2} ]_{z_2} C_{{A'}_1 {B'}_2 {H'}} \nonumber \\
=&
\left \{
{\delta^{B_2 {A'}_1}\over z_{12}}
[(\partial_{z} \varphi_L^{A_1}) \varphi_L^{{B'}_2} ]_{z_2}
-
{\delta^{{B'}_2 {A'}_1}\over z_{12}}
[ \varphi_L^{B_2} (\partial_{z}\varphi_L^{{A'}_1}) ]_{z_2}
\right \}
\times C_{A_1 B_2 H} C_{{A'}_1 {B'}_2 {H'}}+ ...
\nonumber \\
=& {1 \over z_{12}}
[(\partial_{z} \varphi_L^{A_1}) \varphi_L^{{B'}_2} ]_{z_2}\times
\left \{
C_{a_1, B_2, H} C_{B_2 {B'}_2 {H'}}
- C_{A_1 B_2 {H'}}
C_{B_2 {B'}_2 H}
\right \} +...
\nonumber \\
=&
{1 \over z_{12}}
[(\partial_{z} \varphi_L^{A_1}) \varphi_L^{{B'}_2} ]_{z_2}
\ \i^2 C_{A_1 {B'}_2 {H''}} C_{{H''} H {H'}} + .... \label{LabelEq-BulkBulk-OPE-holomorphic}
\end{align}
The last equality follows because,
upon making use of
Eq.~(\ref{LabelEq-Adjoint-Generators}),
the curly bracket on its left hand side can be written
as a commutator of the matrices $T^{adj}$ in the adjoint representation,
\begin{equation}
\i^2 \{ 
 (T^{adj}_H )_{A_1 B_2}
 (T^{adj}_{H'} )_{B_2 {B'}_2}
-
 (T^{adj}_{H'} )_{A_1 B_2}
 (T^{adj}_{H} )_{B_2 {B'}_2}
\}=
\i^2 \{
\i C_{H {H'} {H''}}
(T^{adj}_{H'} )_{A_1 {B'}_2}\}.
\label{LabelEq-Commutator}
\end{equation}
The result on the right hand side of the last equality in Eq.~(\ref{LabelEq-BulkBulk-OPE-holomorphic})
now follows upon making once again use of Eq.~(\ref{LabelEq-Adjoint-Generators}).

\vspace{0.3cm}

\noindent ({\it ii}): The OPE of two of the anti-holomorphic parts of the bulk operator $O({\vec r})$ at two positions $z^*_1$ and $z^*_2$
is obtained in the analogous way with the result:
\begin{align}
\nonumber
\MoveEqLeft[4]
[\varphi_R^{A_2} (\partial_{z^*} \varphi_R^{B_1}) ]_{z^*_1}C_{H A_2 B_1}
\ \ 
 [ \varphi_R^{{A'}_2} (\partial_{z^*}\varphi_R^{{B'}_1}) ]_{z^*_2} C_{{H'}{A'}_2 {B'}_1}
 \qquad 
\\ \nonumber
=&[(\partial_{z^*} \varphi_R^{B_1}) \varphi_R^{A_2} ]_{z^*_1}
\ \ 
 [(\partial_{z^*} \varphi_R^{{B'}_1}) \varphi_R^{{A'}_2} ]_{z^*_2} 
 C_{B_1 A_2 H}
 C_{{B'}_1{A'}_2 {H'}}
 \qquad 
\\
\nonumber
=&
{1 \over z^*_{12}}
[(\partial_{z} \varphi_L^{B_1}) \varphi_L^{{A'}_2} ]_{z^*_2}
\ \i^2 C_{B_1 {A'}_2 {H''}} C_{{H''} H {H'}} + ....
\\ 
=&
{1 \over z^*_{12}}
[\varphi_L^{{A}_2}
(\partial_{z} \varphi_L^{B_1})  ]_{z^*_2}
\ \i^2 (-1) C_{A_2 B_1  {H''}} C_{{H''} H {H'}} + .... 
\label{LabelEq-BulkBulk-OPE-anti-holomorphic}
\end{align}

\vspace{0.3cm}

\noindent ({\it iii}):
Combining the last two equations
we obtain the first OPE
listed in Eq.(\ref{LabelEq-Result-OPE-Coefficients}) with
$(-b)= - {\cal C}^{(2)}_{adj}$ 
upon making use of Eq. (\ref{LabelEq-Adjoint-Casimir}).

\subsubsection{Bulk-Boundary OPE}
\label{LabelSubsubsection-Bulk-Boundary-OPE}

In this section we discuss the renormalization of the boundary operator
${\Phi}_s(x)=$ $(\partial_z \varphi_L^A)_{x}$
whose 2-point function appears in Eq. \eqref{LabelEq-Boundary-2-point-function}, by computing its OPE with the bulk operator ${\Phi}$
appearing in the perturbation, Eqs. \eqref{LabelEq-InteractionOperator} and \eqref{LabelEqSmall-g-Expansion-Class-DIII-complex}. It proves convenient to place the boundary operator ${\Phi }_s$
at the origin $z=0$ on the real axis, as in the second of Eq. (\ref{LabelEq-Result-OPE-Coefficients}).
Using Wick's Theorem and Eq. (\ref{LabelEqFreeScalarCorrelators}) yields
\begin{eqnarray}
\nonumber
{\Phi}(z,z^*)
{\Phi}_s(z=0)
&=& \left (
[\varphi_L^{B_2}
(\partial_z\varphi_L^{A_1})]_z
(-1) C_{B_2 A_1 H}\ 
[\varphi_L^{A_2} (\partial_z \varphi_L^{B_1})]_{z^*}
C_{A_2 B_1 H}
\right )
\times
(\i\partial_z\varphi_L^E)_{z=0}
\\ \nonumber
&=&
{-\delta^{B_2 E}\over z}
[\varphi_L^{A_2} (\partial_z \varphi_L^{B_1})]_{z^*}
(\i\partial_z \varphi_L^{A_1})_{z=0}C_{B_2 A_1 H} C_{A_2 B_1 H} + ...
\\ \nonumber
&=&
{
-\delta^{B_2 E} \delta^{A_2 A_1}
\over z z^*}
(\i\partial_z \varphi_L^{B_1})_{z=0}
(-1) C_{E A_1 H} C_{A_1 B_1 H} + ...
\\
\label{LabelEq-BulkBoundaryOPE-Calculation}
&=&
{ {\cal C}^{(2)}_{adj}
\over
z z^*}
(\i\partial_z \varphi_L^E)_{z=0} + ...,
\end{eqnarray}
where in the last line we used the cyclic property  and the total asymmetry of the structure constants, together
with Eq. (\ref{LabelEq-Adjoint-Casimir}),
to obtain
$(-1) C_{E A_1 H} C_{A_1 B_1 H}=$
$
C_{E A_1 H} C_{B_1 A_1 H}=
$
${\cal C}^{(2)}_{adj} \ \delta_{E B_1}
$. Thus, in Eq. (\ref{LabelEq-BulkBoundaryOPE-Calculation}) we have obtained the result quoted in Eq.
(\ref{LabelEq-Result-OPE-Coefficients}) above.

\end{widetext}

\section{Numerical details}
\label{app:numerical-details}

In this section, we provide some additional detail regarding the numerical methods used to obtain the results in Sec.~\ref{sec:numerics}.

\subsection{Efficient contraction}

In the $v$'th step of the evolution, one needs to contract the Gaussian state $\Gamma_v$, defined on $Ld$ Majorana modes, with a row of $L$ tensors with $4\chi$ Majorana modes each, to obtain $\Gamma_{v+1}$. We will label these tensors $\Gamma^{(v,u=1)}$ through $\Gamma^{(v,u=L)}$. In principle, one could perform a sequential contraction, i.e. contract $\Gamma_v$ with $\Gamma^{(v,u=1)}$, then with $\Gamma^{(v,u=2)}$, etc. In each step, one needs to evaluate Eq.~\eqref{eqn:SchurContraction}, which will take $\mathcal{O}(L^2 d^3)$ operations in each step, and thus an entire row takes $\mathcal{O}(L^3 d^3)$.

An alternative contraction scheme is to first contract all the tensors $\Gamma^{(v,u)}$ with a fixed $v$ together to form the transfer operator, and then contract it with $\Gamma_v$ to obtain $\Gamma_{v+1}$ in one step. An efficient way to perform this contraction is in a ``tree-like" fashion (for a system size $L=2^m$): first contract $\Gamma^{(v,u=1)}$ and $\Gamma^{(v,u=2)}$ into $\Gamma^{(v,u=1:2)}$, and likewise for $\Gamma^{(v,u=2n-1)}$ and $\Gamma^{(v,u=2n)}$ into $\Gamma^{(v,u=2n-1:2n)}$. In the next step, contract $\Gamma^{(v,u=1:2)}$ and $\Gamma^{(v,u=3:4)}$ into $\Gamma^{(v,u=1:4)}$, and likewise for the rest of the system, and then repeating this procedure until one has the full tensor $\Gamma^{(v,u=1:L)}$ for the $v$'th row. There are $m = \log_2(L)$ such steps, and the $n$'th one, e.g. contracting $\Gamma^{(v,u=1:l)}$ with $\Gamma^{(v,u=l+1:2l)}$, for $l=2^n$, takes $\mathcal{O}(l^2 d^3) = \mathcal{O}(2^{2n} d^3)$ operations; therefore, all the operations to construct $\Gamma^{(v,u=1:L)}$ take $\sum_{n=0}^{m-1} 2^{m-n} 2^{2n} d^3 = \mathcal{O}(d^3 L^2)$ operations. The final step of contracting $\Gamma_v$ with $\Gamma^{(v,u=1:L)}$ is the most expensive one, taking $\mathcal{O}(L^3 d^3)$ operations.

Therefore, the scaling of the two approaches is asymptotically the same, $\mathcal{O}(L^3 d^3)$. However, in practice we find the second approach to be faster by a constant, yet significant factor.

\subsection{Numerical stabilization}

Due to numerical roundoff, the pure-state property $\Gamma_v^2 = -\openone$ can be destroyed after many layers of the contraction, i.e. for large $v$. To remove such numerical stability issues, we periodically reset the state to the closest pure state. To this end, we find the orthogonal transformation such that
\begin{equation}
\Gamma_v = O \left(\bigoplus \lambda_n
\begin{psmallmatrix}
0 &1 \\
-1 &0
\end{psmallmatrix} \right) O^\tr
\end{equation}
with $\lambda_n \geq 0$ and $O$ an orthogonal matrix~\cite{wimmer2012algorithm}. If $\Gamma_v$ is pure, $\lambda_n = 1$ for all $n$; however, in practice, we find that some $\lambda_n < 1$ by a small amount. In such cases, we replace $\Gamma_v$ by $\Gamma_v'$ defined by
\begin{equation}
\Gamma_v' = O \left(\bigoplus
\begin{psmallmatrix}
0 &1 \\
-1 &0
\end{psmallmatrix} \right) O^\tr,
\end{equation}
i.e. we set all $\lambda_n = 1$.

\section{Level statistics of the entanglement spectrum}
\label{sec:entanglement-spectrum}

In addition to the entanglement entropies, we can also examine the full entanglement spectrum of the state $\Gamma_v$ (as defined in Sec.~\ref{sec:numerics-setup}). We expect to find level statistics consistent with the Gaussian unitary ensemble; one way to see this is that the underlying non-unitary circuit that describes our ensemble has no symmetries (except for the particle-hole symmetry implied by the Majorana description), and in particular has no time-reversal symmetry (which would lead to the Gaussian orthogonal ensemble). In related work, the entanglement spectrum across the measurement-driven phase transition in interacting quantum circuits was studied in Ref.~\onlinecite{zhang2020nonuniversal}, where a non-universal distribution interpolating between Poisson and GUE was found.

Here, similar to the entanglement entropy, we partition the system into two halves and consider the covariance matrix of the left $Ld/2$ Majorana modes, which we will denote as 
$\Gamma_v^{\rm half}$. There exists an orthogonal transformation $O$ such that $\Gamma_v^{\rm half} = O \left( \bigoplus \lambda_n 
\begin{psmallmatrix}
0 & 1 \\ -1 & 0
\end{psmallmatrix} \right) O^\tr$, where $\lambda_n \in [0,1]$ is the entanglement spectrum. Modes with $\lambda_n=1$ correspond to unentangled modes, while modes with $\lambda_n=0$ are maximally entangled (i.e., contribute $\log 2$ to the entanglement entropy).

A convenient way to characterize the entanglement spectrum is through the ratios of consecutive gaps in the entanglement spectrum. Assuming that $\lambda_n \geq \lambda_{n-1}$, let $\delta_n = \lambda_n - \lambda_{n-1}$. Then, we define~\cite{Oganesyan07,atas2013distribution}
\begin{equation} \label{eqn:es-r}
\tilde{r} = \frac{\min (\delta_n, \delta_{n+1})}{\max (\delta_n, \delta_{n+1})}
\end{equation}
This quantity is well-characterized for several random matrix ensembles. In particular, for the Gaussian unitary ensemble, it is known to follow the Wigner surmise, or more precisely
\begin{equation} \label{eqn:gue-surmise}
    P(\tilde{r}) = \frac{1}{Z_\beta} \frac{(\tilde{r}+\tilde{r}^2)^\beta}{(1+\tilde{r}+\tilde{r}^2)^{1+3\beta/2}},
\end{equation}
with $\beta=2$ and $Z_\beta = \frac{4 \pi}{81 \sqrt{3}}$ (see Ref.~\onlinecite{atas2013distribution}, in particular for other ensembles).

\begin{figure}
    \centering
    \includegraphics{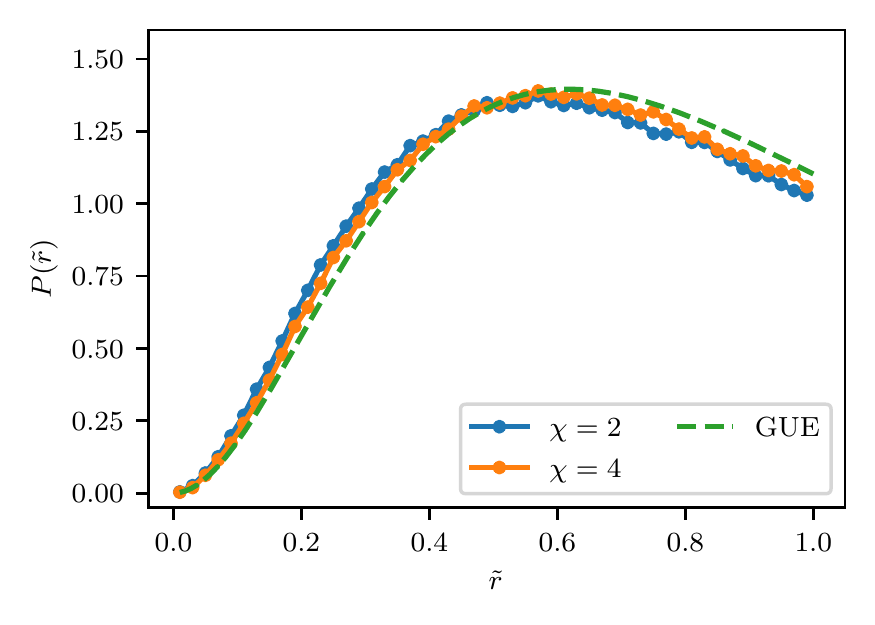}
    \caption{
    Probability distribution of the ratio of consecutive entanglement spectrum gaps, see Eq.~\eqref{eqn:es-r}, for system size $L=256$. The green line shows the expected distribution for the Gaussian unitary ensemble (GUE), Eq.~\eqref{eqn:gue-surmise}.}
    \label{fig:ES-levelstat}
\end{figure}

To perform a numerical comparison, we (similar to the computation of the entanglement entropy) take $\Gamma_v$ for $v > 20$, and average over 100 independent runs. To remove non-universal contributions to the entanglement spectrum, we exclude all $\lambda_n > 0.75$, and then perform the analysis outlined above. Our results are shown in Fig.~\ref{fig:ES-levelstat}. We find excellent agreement between our numerical observation and the Gaussian unitary ensemble. The agreement improves as the system size and bond number are increased.

\section{Relation between Haar-random GTN and quantum dynamics with unitary evolution and generalize measurements}
\label{app:random_GTN_vs_MeasurementProblem}
As discussed Sec.~\ref{sec:UnitaryTimeEvolutionAndMeasurements}, for a physical system undergoing unitary evolution and generalized measurements, the ensemble of different quantum trajectories of the physical system is characterized by an ensemble of quantum operators $\{C_m\}$ (acting on the physical system) that satisfies the normalization condition of the positive operator-valued measure, i.e. $\sum_m C^\dag_m C_m = \mathds{1}$. This normalization condition is equivalent to the trace-preserving condition, i.e. that the map from any density operator $\hat{\rho}$ (with $\Tr \hat{\rho}  = 1$) to the operator $\sum_m C_m \hat{\rho} C_m^\dag$ preserves the operator trace, i.e. $\Tr \left(\sum_m C_m \hat{\rho} C_m^\dag\right) = \Tr \hat{\rho} = 1$. Here, each term $C_m^\dag \hrho C_m$ in the summation should be viewed as the un-normalized density operator obtained from the circuit $C_m$ acting on the initial density operator $\hrho$. The Haar-random ensemble of square-lattice pure-state GTNs introduced in Sec.~\ref{sec:numerics} can be viewed as an ensemble of quantum circuits following the correspondence between GTNs and quantum circuits discussed in Sec.~\ref{sec:TN_QC_relation}. In this appendix, we show that the ensemble of quantum circuits obtained from the Haar-random square-lattice pure-state GTN satisfies the trace-preserving condition discussed above
and hence can be interpreted as the operator ensemble that governs the dynamics of a quantum system undergoing both unitary evolution and generalized measurements.

Following the discussion in Sec.~\ref{sec:TN_QC_relation}, a square-lattice pure-state GTN with Majorana bond number $\chi$ on each leg can be viewed as a quantum circuit that acts on a chain along the $x$-direction with $\chi$ Majorana modes on each site (as shown in Fig.~\ref{fig:circuit}~(a)). This quantum circuit evolves a generic initial state $|\psi_0\rangle$, which is not necessarily a Gaussian state, to the (un-normalized) final state $P(
|\psi_0\rangle \otimes |\Gamma_{\rm tot}\rangle)$, where $|\Gamma_{\rm tot}\rangle \equiv \otimes_n |\tG{n}\rangle$ is the tensor product of all the Gaussian states $|\tG{n}\rangle$ associated with each Gaussian tensors $\tG{n}$ in the pure-state GTN and $P$ is the projection operator that projects the Majorana modes on the contracted legs into maximally-entangled-pair states. The projection given by $P$ essentially implements all the contractions in the whole tensor network as discussed in Sec.~\ref{sec:TN_QC_relation}. Here, we've viewed the initial state $|\psi_0\rangle $ also as a generic (non-Gaussian) tensor which is contracted with the GTN. The contractions within the GTN and the contraction between the GTN and the initial state $|\psi_0\rangle $ yield the final state $P(
|\psi_0\rangle \otimes |\Gamma_{\rm tot}\rangle)$. 

At the level of quantum state, the quantum circuit given by the GTN transforms an initial state $|\psi_0\rangle$ into the final state $P(
|\psi_0\rangle \otimes |\Gamma_{\rm tot}\rangle)$. 
At the level of density operators, the same quantum circuit transforms the initial density operator $\hrho_0 \equiv \ket{\psi_0}\bra{\psi_0}$ into an (un-normalized) density operator $P\left(\hrho_0 \otimes \ket{\Gamma_{\rm tot}}\bra{\Gamma_{\rm tot}} \right) P$. In fact, for any initial (pure-state or mixed-state) density operator $\hrho_0$ (with $\Tr\hrho_0 = 1$), the same quantum circuit transforms it into the (un-normalized) density operator $P\left(\hrho_0 \otimes \ket{\Gamma_{\rm tot}}\bra{\Gamma_{\rm tot}} \right) P$. 

To show that the ensemble of quantum circuit given by the Haar-random ensemble of GTNs satisfies the tracing-preserving condition, we need to show that $\Tr\left(\frac{1}{\mathcal{N}}\sum_{\{\tG{n}\}} P\left(\hrho_0 \otimes \ket{\Gamma_{\rm tot}}\bra{\Gamma_{\rm tot}} \right) P \right) = 1$ for any initial density operator $\hrho_0$. Here $\mathcal{N}$ is an overall normalization constant and the summation $\sum_{\{\tG{n}\}}$ is the summation over the fermion-parity-preserving Haar-random ensemble defined in Sec.~\ref{sec:numerics} for each individual Gaussian tensor $\tG{n}$ in the GTN. Since every Gaussian tensor in the GTN is independently random, we can perform the summation over the Haar random ensemble independently for each pure-state Gaussian tensor $\tG{n}$ in the GTN. For a single pure-state Gaussian tensor $\tG{n}$, the summation over the fermion-parity-preserving Haar-random ensemble yields 
\begin{align}
    \sum_{\tG{n}} \ket{\tG{n}} \bra{\tG{n}} = \frac{1 + (-1)^{\hat{F}_n}}{2},
    \label{eqn:Haar_Average_Single_Site}
\end{align}
where $(-1)^{\hat{F}_n}$ is the many-body fermion parity operator of all the Majorana modes residing on the tensor $\tG{n}$. The right hand side of Eq.~\eqref{eqn:Haar_Average_Single_Site} is exactly the many-body fermion-parity projection operator acting on the local fermionic Hilbert space associated with the tensor $\tG{n}$.

Applying Eq.~\eqref{eqn:Haar_Average_Single_Site} to every tensor in the GTN, we can obtain that
\begin{align}
    &\Tr\left(\frac{1}{\mathcal{N}}\sum_{\{\tG{n}\}} P\left(\hrho_0 \otimes \ket{\Gamma_{\rm tot}}\bra{\Gamma_{\rm tot}} \right) P \right) 
    \nonumber \\
    & = 
   \frac{1}{\mathcal{N}} \Tr\left\{ P\left[\hrho_0 \otimes \left(\otimes_n \frac{1 + (-1)^{\hat{F}_n}}{2} \right)\right] P \right\}.
\end{align}
By expanding all the operators on the second line of this equation using Majorana modes in the tensor network, we can show that, with a properly chosen constant $\mathcal{N}$ that is independent of $\hrho_0$, the expression on the second line of the equation above always evaluates to $1$. Therefore, the ensemble of random quantum circuits obtained from the Haar-random ensemble of square-lattice pure-state GTN defined in Sec.~\ref{sec:numerics} satisfies the trace-preserving condition and, hence, can be associated with the dynamics of a quantum system whose dynamics are governed by unitary evolution and generalized measurements.

In the discussion of the trace-preserving condition above, each square-lattice GTN is treated as a quantum circuit acting on a one-dimensional fermion chain along the $x$-direction (as shown in Fig.~\ref{fig:circuit}~(a)). In fact, we can also alternatively treat the same square-lattice pure-state GTN as a different quantum circuit acting on a one-dimensional fermion chain along the $u$-direction (as shown in Fig.~\ref{fig:Numerics_Illutstration}). The ``physical time" for this alternative type of quantum circuit is along the $v$-direction of the GTN. Under this alternative treatment, the Haar-random ensemble of the square-lattice pure-state GTNs gives rise to a different random ensemble of quantum circuit which also satisfies the trace-preserving condition. The proof of the trace-preserving condition for this different ensemble of quantum circuits is completely parallel to the discussions given in the earlier paragraphs of this appendix. Therefore, when the Haar-random ensemble of square-lattice pure-state GTN is viewed as a random ensemble of quantum circuits acting on a fermion chain along the $u$-direction, this ensemble of quantum circuit can also describe the dynamics of this fermion chain induced by both unitary evolution and generalized measurements.

The discussion above shows that the ensemble of quantum circuits obtained from the Haar-random square-lattice GTN can be used to described the dynamics of a fermion chain undergoing both the unitary evolution and generalized measurements. However, one needs to be careful that, within the ensemble of quantum circuits, the probability of each quantum circuit to appear in this fermion chain system whose non-unitarity is solely due to generalized measurements should follow Born's rule as discussed in Sec.~\ref{sec:UnitaryTimeEvolutionAndMeasurements}. Hence, generically, the probabilities for different quantum circuits to appear are different from one another. In contrast, in the problem of the random ensemble of GTNs studied in the main text, each GTN and its corresponding quantum circuit appears with the same probability. Therefore, the problem of Haar-random GTNs is not exactly equivalent to the problem of a quantum system undergoing both unitary evolution and generalized measurements. What effect this difference has on corresponding universal critical behavior requires further investigation.

\section{Transfer Matrix of Pure-State GTN}
\label{app:TranferM_pGTN_Derivation}
As stated in the Sec.~\ref{sec:TransferM_Intro}, a two-leg pure-state Gaussian tensor $\Gamma$ (as shown in~\ref{fig:1dGTN}~(b) for instance) with Majorana bond number $\chit$ can be viewed as a quantum gate $g_\Gamma$ acting on the Hilbert space of the $\chit$ Majorana fermion $\hat{\alpha}_{i=1,2,...,\chit}$. The P-sector transfer matrix $\ft_p[\Gamma]$ is defined via $\hat{\alpha}_i \rightarrow g_\Gamma \hat{\alpha}_i g_\Gamma^{-1} = \sum_j \ft_p[\Gamma]_{ij}\hat{\alpha}_j$, namely the evolution of the Majorana fermion operators $\hat{\alpha}_i$ under the quantum gate $g_\Gamma$. We will derive the expression of the P-sector transfer matrix $\ft_p[\Gamma]$ shown in Eq.~\eqref{eqn:transferM_Def} in this appendix. We will also introduce the H-sector transfer matrix $\ft_h[\Gamma]$ and the full transfer matrix $\ft[\Gamma]$. In particular, we will discuss a formalism where the P-sector and the H-sector transfer matrices are treated in equal footing and where the full transfer matrix naturally arises. 

\begin{figure}
    \centering
    \includegraphics[width=1\columnwidth]{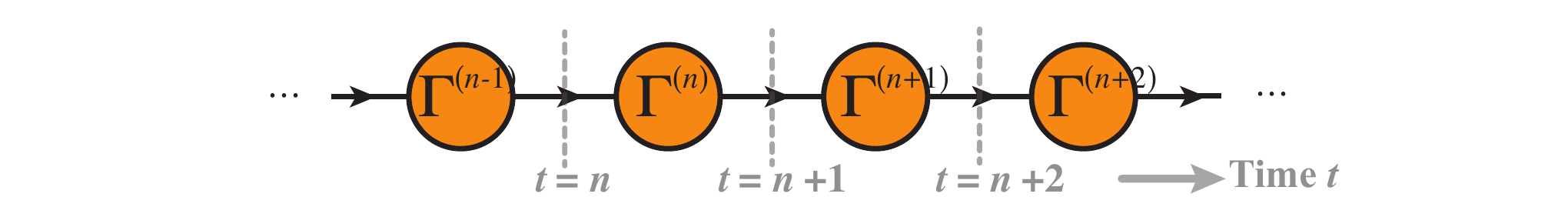}
    \caption{A one-dimensional GTN consists of tensors $\{\Gamma^{(n)}\}$ is shown. The tensor $\Gamma^{(n)}$ of the GTN is located in between the coordinates $t=n-1$ and $t=n$. }
    \label{fig:1dGTN_app}
\end{figure}
 
Consider a one-dimensional pure-state Gaussian tensor network of the form shown in Fig.~\ref{fig:1dGTN_app}. This GTN also describes the evolution of a quantum system under the quantum circuit $\prod_n g_{\tG{n}}$ where $g_{\tG{n}}$ is the quantum gate associated with the tensor $\tG{n}$. Furthermore, as indicated in Fig.~\ref{fig:1dGTN_app}, we can view $g_{\tG{n}}$ as the quantum gate that evolves the system from the (discrete) time step $t=n$ to the time step $t=n+1$. 

Let's denote the Majorana bond number of each leg of this one-dimensional pure-state GTN as $\chit$. Each two-leg pure-state Gaussian tensor $\Gamma^{(n)}$ here is associated with a Gaussian state $|\Gamma^{(n)}\rangle$. We organize the $\chit$ Majorana modes residing on the left (right) leg of the tensor $\tG{n}$ into a $\chit$-component column vector $\hgamma^{(n)}_{L,i=1,2,...,\chit}$ ($\hgamma_{R,i=1,2,...,\chit}^{(n)}$). 
The whole one-dimensional pure-state GTN shown in Fig.~\ref{fig:1dGTN_app} yields the Gaussian state $P |\Gamma_{\rm tot}\rangle $ where $\Gamma_{\rm tot} \equiv \otimes_n |\tG{n}\rangle$ is the tensor product of all the independent Gaussian states $|\tG{n}\rangle$ and $P \equiv \prod_n P_{n,n+1}$ is the product of projections $P_{n,n+1} \equiv \prod_{i=1}^{\chit} \frac{1 + \i \hgamma^{(n)}_{R,i} \hgamma^{(n+1)}_{L,i} }{2}$ that implement the contractions between the tensors $\tG{n}$ and $\tG{n+1}$. As mentioned earlier, the Gaussian state $P |\Gamma_{\rm tot}\rangle $, which represents the GTN shown in Fig.~\ref{fig:1dGTN_app}, captures the quantum evolution given by the product of quantum gates $\prod_n g_{\tG{n}}$. As a generalization, for the same quantum evolution but with an extra Majorana fermion operator inserted at the discrete time $t = n$, the representing Gaussian state can be given by $P\hgamma^{(n)}_L |\Gamma_{\rm tot}\rangle $ instead. Here, since $-\i P\hgamma^{(n)}_L =  P\hgamma^{(n-1)}_R$, there is no need to separately consider the Gaussian state of the form $P\hgamma^{(n)}_R |\Gamma_{\rm tot}\rangle $. The P-sector transfer matrix $\ft_p[\tG{n}]$, which is the single-particle version of the gate $g_{\tG{n}}$, can therefore be equivalently defined by the linear relation between $P\hgamma^{(n)}_L |\Gamma_{\rm tot}\rangle$ and $P\hgamma^{(n+1)}_L |\Gamma_{\rm tot}\rangle$, i.e. $\ft_p[\tG{n}] P\hgamma^{(n)}_L |\Gamma_{\rm tot}\rangle =  P\hgamma^{(n+1)}_L |\Gamma_{\rm tot}\rangle$ (or equivalently $\sum_{j} \ft_p[\tG{n}]_{ij} P\hgamma^{(n)}_{L,j} |\Gamma_{\rm tot}\rangle =  P\hgamma^{(n+1)}_{L,i} |\Gamma_{\rm tot}\rangle$). Here, $P\hgamma^{(n)}_L |\Gamma_{\rm tot}\rangle$ is viewed as a $\chit$-component vector of ket states where the components are given by $P\hgamma^{(n)}_{L,i} |\Gamma_{\rm tot}\rangle$ with $i=1,2,...,\chit$.

Now, we derive the expression of the P-sector transfer matrix $\ft_p[\Gamma^{(n)}]$. Similar to our treatment in Sec.~\ref{sec:TransferM_Intro}, we can write each of the covariance matrices
$\tG{n}$ in block matrix form
$\begin{psmallmatrix}
\tG{n}_{LL} & \tG{n}_{LR} \\ \tG{n}_{RL} & \tG{n}_{RR}  
\end{psmallmatrix}$  with each block a $d\times d$ matrix. The subscripts of each block indicate the type of correlation it captures. For example, the block $\tG{n}_{LR}$ captures the correlation between Majorana modes $\hgamma_L^{(n)}$ and $\hgamma_R^{(n)}$. The block-matrix form of Eq.~\eqref{eqn:GS_Def_Prop} applied to the Gaussian state $|\tG{n}\rangle$ reads
\begin{align}
\left[ 
\left(\begin{array}{c}
\hgamma_L^{(n)}       \\ \hgamma_R^{(n)}
\end{array}\right) - \i \left( \begin{array}{cc}
\tG{n}_{LL} & \tG{n}_{LR} \\ \tG{n}_{RL} & \tG{n}_{RR}  
\end{array}\right)
\left(\begin{array}{c}
\hgamma_L^{(n)}       \\ \hgamma_R^{(n)} 
\end{array}\right)
\right]|\tG{n} \rangle = 0.
\label{eqn:GS_Def_prop_Blocks}
\end{align}
Based on the first row of this block-matrix equation, we obtain that 
\begin{align}
    & (\tG{n}_{LR})^{-1} (\mathds{1}- \i \tG{n}_{LL} )  \left(P \hgamma^{(n)}_L |\Gamma_{\rm tot}\rangle \right) 
    \nonumber \\
    & = \i \left(P \hgamma^{(n)}_R |\Gamma_{\rm tot}\rangle \right)
     = \left(P \hgamma^{(n+1)}_L |\Gamma_{\rm tot}\rangle \right).
    \label{eqn:transferM_p_action}
\end{align}
Hence, we conclude that 
\begin{align}
    \ft_p[\tG{n}] = (\tG{n}_{LR})^{-1} (\mathds{1}- \i \tG{n}_{LL} ),
\end{align}
which is exactly the result shown in Eq.~\eqref{eqn:transferM_Def}. Since the contraction of two neighboring pure-state Gaussian tensors captures the multiplication of their corresponding quantum gates, the same tensor contraction gives rise to the multiplication of their corresponding P-sector transfer matrices as shown by Eq.~\eqref{eqn:transferM_Multiply}.

The quantum circuit $\prod_n g_{\tG{n}}$ evolves an initial ket state $|\psi_i\rangle$ by $|\psi_i\rangle \rightarrow \left(\prod_n g_{\tG{n}}\right) |\psi_i\rangle $. In the same time, the initial bra state $\bra{\psi_i}$ evolves as $\bra{\psi_i} \rightarrow \bra{\psi_i} \left(\prod_n g_{\tG{n}}\right)^\dag$. In the tensor network language, the bra-state evolution is given by the (bra) Gaussian state $\bra{\Gamma_{\rm tot}}P $. Similarly, the bra state evolution with a Majorana fermion operator insertion at the time step $t=n$ is given by the $\chit$-component vector of the Gaussian bra states $\bra{\Gamma_{\rm tot}} \hgamma_{L}^{(n)} P$. The H-sector transfer matrix $\ft_h[\tG{n}]$ is defined as the linear relation:
\begin{align}
        \sum_{j} \ft_h[\tG{n}]_{ij} \bra{\Gamma_{\rm tot}} \hgamma_{L,j}^{(n)} P = \bra{\Gamma_{\rm tot}} \hgamma_{L,i}^{(n+1)} P,
        \label{eqn:transferM_h_action}
\end{align}
which the single-particle version of the quantum gate $g_{\tG{n}}$ as the evolution of a bra state. We can easily obtain that
\begin{align}
    \ft_h[\tG{n}] = \ft_p[\tG{n}]^*.
\end{align}

In fact, even though the P-sector and the H-sector transfer matrices describe the ket-state and bra-state respectively, they can be treated in a unified way. Notice that Eq.~\eqref{eqn:transferM_p_action} still holds when we replace the $\chit$-component vector of ket states $P \hgamma^{(n)}_{L/R} |\Gamma_{\rm tot}\rangle$ by the $\chit$-component vector of operators $P \hgamma^{(n)}_{L/R} \ket{\Gamma_{\rm tot}}  \bra{\Gamma_{\rm tot}} P$. Similarly, Eq.~\eqref{eqn:transferM_h_action}
still holds when we replace the vector of bra states $\bra{\Gamma_{\rm tot}} \hgamma_{L}^{(n)} P$ by the vector of operators $P  \ket{\Gamma_{\rm tot}}  \bra{\Gamma_{\rm tot}} \hgamma^{(n)}_{L} P$. Using the operator version of Eq.~\eqref{eqn:transferM_p_action} and Eq.~\eqref{eqn:transferM_h_action}, we can obtain the operator relation associated with the full transfer matrix $\ft[\tG{n}] $:
\begin{multline}
    \ft[\tG{n}] 
    \left(
    \begin{array}{c}
     P \hgamma^{(n)}_L |\Gamma_{\rm tot}\rangle\langle \Gamma_{\rm tot}| P \\ P |\Gamma_{\rm tot}\rangle\langle \Gamma_{\rm tot}| \hgamma^{(n)}_L P
    \end{array}
    \right) \\
    =
    \left(
    \begin{array}{c}
     P \hgamma^{(n+1)}_L |\Gamma_{\rm tot}\rangle\langle \Gamma_{\rm tot}| P \\ P |\Gamma_{\rm tot}\rangle\langle \Gamma_{\rm tot}| \hgamma^{(n+1)}_L P
    \end{array}
    \right)
    \label{eqn:transferM_full_action}.
\end{multline}
From the perspective of this operator relation, the decoupling of the full transfer matrix $\ft$ into the P-sector transfer matrix $\ft_p$ and the H-sector transfer matrix $\ft_h$ is the natural consequence of the fact that ket states and bra states do not mix under the quantum circuit evolution given by the product $\prod_n g_{\tG{n}}$ of quantum gates. The decoupling between the P sector and the H sector in the full transfer matrix $\ft$ directly results in the chiral symmetry Eq.~\eqref{eqn:full_transferM_chiral_sym} of $\ft$. Interestingly, the P sector and the H sector are not just simply decoupled, they are also related by the TR and the PH symmetries as shown in Eq.~\eqref{eqn:full_transferM_TR_sym} and Eq.~\eqref{eqn:full_transferM_PH_sym}.

In App.~\ref{app:TranferM_mGTN_Derivation}, we will see that Eq.~\eqref{eqn:transferM_full_action} can be generalized to one-dimensional GTNs that involves mixed-state tensors $\Gamma$ with $\Gamma^2 \neq \mathds{1}$. Therefore, the full transfer matrix $\ft[\Gamma]$ can be defined even for mixed-state Gaussian tensor $\Gamma$. However, the contraction with a mixed-state Gaussian tensor $\Gamma$ cannot be described as an evolution induced by a single quantum gate. Therefore, there is no decoupling between the P sector and the H sector in a mixed-state GTN. In other words, in a mixed-state GTN, the P-sector transfer matrix $\ft_p$ and the H-sector transfer matrix $\ft_h$ become ill-defined while the full transfer matrix $\ft$ still remains a valid notion. The absence of such decoupling between the P and the H sector changes the symmetry class of the full transfer matrix $\ft$ when  $\ft$ is interpreted as the transfer matrix in a unitary scattering problem (with a static Hamiltonian).

\section{Mixed-state GTN and its transfer matrix}
\label{app:TranferM_mGTN_Derivation}
In this appendix, we discuss the mixed-state GTN and its transfer matrix. A mixed-state tensor $\Gamma$ is fully characterized by its covariance matrix $\Gamma$ with the conditions that (1) $\Gamma^\T =-\Gamma$, (2) $\Gamma^* =\Gamma$, and (3) $\Gamma^2 \succeq -\mathds{1}$, namely no eigenvalues of $\Gamma^2$ are smaller than $-1$. For a mixed-state tensor $\Gamma$ with Majorana modes $\hgamma_i$, we can represent $\Gamma$ by a Gaussian density matrix $\hrho_{\Gamma}$ such that 
\begin{align}
\Gamma_{ij} = \Tr\left( {\frac{\i}{2} [\hgamma_i, \hgamma_j] \hrho_{\Gamma}} \right).
\end{align}
A Gaussian density matrix $\hrho_{\Gamma}$ is completely determined by its two-point correlation functions, namely its covariance matrix $\Gamma_{ij}$. All multi-point correlation functions can be obtained from the two-point functions via Wick's theorem. Eq.~\eqref{eqn:GS_Def_Prop} which is applicable for a pure-state tensor can be generalized to the case of a mixed-state Gaussian tensor:
\begin{align}
\left(\hgamma_i - \i \sum_j \Gamma_{ij} \hgamma_j \right) \hrho_{\Gamma} = \hrho_{\Gamma} \left(\hgamma_i + \i \sum_{j'} \Gamma_{ij'} \hgamma_{j'} \right)  ,
\label{eqn:GS_Def_Prop_mixed}
\end{align}
which can be viewed as the defining relation of the Gaussian density matrix $\hrho_{\Gamma}$ based on the covariance matrix $\Gamma_{ij}$. 

The contraction of a mixed-state GTN can be described using the Gaussian density matrices. Consider a mixed-state GTN with the set of tensors $\{\tG{n}\}$. The contraction of these tensors in the GTN produces a new Gaussian density matrix $P\left(\otimes_n \hrho_{\tG{n}}\right) P$ where $P$, as introduced in Sec.~\ref{sec:gtn_contraction} for the case of pure-state GTNs, is still the projection onto the maximally-entangled-pair states on all of the contracted legs in the GTN. Similar to the pure-state case, we can study the contraction of mixed-state Gaussian tensors directly at the level of covariance matrices. In Sec.~\ref{sec:gtn_contraction}, we've discussed the contraction of the two Gaussian tensors $\Gamma$ and $\Upsilon$ that are of the forms shown in Eq.~\eqref{eqn:Gamma_Upsilon} and in the configuration shown in Fig.~\ref{fig:FermionGTN_Contraction}. Their contraction gives rise to a new Gaussian tensor/covariance matrix $\Psi$ shown in Eq.~\eqref{eqn:SchurContraction}. In fact, Eq.~\eqref{eqn:SchurContraction} holds even if $\Gamma$ and $\Upsilon$ are mixed-state Gaussian tensors and Eq.~\eqref{eqn:SchurContraction} is consistent with the formulation of tensor contractions using Gaussian density matrices and the projection onto maximally-entangled-pair states on the contracted legs. 

A one-dimensional mixed-state GTN that takes the same geometry as the one shown Fig.~\ref{fig:1dGTN_app} can be represented by the Gaussian density matrix $P\left(\otimes_n \hrho_{\tG{n}}\right) P$. In this GTN, similar to the discussion in App.~\ref{app:TranferM_pGTN_Derivation}, each tensor $\Gamma^{(n)}$ has two legs with the Majorana modes residing on them denoted as $\hgamma^{(n)}_L$ and $\hgamma^{(n)}_R$ respectively. This mixed-state GTN can no longer be interpreted as the product a sequence of quantum gates. We can still generalize Eq.~\eqref{eqn:transferM_full_action} and define the full transfer matrix $\ft[\tG{n}]$ of the Gaussian tensor $\tG{n}$ by the linear relation
\begin{align}
    \ft[\tG{n}] 
    \left(
    \begin{array}{c}
     P \hgamma^{(n)}_L \hrho_{\rm tot} P \\
     P \hrho_{\rm tot} \hgamma^{(n)}_L P.
    \end{array}
    \right)
    = 
    \left(
    \begin{array}{c}
     P \hgamma^{(n+1)}_L \hrho_{\rm tot} P \\
     P \hrho_{\rm tot} \hgamma^{(n+1)}_L P.
    \end{array}
    \right)
    \label{eqn:transferM_full_action_mixed},
\end{align}
where $\hrho_{\rm tot} \equiv \otimes_n \hrho_{\tG{n}}$. In the special case when $\hrho_{\rm tot}$ is the tensor product of pure Gaussian state density matrices $\hrho_{\tG{n}}$,  the whole GTN becomes a pure-state GTN and Eq.~\eqref{eqn:transferM_full_action} is immediately restored from Eq.~\eqref{eqn:transferM_full_action_mixed}. Without assuming the purity of the tensors, we can obtain the general expression of the transfer matrix $\ft[\Gamma]$ for a two-leg Gaussian tensor $\Gamma$ using Eq.~\eqref{eqn:GS_Def_Prop_mixed} and the property of the projection operator $P$:
\begin{widetext}
\begin{align}
    \ft[\Gamma] 
=
\left(
\begin{array}{cc}
    \frac{1}{\sqrt{2}} \mathds{1} & \frac{1}{\sqrt{2}} \mathds{1} \\
    \frac{\i}{\sqrt{2}} \mathds{1} & \frac{-\i}{\sqrt{2}} \mathds{1}
\end{array}
\right)^\dag 
\left(
\begin{array}{cc}
 \Gamma_{LR}^{-1} & - \Gamma_{LR}^{-1}  \Gamma_{LL} \\
 -\Gamma_{RR}\Gamma_{LR}^{-1}  & ~~~ \Gamma_{RR} \Gamma_{LR}^{-1} \Gamma_{LL} - \Gamma_{RL}\\
\end{array}
\right)
\left(
\begin{array}{cc}
    \frac{1}{\sqrt{2}} \mathds{1} & \frac{1}{\sqrt{2}}\mathds{1}\\
    \frac{\i}{\sqrt{2}} \mathds{1} & \frac{-\i}{\sqrt{2}} \mathds{1}
\end{array}
\right),
\label{eqn:TransferM_Def_full_mixed}
\end{align}
\end{widetext}
where we have used the block-matrix form $\Gamma= \begin{psmallmatrix}
\Gamma_{LL} & \Gamma_{LR} \\ \Gamma_{RL} & \Gamma_{RR}  
\end{psmallmatrix}$ of the two-leg tensor $\Gamma$ in the one-dimensional GTN like what we've done in App.~\ref{app:TranferM_pGTN_Derivation}. 

We can check that for any covariance matrix $\Gamma$, the following relation always holds
\begin{align}
    \ft[\Gamma]^\dag \cdot \Jp' \cdot \ft[\Gamma] = \Jp',
    \label{eqn:mGTN_TransM_Jconserve}
\end{align}
where 
\begin{align}
    \Jp' = 
    \left( 
    \begin{array}{cc}
    0 & -\i \mathds{1} \\
    \i \mathds{1} & 0 
    \end{array}
    \label{eqn:mGTN_J}
    \right).
\end{align}
When we identify the full transfer matrix $\ft[\Gamma]$ of a generic (mixed-state or pure-state) Gaussian tensor $\Gamma$ as the transfer matrix in a unitary scattering problem (with a static Hamiltonian), Eq.~\eqref{eqn:mGTN_TransM_Jconserve} should be viewed as the conservation probability current. We emphasize that the conservation of the probability current is exactly what guarantees the unitarity of the corresponding scattering problem. Note the probability current $\Jp'$ defined in Eq.~\eqref{eqn:mGTN_J} is different from the probability current $\Jp$ in Eq.~\eqref{eqn:DIII_conserved_current} which is defined for the pure-state GTN in the main text. In fact, Eq.~\eqref{eqn:mGTN_TransM_Jconserve} holds for {\it any}
two-leg tensor $\Gamma$ regardless of its purity, while the current $\Jp$ defined in Eq.~\eqref{eqn:DIII_conserved_current} is 
conserved only in the case of a pure-state GTN.
The reason that we've chosen $\Jp$ defined in Eq. \eqref{eqn:DIII_conserved_current} instead of $\Jp'$ defined above for the pure-state GTN is that the probability current $\Jp$ further enables us to identify at the microscopic level the absorbing boundary condition, which was introduced in App. \ref{app:CC_intro} to understand the form of the averaged squared two-point function in  Eq.~ \eqref{eqn:Cd} of the Chalker-Coddington network model that corresponds to the pure-state GTN. In the current context of mixed-state GTNs, the only conserved probability current is given by $\Jp'$ in Eq.~\eqref{eqn:mGTN_J}, which is enough for
the justification of the unitarity of the scattering problem defined by the full transfer matrix $\ft[\Gamma]$ (regardless of the purity of the tensor $\Gamma$). 

With the conserved probability current defined by $\Jp'$, one can calculate the corresponding scattering $S$-matrix of the scattering problem that corresponds to the (mixed-state or pure-state) Gaussian tensor $\Gamma$. The $S$-matrix expression given in
Eq.~\eqref{eqn:1d_scacttering_Smatrix} will no longer hold as we have used a different definition of the probability current.
Nevertheless, the scattering $S$-matrix can be shown to be unitary even for a general mixed-state tensor $\Gamma$.

Also, for any covariance matrix $\Gamma$, the unitary scattering problem with its transfer matrix given by $\ft[\Gamma]$  always has a PH symmetry:
\begin{align}
    \Xi_{ph} \cdot \ft[\Gamma]^* \cdot \Xi_{ph} = \ft[\Gamma]
\end{align}
with 
\begin{align}
    \Xi_{ph} = 
    \left( 
    \begin{array}{cc}
    0 &  \mathds{1} \\
     \mathds{1} & 0 
    \end{array}
    \right).
\end{align}
Here, we note that the PH symmetry exchanges the P-sector and the H-sector of the transfer matrix. Also, we have $\Xi_{ph}^2 = 1$. For a generic Gaussian tensor $\Gamma$, the unitary scattering problem with its transfer matrix given by $\ft[\Gamma]$ does not have any other extra symmetries and therefore should correspond to symmetry class D in the Altland-Zirnbauer ten-fold classification. It is straightforward to see that under matrix multiplication, the group of all full transfer matrices $\ft[\Gamma]$ obtained from all (pure- and mixed-state) covariance matrices $\Gamma$ forms a subgroup of $\SO(\chit,\chit)$ where $\chit$ is the Majorana bond number of each of the legs of the two-leg Gaussian tensor $\Gamma$. As is shown in Ref. \onlinecite{SchnyderRyuFurusakiLudwig}, the space of transfer matrices in a class-D unitary scattering problem is indeed given by $\SO(\chit,\chit)$. As a real
manifold, this subgroup formed by $\ft[\Gamma]$ has the same dimension as the group manifold $\SO(\chit,\chit)$. However, not every element of $\SO(\chit,\chit)$ corresponds to a physical covariance matrix. In fact, any group element in $\SO(\chit,\chit)$ corresponds to a skew-symmetry matrix $\Gamma$ via Eq.~\eqref{eqn:TransferM_Def_full_mixed}. However, it is not guaranteed that so-obtained $\Gamma$ satisfies the physical condition (of a covariance matrix) that no eigenvalues of $\Gamma^2$ are smaller than $-1$. The set of group elements in $\SO(\chit,\chit)$ that does not correspond to a physical covariance matrix in fact has a finite measure in the non-compact group manifold $\SO(\chit,\chit)$.

\section{Random GTN and Clifford algebra extension problem with positive generators}
\label{app:clifford_extension}
In Sec.~\ref{sec:real_classes_general_construction}, we've discussed how to utilize the Clifford algebra extension problem with negative generators to restrict the space of permissible pure-state Gaussian tensor $\Gamma$ so that it matches the desired symmetric space $R_p$.
In this appendix, we provide an alternative procedure motivated by the Clifford algebra extension problem with positive generators. 

To realize $R_p$ as the space of permissible pure-state Gaussian tensor $\Gamma$, we can start with an positive integer $p$. Then, we write down $p$ operators $\Lambda'_{i=1,2,...,p}$ in the real matrix algebra such that 
\begin{align}
    & {\Lambda'_{i}}^2 =  \mathds{1},~~~~~~{\Lambda_{i}'}^{\T} =  \Lambda'_{i}~~~~~~{\rm for}~i=1,2,...,p, \nonumber \\ 
    & \Lambda'_{i}\Lambda'_{j} = -\Lambda'_{j}\Lambda'_{i}~~~~~~~~~~~~~~~~~~ ~{\rm for}~i\neq j.
\end{align}
The operators $\Lambda_{1,2,..,p}'$ generate the real Clifford algebra $\CLR{p,0}$ with $p$ positive generators. The conditions we impose on the Gaussian tensor $\Gamma$ are given by
\begin{align}
    & \Gamma \, \Lambda_1' =  - \Lambda_1'  \, \Gamma \nonumber \\
    & \Gamma \, \Lambda'_{i} = \Lambda'_{i} \, \Gamma,~~~ ~{\rm for}~i=2,3,...p.
    \label{eqn:all_real_symmetry_class_constraints2}
\end{align}
We notice that the operator $\Gamma \Lambda_1'$ is a real and symmetric operator that squares to $\mathds{1}$. Therefore, the operators $\Lambda_{1,2,..,p}'$ together with the operator $\Gamma \Lambda_1'$ generate the real Clifford algebra $\CLR{p+1,0}$ with $p+1$ positive generators. Therefore, the space of all pure-state Gaussian tensors $\Gamma$ that satisfy the conditions Eq.~\eqref{eqn:all_real_symmetry_class_constraints2} is the same as the classifying space of extensions from the real Clifford algebra $\CLR{p,0}$ to $\CLR{p+1,0}$, which is given by $R_p$. The case with $p=1$ which corresponds to symmetry class D was discussed earlier in greater detail in Sec. \ref{sec:class_D}.

\bibliography{grtn}

\end{document}